\documentclass[preprint2]{pp7}
\usepackage{natbib,color}
\usepackage{xcolor}
\usepackage[switch]{lineno}
\definecolor{red}{rgb}{1.0,0.0,0.0}

\input{pp7.h}
\begin{document}

\title{\textbf{\LARGE Direct Imaging and Spectroscopy of Extrasolar Planets }} %

\author {\textbf{\large Thayne Currie}}
\affil{\small\em Subaru Telescope, National Astronomical Observatory of Japan, 
650 North A`oh$\bar{o}$k$\bar{u}$ Place, Hilo, HI  96720, USA}
\affil{\small\em Department of Physics and Astronomy, University of Texas-San Antonio, One UTSA Circle, San Antonio, TX, USA}
\author {\textbf{\large Beth Biller}}
\affil{\small\em SUPA, Institute for Astronomy, University of Edinburgh, Blackford Hill, Edinburgh EH9 3HJ, UK}
\author {\textbf{\large Anne-Marie Lagrange}}
\affil{\small\em LESIA, Observatoire de Paris, Université PSL, CNRS, Sorbonne
Université, Université Paris Cité, 5 place Jules Janssen, 92195 Meudon, France}
\author {\textbf{\large Christian Marois}}
\affil{\small\em National Research Council of Canada, Herzberg Astronomy \& Astrophysics, Victoria, BC, Canada}
\author {\textbf{\large Olivier Guyon}}
\affil{\small\em Subaru Telescope, National Astronomical Observatory of Japan, 
650 North A`oh$\bar{o}$k$\bar{u}$ Place, Hilo, HI  96720, USA}
\author {\textbf{\large Eric L. Nielsen}}
\affil{\small\em Department of Astronomy, New Mexico State University, Las Cruces, NM, USA}
\author {\textbf{\large Mickael Bonnefoy}}
\affil{\small\em IPAG, Univ. Grenoble Alpes, CNRS, IPAG, 38000, Grenoble, France}
\author {\textbf{\large Robert J. De Rosa}}
\affil{\small\em European Southern Observatory, Alonso de C\'{o}rdova 3107, Vitacura, Santiago, Chile}

\begin{abstract}
\baselineskip = 11pt
\leftskip = 1.5cm 
\rightskip = 1.5cm
\parindent=1pc
{\small \textbf{Abstract}--
Direct imaging and spectroscopy is the likely means by which we will someday identify, confirm, and characterize an Earth-like planet around a nearby Sun-like star.  
This Chapter summarizes the current state of knowledge regarding discovering and characterizing exoplanets by direct imaging and spectroscopy. We detail instruments and software needed for direct imaging detections and summarize the current inventory of confirmed and candidate directly-imaged exoplanets.  Direct imaging and spectroscopy in the past decade has provided key insights into jovian planet atmospheres, probed the demographics of the outskirts of planetary systems, and shed light on gas giant planet formation.   We forecast the new tools and future facilities on the ground and in space that will enhance our capabilities for exoplanet imaging and will likely image habitable zone rocky planets around the nearest stars.
 \\~\\~\\~}
\end{abstract}  

\section{\textbf{Introduction}}

Identifying, confirming, and characterizing an Earth-like planet around a nearby Sun-like star is a key goal of exoplanet\index{exoplanet} science.
In the coming decades, direct imaging\index{direct imaging} and spectroscopy are the likely means by which this goal will be achieved.

\begin{figure}[h]
 \epsscale{1.0}
  \plotone{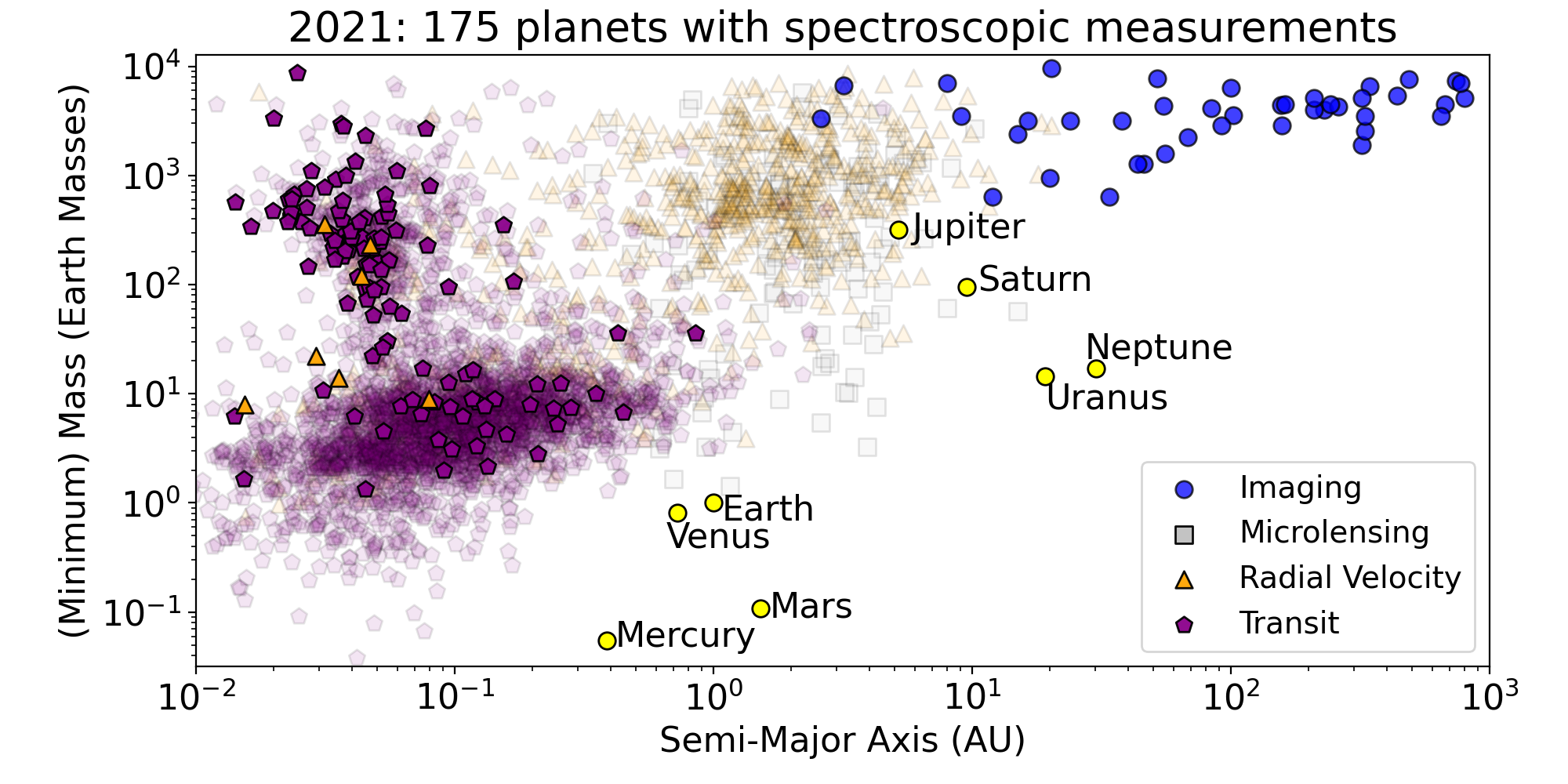}
 \vspace{-0.375in}
\caption{\small Demographics of planetary-mass companions detected with various methods.   Bold symbols denote the 175 companions with spectroscopic measurements constraining planet atmospheres: directly imaged companions comprise a substantial fraction of this population.  Figure courtesy of Dmitry Savransky, using data from the NASA Exoplanet Archive.}  
\vspace{-0.05in}
\label{fig1}
\end{figure}

Over 5000 extrasolar planets\index{extrasolar planets} and candidates have been detected through indirect means.  In contrast, 20--25 exoplanets have been directly imaged \citep[e.g.][see~Fig.~\ref{fig1}]{Marois2008,Lagrange2010}.   
Despite this relatively low yield of discoveries, direct imaging comprises a large fraction of the known exoplanet population amenable to atmospheric characterization, since the method provides photons from the planets themselves instead of just inferring their presence \citep[e.g.][]{Currie2011,Konopacky2013}.

Almost all of the directly imaged planets known thus far are young, self-luminous gas giants detected in thermal emission\index{thermal emission}.  They represent the extremes of planet formation \citep{Nielsen2019,Vigan2021}: planets with masses of 2 $M_{\rm J}$ or more orbiting at moderate to wide separations ($\sim$10--250 au). A subset of directly detected planets appear to be in or just finished with the final stages of assembly (``protoplanets") \citep{Keppler2018}. The direct imaging field currently also provides key insights into jovian planet formation and the demographics of the outskirts of planetary systems.

Remarkably, the direct imaging method 
enables the acquisition of hundreds to thousands of spectroscopic datapoints on exoplanets in a few hours of telescope time. This wealth of information can critically constrain individual planet atmospheric properties -- e.g. temperature, clouds, chemistry, and gravity -- as well as 
the atmospheric evolution of gas giants as a population.

The first direct images and spectra of planet candidates relied on facility, general-use adaptive optics\index{adaptive optics} (AO) systems to deblur starlight or space telescopes (i.e. the \textit{Hubble Space Telescope}) coupled with simple coronagraphs\index{coronagraphs} \citep[e.g.][]{Chauvin2004, Kalas2008}.   The past decade has seen the development, demonstration, and honing of dedicated extreme AO systems coupled with advanced coronagraphs and sophisticated post-processing methods, allowing the detection of planets that are fainter and less massive and/or located at smaller angular separations that probe tighter orbits
\citep[e.g.][]{Macintosh2014,Soummer2012,Mawet2012}.   In-development extreme AO systems and upgrades to first-generation extreme AO systems will provide images of many exoplanets near the ice line \citep{Guyon2020,Males2020}.   The  Coronagraphic Instrument on NASA's \textit{Roman Space Telescope} (Roman-CGI) could provide the first detections of mature planets in reflected light\index{reflected light} \citep{Spergel2013} (Figure \ref{fig2}).

Planned ground-based extremely large telescopes and proposed space missions promise to make the discovery and confirmation of a habitable, Earth-like exoplanet around a Sun-like star a reality within the next 25 years \citep[e.g.][]{LopezMorales2019,Gaudi2021,LUVOIR2019,Quanz2021}.   These facilities will endeavor to reveal biomarkers -- e.g. water, oxygen, ozone -- in individual systems.   Their surveys will provide the first assessment of potential habitability around stars of different masses and thus the true context for life on Earth.   Technological innovation, atmospheric characterization, and demographic studies of young jovian planets over the past decade provide first, key steps towards this goal.

In this Chapter, we provide an updated description of the state of our knowledge about detecting and characterizing exoplanets by direct imaging.  The last dedicated direct imaging review chapter in \textit{Protostars and Planets} was written in 2007 \citep{Beuzit2007}, before the first incontrovertible exoplanet imaging detections: superjovian planets around HR 8799\index[obj]{HR8799} and $\beta$ Pic\index[obj]{$\beta$ Pic} \citep{Marois2008,Marois2010a,Lagrange2010}.   Previous reviews from \citet{Traub2010} and \citet{Bowler2016} bracket the start and end of the era of direct imaging surveys with facility, conventional AO systems.  Now, $>$5 years later, the first generation of extreme AO surveys have ended, bringing new discoveries, fundamentally new insights into atmospheric properties of individual systems and imaged planet demographics, motivating substantially more powerful instrumentation and tangible plans for directly detecting true solar system analogues, including Earths.  

This Chapter is pedagogical in nature and is organized as follows.   In Section 2, we outline key direct imaging instrumentation and methods, describing the challenges in achieving planet detections by direct imaging, and detailing critical, novel hardware and software needed to image planets.  Section 3 summarizes our current inventory of directly imaged exoplanets, discusses challenges with interpreting direct imaging detections, and outlines synergies with other detection techniques. Section~\ref{sec:characterization} focuses on atmospheric characterization of exoplanets by direct imaging and spectroscopy, using other substellar objects (i.e. brown dwarfs\index{brown dwarfs}) as anchors to provide empirical constraints on exoplanet atmospheres and employing atmospheric models to infer intrinsic properties, including clouds, chemistry, and gravity.  Next we overview the architecture of directly imaged planetary systems gleaned from astrometric monitoring, dynamical models, and planet-disk interactions (Section 5).   Section 6 summarizes recent direct imaging surveys, the occurrence rates and demographics derived from them, and what these results mean for models of jovian planet formation.  Finally, in Section 7 we forecast the future of direct imaging, describing how technological innovations and new, vastly more powerful facilities may provide humanity with a glimpse of a habitable world for the first time.


\section{\textbf{Direct Imaging Instrumentation and Methods} }
Direct imaging detections require separating the halo of bright, highly structured scattered starlight from faint exoplanet light.    The key metric used to determine the detectability of extrasolar planets is the planet-to-star contrast ratio at a given off-axis angular separation: the contrast ratio needed for a detection varies with planet properties (Sect 2.1).   Critical hardware like AO -- or (more generally) wavefront control systems consisting of sensors and deformable mirrors -- and coronagraphs sharpen and then suppress scattered starlight (Sect 2.2) to achieve deep raw contrasts.   Novel observing techniques allow post-processing algorithms to further remove residual starlight, increasing achievable contrast ratios and thus improving planet detection capabilities (2.3).   

\subsection{Detectability of Planets by Direct Imaging}
\begin{figure}[h]
 \epsscale{1.0}
  \plotone{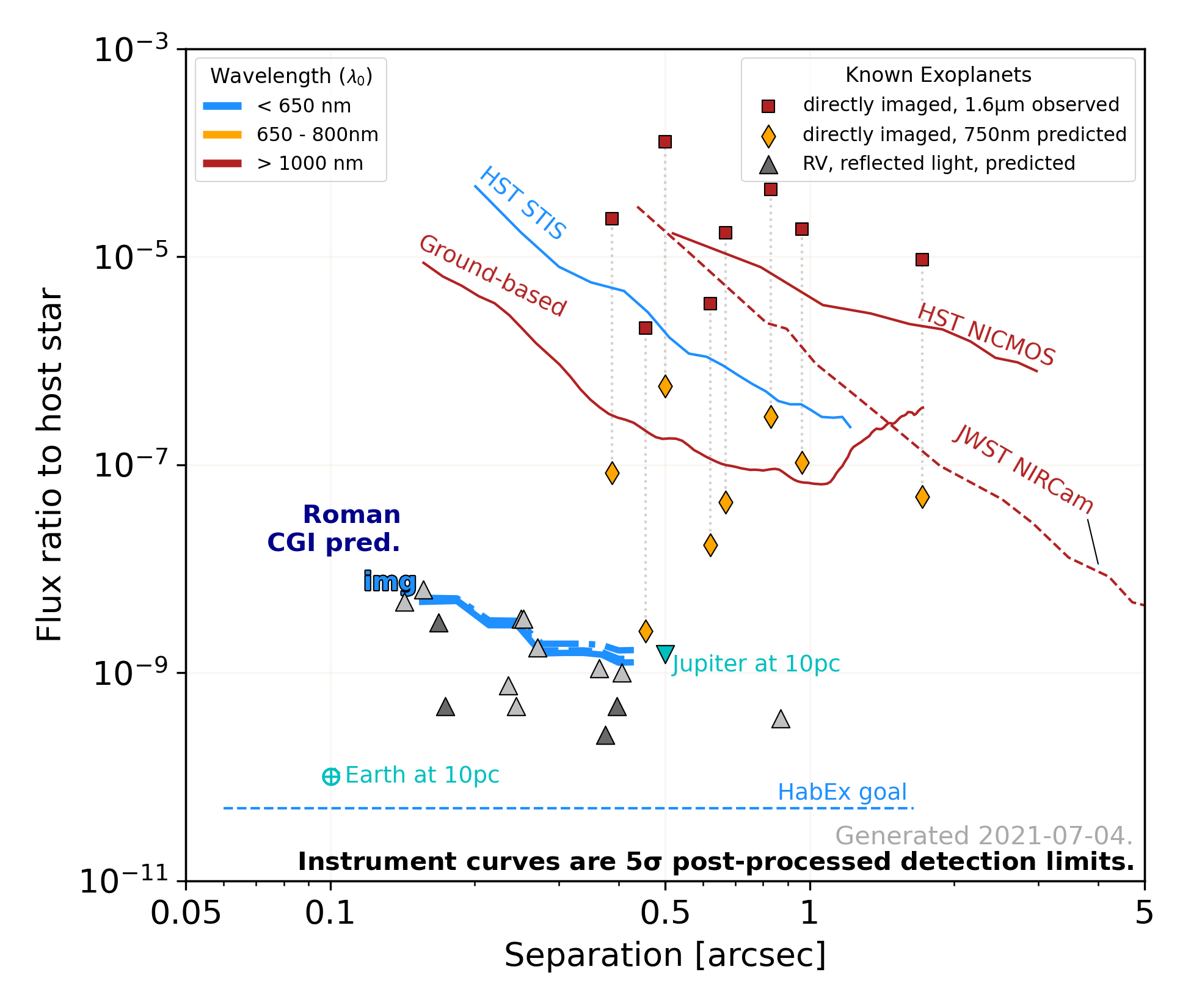}
 \vspace{-0.35in}
\caption{\small Contrast vs. angular separation for current and future ground and space-based high-contrast imaging platforms vs. measured and predicted contrasts for selected directly imaged planets, radial-velocity (RV) detected planets, and Earth and Jupiter analogues. Generated using tools developed by Vanessa Bailey, which are available at: https://github.com/nasavbailey/DI-flux-ratio-plot.}  
\vspace{-0.05in}
\label{fig2}
\end{figure}

Even in the absence of an atmosphere to blur starlight, 
an astronomical image of a point source (e.g. a distant star) will not be a single point: its intensity distribution follows from the  Fourier transform of the telescope pupil function.  
For a simple unobscured circular aperture, the image intensity follows an Airy pattern whose values compared to the peak intensity ($I_{\rm o}$) at angular separation $\rho$, effective telescope aperture diameter $D_{\rm tel}$, and wavenumber $k$ = 2$\pi$/$\lambda$ is $I = I_{\rm o}[\frac{4J_{\rm{1}}(0.5k D_{\rm tel} sin\rho)}{D_{\rm tel} sin\rho}]^{2}$
, where $J_{\rm 1}$
is the Bessel function of the first kind of order one.  The first null of the Airy function occurs at $\sim$ 1.22 $\lambda$/$D_{\rm tel}$.
The \textit{full-width at half-maximum} (FWHM) of this point source on the sky, defining the telescope diffraction limit, is given by
\begin{equation}
    \theta(\arcsec{})_{\rm FWHM} \sim 0.21 \left(\frac{\lambda(\mu m)}{D_{\rm tel} (m)}\right).
\end{equation}
Even current telescopes in principal have the angular resolution needed to image planets on solar system-like scales orbiting nearby stars.   An Earth twin orbiting at a projected separation of 1 au from a Sun-like star at 10 $pc$ subtends an angle of 0\farcs{}1.   Diffraction limited imaging with the \textit{Hubble Space Telescope} at optical wavelengths or with 8-10m class telescopes like VLT, Subaru, and Keck at near-infrared (IR) wavelengths is sufficient to resolve objects at Earth-Sun projected physical separations out to a distance of 10--20 $pc$.  However, the star's halo light must be reduced through hardware and software to a level sufficient to make the planet's light detectable.   The starlight suppression level required to detect a planet by direct imaging depends on whether the detection is in reflected light or thermal emission and varies with the planet angular separation, age, size, temperature, and other properties (see Figure~\ref{fig2}).

\subsubsection{Reflected Light}
Planets reflect the light of the stars they orbit.  At optical wavelengths, the contrast ratio for a planet in reflected light can be approximated as
\begin{equation}
C_{\rm optical, \lambda} \sim A_{\rm g}(\lambda)~\phi(\lambda, \alpha)\left(\frac{r_{\rm p}}{a_{\rm p}}\right)^{2},
\end{equation}
where $r_{\rm p}$ is the planet radius, $a_{\rm p}$ is the planet-to-star physical separation, $A_{\rm g}$ is the visible geometric albedo spectrum, $\phi$ is the phase function as a function of  $\alpha$, the phase angle, which is the angle between the star, planet, and observer \citep{Greco2015,Traub2010}. $A_{\rm g}$ and $\phi$ depend on the planet's atmospheric properties.  For a Lambertian phase function, valid for high albedo atmospheres, the phase function follows a simple relation of $\phi(\alpha)$ = [sin($\alpha$)+($\pi$-$\alpha$)cos($\alpha$)]/$\pi$,
where $\phi$ = 1/$\pi$ at maximum elongation ($\alpha$ = $\pi$/2).   Other potential scattering phase functions include isotropic scattering and Rayleigh scattering. 
The value of $\phi$ is considerably lower for Rayleigh scattering appropriate for jovian atmospheres at angles of 60--90$^{o}$ \citep{Madhusudhan2012}. 

For an exo-Jupiter and exo-Earth emitting as Lambertian spheres with measured geometric albedos of 0.52 and 0.367, these contrasts at maximum elongation reduce to:
\begin{equation}
C_{\rm optical, J} \sim 1.4\times10^{-9}\left(\frac{r_{\rm p}}{r_{\rm J}}\right)^{2}\left(\frac{5.2~au}{a_{\rm p}}\right)^{2}
\end{equation}
and
\begin{equation}
C_{\rm optical, \oplus} \sim 2.1\times10^{-10}\left(\frac{r_{\rm p}}{r_{\oplus}}\right)^{2}\left(\frac{1~au}{a_{\rm p}}\right)^{2}, 
\end{equation}

where $r_J$ and $r_{\oplus}$ are the radii of Jupiter and Earth respectively.  At $V$ band, a Jupiter and Earth at 10 pc will then have apparent magnitudes of slightly greater than 27 and 29, respectively, and require starlight removal at angular separations of 0\farcs{}5 and 0\farcs{}1 better than one part in 10$^{-9}$ and 10$^{-10}$, respectively.   These required contrasts are well beyond the capabilities of current ground and space-based high-contrast imaging instruments.   Thus far, exoplanet direct imaging has therefore focused on the detection of thermal emission from self-luminous jovian planets.

\subsubsection{Thermal Emission}
\begin{figure}[h]
 \includegraphics[angle=0,width=1.025\columnwidth,trim=3mm 5mm 45mm 15mm,clip]{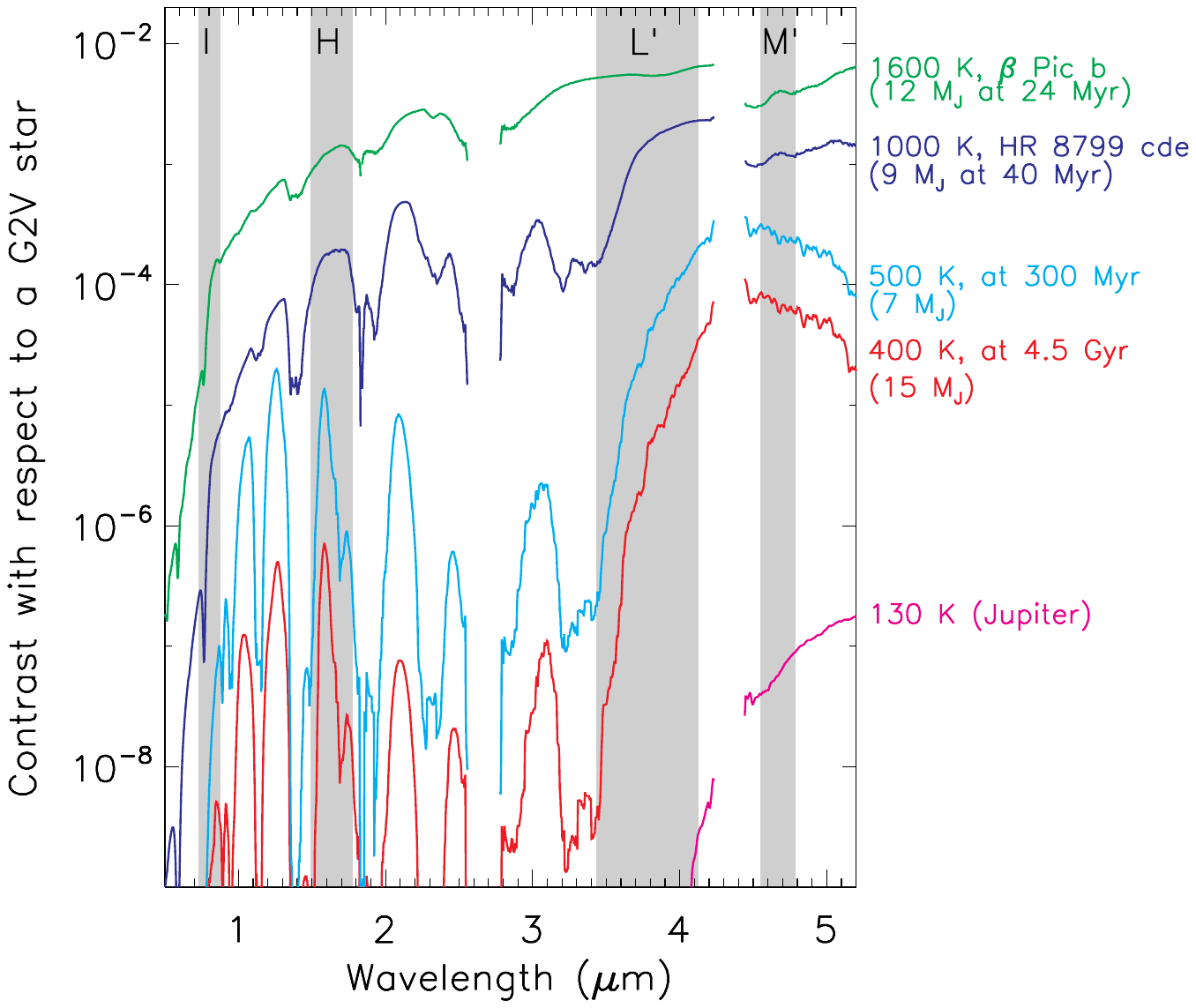}
 \vspace{-0.35in}
\caption{\small Contrast ratio vs. selected optical, near IR, and mid-IR passbands for models for known directly imaged planets ($\beta$ Pic b and HR 8799 cde), models for older cooler exoplanets, and Jupiter assuming a G2V primary in all cases.  The actual planet-to-star contrast ratios for $\beta$ Pic b and HR 8799 cde are 5--10 times higher as they orbit intrinsically brighter, hotter stars.  Plot adapted and modified from \citet{Skemer2014}.   Models drawn from \citet{Currie2011,Madhusudhan2011,Burrows2003}.}  
\vspace{-0.05in}
\label{fig3}
\end{figure}

Like brown dwarfs, jovian planets cool and contract with time, releasing gravitational potential energy as thermal emission \citep[e.g.][]{Burrows2001}.   At ages of 1--10 Myr, models for the luminosity evolution of 1--10 $M_{\rm J}$ exoplanets predict temperatures of $\approx$ 500--3000 $K$ and radii up of $r_{\rm p}$ $\sim$ 2.5--3 $R_{\rm J}$.  By 1 Gyr, these models predict that these planets cool to temperatures $T$ $\lesssim$ 500 $K$ and contract to Jupiter-like radii ($r_{\rm p}$ $\sim$1--1.2 R$_{\rm J}$)\citep[e.g. Fig. 5 in][]{Spiegel2012}. Thus, thermal emission from young jovian exoplanets peaks at near-to-mid IR wavelengths at the youngest ages, moving to well into the mid-IR at older ages (Figure~\ref{fig3}). The contrast ratio for planets in thermal emission depends on the radius and effective temperature of the planet and star and other properties:
\begin{equation}
C_{\rm IR, \lambda} \sim \frac{F_{\lambda,p}(T,X)~r_{\rm p}^{2}}{F_{\lambda,s}(T)~r_{\rm s}^{2}},
\end{equation}
where $F_{\lambda,p}(T,X)$ is the thermal flux from the planet as a function of wavelength, $F_{\lambda,s}(T)$ is the thermal flux from the star as a function of wavelength, and $X$ depends on the planet's atmospheric characteristics, such as clouds, chemistry, and gravity.

Fig. \ref{fig3} shows how the contrast ratios vary for various types of exoplanets orbiting a G2V star.   In the near-IR, young superjovian ($\sim$10 $M_{\rm J}$) exoplanets similar to $\beta$ Pic b and HR 8799 cde are roughly a factor of 10$^{-3}$ to 10$^{-5}$ fainter than a Sun-like star.  In the red optical, predicted contrasts from these planets are 100 times larger; however, mid-IR contrasts are a factor of 10 smaller.  Jovian exoplanets of similar masses around much older stars are far cooler -- another factor of 100--1000 fainter in the red optical and near-IR -- but remain bright in the mid IR.   A true Jupiter analogue emits negligible thermal emission shortward of $\lambda$ $\sim$4--5 $\mu m$.   Rocky terrestrial planets also emit thermally.  An exo-Earth analogue with $T_{\rm eff}$ $\approx$ 260 $K$ has blackbody emission peaking at 10 $\mu m$: its contrast ratio is $\approx$10$^{-7}$, a factor of $\approx$1000 shallower than its reflected light contrast.

 

\subsection{\textbf{Wavefront Control and Coronagraphy}} 


\subsubsection{Atmospheric Wavefront Control for Direct Imaging}
\label{sssec:atmWFC}
Wavefront errors, whether induced by atmospheric turbulence on the ground or intrinsic to an optical imaging system on the ground or in space, substantially limit an image system's achievable contrast.  Left unmitigated, these errors preclude the direct detection of exoplanets.  In the past decade, the direct imaging field has made key strides in 
wavefront control hardware and software to drastically reduce wavefront errors and has developed sophisticated coronagraph designs to further suppress diffracted starlight.

For ground-based imaging systems, turbulence arising from many atmospheric layers along the path from the star to the telescope induces changes in the optical path length difference of starlight \citep{Guyon2005}, blurring images.   These aberrations must be corrected by an AO system consisting of wavefront sensors and deformable mirrors (DMs).  An AO system splits incoming light from a guide star between the science detector and a wavefront sensor to measure and then correct -- with a DM(s) -- atmospheric turbulence distorting the incoming wavefront, sharpening starlight at the image plane.  

Key terms driving the wavefront error budget for an AO system include 1) measurement error ($\sigma_{m}$) which depends on the noise properties of a wavefront sensor and the guide star brightness, 2) temporal bandwidth error ($\sigma_{t}$) which depends on the AO system time lag $\tau$ compared to the atmospheric coherence time $\tau_{\rm o}$, and 3) the fitting error ($\sigma_{\rm f}$) which depends on the coherence length $r_{\rm o}$ compared to the DM actuator density.   The quadrature-added sum of these terms sets the Strehl ratio -- a measure of the optical quality of the image -- and the raw contrast vs. angular separation \citep[e.g.][]{Tyson1998}.       At the angular separations relevant for most current planet searches ($\theta$ $\sim$ 0\farcs{}2--1\farcs{}0), wavefront measurement error and temporal bandwidth error set the contrast floor.  Wavefront chromaticity and non-common path aberrations due to instrument optics can also limit contrast \citep{Guyon2005,Currie2020a}.  

Facility, conventional AO systems were used for the first direct imaging searches.  They typically sample and correct the incoming wavefront at frequencies of 0.2--1 kHz and use DMs with a few hundred actuators: aberrations are measured by standard Shack-Hartmann wavefront sensors \citep[e.g. see][]{Tyson1998}.   These systems yielded partial corrections, achieving modest Strehl ratios (SR $\sim$ 0.1--0.4) at near-IR wavelengths focused on by most exoplanet imaging searches \citep[e.g.][]{Lafreniere2007b}.  

The past 5-10 years have seen the deployment of numerous extreme AO systems on 5-10m telescopes, which have wavefront control loop speeds of $\gtrsim$1 kHz and DMs with $\gtrsim$1000 actuators.  Examples of extreme AO systems include the \textit{Large Binocular Telescope Adaptive Optics} system (LBTAO), PALM-3000 on the Hale telescope at Palomar Observatory, the \textit{Gemini Planet Imager} on Gemini-South, and the \textit{Spectro-Polarimetric High-contrast Exoplanet REsearch} instrument (SPHERE) at the VLT, the \textit{Subaru Coronagraphic Extreme Adaptive Optics} project (SCExAO) on the Subaru Telescope, and MagAO-X on the  Clay telescope at the Magellan Observatory  \citep{Esposito2011,Dekany2013,Macintosh2014,Beuzit2019,Jovanovic2015,Males2020}.   These systems obtain higher-quality AO corrections -- Strehl ratios of 0.7--0.95 at 1.6 $\mu m$ -- and yield a factor of 10-100 deeper contrast at sub-arcsecond separations than conventional AO systems \citep[e.g][]{Macintosh2014}.   

Extreme AO systems have relied on hardware developments in three key areas to operate high-performance AO correction: detectors, deformable mirrors, and computing hardware.  Most extreme AO platforms utilize fast ultra-low noise detectors like \textit{Electron-Multiplying Charge-Coupled Devices} (EMCCDs) to record stellar photons used for wavefront sensing with reduced measurement errors \citep[e.g.][]{Jovanovic2015,Beuzit2019,Males2020}. High-actuator count DMs are now available. For example, the microelectromechnical-type DMs (MEMS) offer unprecedented actuator density ($\approx$10 act per square mm), allowing for fairly compact extreme AO instruments \citep[e.g.][]{Macintosh2014,Jovanovic2015,Males2020}. Other systems employ adaptive secondary mirrors (usually voice-coil DMs) whose low emissivity makes them especially well suited for exoplanet imaging in thermal IR \citep{Esposito2011}. High performance computing hardware with low latency can handle the demanding computation requirements of fast, high order extreme AO systems.

\begin{figure}[h]
 \includegraphics[angle=0,width=0.95\columnwidth,trim=0mm 0mm 0mm 0mm,clip]{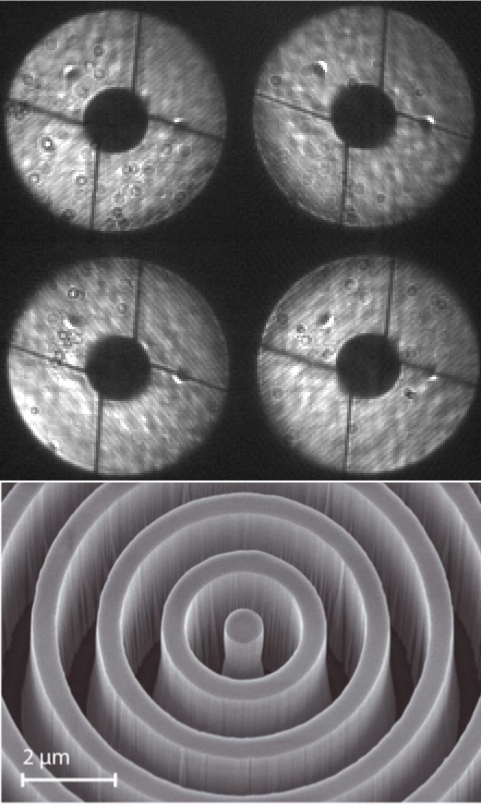}
 \vspace{-0.175in}
\caption{\small Some key high-contrast imaging hardware components.  (top panel) Wavefront control with a Pyramid WFS and high-performance camera: images of the 4 pupils created by a pyramid optic and acquired by a photon counting, high speed, low-latency EMCCD \citep{Jovanovic2015}. (bottom panel) Advanced starlight suppression masks: electron scanning microscopic image of an annular groove phase mask (``vector vortex" coronagraph) \citep{Mawet2012,Delacroix2013}.  }  
\vspace{-0.075in}
\label{fig:hardware}
\end{figure}

Extreme AO systems also differ from conventional AO systems in their optical design and system architecture. Wavefront sensors (WFS) optimized for speed, sensitivity, accuracy and precision such as the Pyramid wavefront sensor have been adopted by many leading extreme AO systems \citep[top panel of Figure \ref{fig:hardware},][]{Esposito2011,Jovanovic2015,Males2020}, while the more established Shack-Hartmann WFS approach has been upgraded with spatial filtering to improve performance. Wavefront correction is often performed in two steps: a conventional coarse ``woofer" correction is followed by a faster and more accurate ``tweeter" correction \citep[e.g.][]{Macintosh2014}. Extreme AO systems also include dedicated sensing and control of pointing for precise co-alignment of the star and coronagraph mask \citep{singh15,shi17,huby2017}, which is essential to maintain high contrast, especially at the smallest angular separations. 

Advances in wavefront control algorithms are also enabling significant improvements in high-contrast imaging performance.  Recent advances include predictive control and sensor fusion.  These and other new software/hardware advances are described in Section 7.1.

\subsubsection{Optical Starlight Suppression}

Starlight suppression is primarily performed by dedicated coronagraph optics: phase and amplitude masks deployed in the beam to remove starlight while preserving planet light. These optical elements include occulters in the image plane to directly block out on-axis starlight, as well as optical elements in the pupil plane to manage the telescope diffraction pattern. Many coronagraph designs and approaches are available \citep[e.g.][]{Kuchner2002, Kasdin2003,Sivaramakrishnan2005,guyon2006,Kenworthy2007,Mawet2010,Otten2017}, providing a wide range of performance characteristics. For example, the well-established conventional Lyot coronagraph combines an occulter in the image plane with a Lyot stop in the pupil plane to deliver robust performance at moderate contrast and large angular separations.  Phase-mask coronagraphs or interferometric designs (e.g. the vector vortex coronagraph) can be highly optimized for deep contrast at small inner working angles (Figure \ref{fig:hardware}, bottom panel). The highest-performing coronagraphs are also the most demanding in terms of wavefront quality and stability. Consequently, extreme AO systems first adopted Lyot coronagraphs, while more recent systems or upgrades deploy higher performance solutions as the corresponding wavefront quality improves.

Wavefront control is also becoming an important component of starlight suppression in the form of speckle control. In speckle control, a feedback loop from the post-coronagraph image to an upstream deformable mirror allows for residual speckles to be measured and canceled. This approach has been used in laboratory testbed to reach deep contrast levels (10$^{-7}$--10$^{-10}$) in support of future space-based missions. Deployment on ground-based systems remains challenging due to the dynamic nature of atmospheric turbulence, but has been successful in removing a fraction of static and slow speckles \citep{codona2013,martinache2016,gerard2018,bos2021}. Sensing and compensation of non-common path aberrations can also be performed using a dedicated sensor, e.g. a Zernike phase-mask sensor \citep{vigan2019}.

 



\subsection{\textbf{Observing, Post-Processing, and Spectral Extraction Methods}}
\begin{figure*}[ht]
  \includegraphics[angle=0,scale=0.485,trim=0mm 0mm 0mm 0mm,clip]{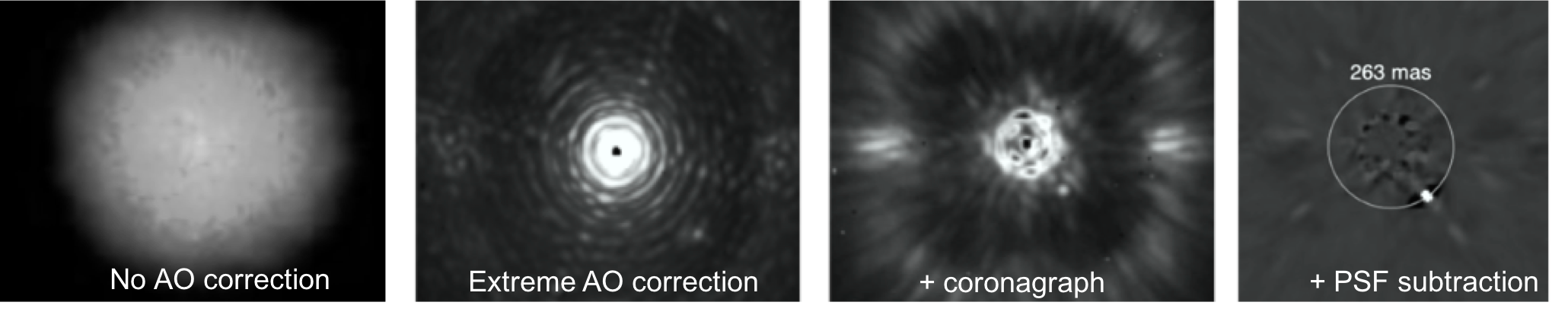}
 \vspace{-0.2in}
\caption{\small A demonstration of high-contrast imaging hardware and software with SPHERE.  Seeing limited point-spread function (PSF) (left) sharpened by an extreme AO system (middle left) with scattered starlight further suppressed by a coronagraph (middle right) and advanced PSF subtraction methods (right).}  
\vspace{-0.05in}
\label{fig:spheresequence}
\end{figure*}
At sub-arcsecond separations, the raw (instrument-delivered) planet-to-star contrasts obtained by the first generation of facility and now extreme AO systems -- 10$^{-3}$--10$^{-4.5}$ -- are still too shallow to yield decisive detections of many young jovian exoplanets. Residual quasi-static speckle noise, due to nanometer-scale imperfections in telescope optics, thermal flexure, etc., limits instrumental contrasts and fails to simply average out over time like photon (``white") noise \citep{Hinkley2007}.  Speckle noise statistics follow a modified Rician distribution whose long positive tail can lead to many false positives \citep{Marois2008b,Soummer2007}.   Therefore, novel observing techniques coupled with advanced post-processing methods are critical to optimizing our ability to image exoplanets by further suppressing stellar halo light by orders of magnitude and making residual noise more Gaussian (``whitening").   

Various forms of differential imaging techniques are utilized to remove speckle noise for direct imaging.  Commonly used techniques include:
\begin{itemize}
    \item \textbf{Angular Differential Imaging} (ADI) \citep{Marois2006} -- ADI exploits the quasi-static nature of residual speckle noise in AO corrected high-contrast images from the ground and diffraction-limited imaging from space. For ground-based AO imaging, quasi-static speckles evolve on a characteristic timescale of $\sim$ 10s of minutes to an hour or more \citep{Hinkley2007}.   For space-based imaging, PSF breathing due to thermal variations and other small mechanical changes cause the quasi-static halo to evolve, but the halo for a given target can remain well correlated for months or years.   By turning off the instrument rotator that keeps north fixed on the detector (on the ground) or obtaining observations at different roll angles (space), astrophysical objects off-axis from a star change position angle on the detector while the speckle halo remains fixed.  Subtracting images obtained in an ADI sequence can then suppress speckle noise without fully suppressing planet signals.    
    
    ADI efficiently suppresses speckles on the ground and in space \citep[][]{Marois2006}.  It is the most widely used observing method in direct imaging.   ADI's advantages are most limited at small angular separations where the displacement of the planet's PSF is smallest for a given parallactic angle change or roll angle change.  Typically, ADI induces signal loss, which must be corrected to achieve absolute spectrophotometric calibration for planets (see below).
    
    \item \textbf{Spectral Differential Imaging} (SDI) \citep{Marois2000,Sparks2002,Biller2004} -- At raw contrasts shallower than the 10$^{-7}$ levels needed to image self-luminous jovian exoplanets, achromatic phase errors dominate the wavefront errors responsible for the speckle halo \citep[e.g.][]{ShaklanGreen2006,Bailey2018}.   For data obtained simultaneously at multiple wavelengths -- e.g. dual-channel imaging or an integral field spectrograph (IFS) -- the speckle halo as a function of wavelength, scaled in radius by the central passband wavelength, is extremely well correlated.   Since the images at different wavelengths are obtained simultaneously with the target, they do not suffer from are temporal decorrelation as with ADI \citep[e.g.][]{Marois2014}.   SDI then suppresses the speckle halo in a given passband by constructing a reference PSF drawn from (rescaled) images at other wavelengths.   
    
    Like ADI, SDI induces signal loss and is least effective at small angular separations, where bandpass rescaling yields small displacements of a (de-)magnified planet PSF.   SDI's efficacy is also limited by non-common path errors, differential sampling of the point-source PSFs across wavelength channels, and the spectrum of the planet to be detected \citep[e.g.][]{Gerard2019, Marois2014}.  Extracting throughput-corrected spectra from data processed with SDI is also more challenging than with other differential imaging techniques \citep{Pueyo2016}.
    
    \item  \textbf{Reference Star Differential Imaging} (RDI) -- RDI subtracts the PSF of a companion-less reference star from the target.  It is 
    widely used in the ultra-stable environment of space.   While  ADI and SDI may yield superior speckle suppression at moderate to wide separations, RDI may be advantageous at very small angular separations where other techniques suffer from self subtraction \citep{Xuan2018}.  RDI requires a very stable PSF to efficiently operate.   Using it on ground-based telescopes may require fast switching between the target and reference star(s) for maximum contrast gain \citep[e.g.][]{Wahhaj2021}; in broad bandpasses, it also requires a reference star that is extremely well color matched \citep[e.g.][]{Krist1998}.
\end{itemize}

These differential imaging techniques are used in combination with PSF subtraction algorithms.   Many widely-used, advanced PSF subtraction methods are different forms of least-squares algorithms, which construct a reference PSF that minimizes the variance when subtracted from a given target image.   The \textit{locally-optimized combination of images} (LOCI) algorithm constructs a reference PSF from a linear combination of reference images weighted by coefficients $\mathbf{c}$ determined from the solution to the matrix equation $\mathbf{c}$ = $\mathbf{A}^{-1}\mathbf{b}$ \citep{Lafreniere2007}, where $\mathbf{A}$ is the (square) covariance matrix for the reference library and $\mathbf{b}$ is the column matrix populated by elements multiplying each reference image by the target image.   Another approach exploits \textit{principal component analysis} (PCA), computing the Karhunen-Lo\`eve transform of the reference image set and projects this set onto the target image to construct a combination of weighted eigenimages to subtract from the target \citep[Karhunen-Lo\`eve~Image~Plane~algorithm,~henceforth~KLIP,][]{Soummer2012,Amara2012}.   Multiple successor algorithms such as TLOCI and A-LOCI draw from the lineage of LOCI and/or KLIP, employing advances such as correlation-based frame selection, pixel masking, various rank-truncations of the covariance matrix, free parameter optimization or exploiting high-performance computing to solve for the variance-minimizing coefficients directly \citep[e.g.][]{Marois2010b,Marois2014,Currie2012,Currie2015,Thompson2021}.  Separate approaches involve  maximum-likelihood methods to model and remove the stellar PSF \citep[e.g.][]{Cantalloube2015} or statistically modeling non-stationary covariances in small regions of images to improve PSF subtraction \citep[e.g. the PACO algorithm][]{Flasseur2018}.

The relative performances of these algorithms vary in the literature and may depend on their exact implementation and suitability for a particular data set, although some work suggests that the newer algorithms descended from LOCI and KLIP can offer significant improvements \citep[e.g.][]{Rameau2013,Thompson2021}.   In general, all of these algorithms enable significant contrast gains over simple, classical methods, especially at small angular separations \citep[][]{Lafreniere2007,Currie2014a}.

ADI, SDI, and RDI combined with PSF subtraction methods can attenuate and distort planet signals.  The first methods to measure these biases and recover true planet flux measurements and astrometry injected synthetic planets into data at other locations or iteratively subtracted negative copies of  planets at their apparent positions \cite[e.g.][]{Lafreniere2007, Lagrange2010}.   Recently, several authors have developed efficient forward-modeling methods to estimate photometric and astrometric biases at the planet's location \citep[e.g.][]{Pueyo2016}.   Planet forward-models in turn can be used as matched filters to improve planet detection capabilities themselves \citep{Ruffio2017}.



Analyses of images whose speckle noise is partially suppressed by advanced PSF subtraction techniques have revised notions of how to quantify detection significances and spectroscopic uncertainties.  For instance, finite sample sizes impact our definitions for contrast limits drawn from images with noise whitened by KLIP, LOCI, and other algorithms, especially at small (1--3 $\lambda$/D) separations \citep{Mawet2014,Pairet2019}.    In IFS data, residual speckle noise may be spatially and spectrally correlated.  Considering the full spectral covariance has a substantial impact on deriving planet atmospheric parameters from model comparisons \citep{Greco2016}.

Figure \ref{fig:spheresequence} shows the combined effect of high-contrast imaging hardware and software on imaging exoplanets, stepping through the successive improvements found from using extreme AO, coronagraphy, and PSF subtraction. 

\section{\textbf{Direct Imaging Detections}}

\subsection{\textbf{Taxonomy of Directly-Imaged Companions}}

Figure~\ref{fig1} shows the demographics of planetary mass companions to stars detected via various methods.  
Community consensus on the planethood of many imaged objects (e.g. HR 8799 bcde, $\beta$ Pic bc, 51 Eri\index[obj]{51 Eri} b, etc.) is clear.  However, identifying the exact criteria needed to distinguish between planets and brown dwarfs is challenging.

Planets have often been identified as objects with masses below the deuterium-burning limit, nominally 13 $M_{\rm J}$ (e.g. the IAU Working Group Definition).  However, this simple criterion is poorly motivated.   In addition to being 
time and metallicity dependent \citep[][]{Spiegel2011}, deuterium burning arguably does not identify a meaningful boundary for the evolution of low-mass objects \citep{Chabrier2007,Luhman2008}, as some objects below the deuterium burning limit have been found in configurations which imply formation by cloud formation and other objects above the deuterium burning limit have been found in configurations which imply formation like a planet in a disk.  
For instance, recent imaging surveys have found objects that are members of quadruple systems, clearly formed by molecular cloud fragmentation, with inferred masses down to 5 $M_{\rm J}$ and free-floating objects with sub-deuterium burning masses as well \citep{Todorov2010,Liu2013}.  In contrast, RV  surveys have identified some systems -- e.g. the 2.7 $M_{\odot}$ star $\nu$ Oph -- with companions at $\sim$ 1 au with masses of 22 and 24 $M_{\rm J}$ that are nevertheless locked in a mean-motion resonance indicating formation in a disk (i.e. like a planet) \citep{Quirrenbach2019}.  The physics of planet formation does not require that gas accretion shuts off once 13 $M_{\rm J}$ of material is accreted. 

An alternate definition 
leverages 
formation processes: a planet is an object formed in a circumstellar disk around a young star.  Demographic analyses of substellar objects can provide empirically-motivated criteria for separating planets from brown dwarfs. Studies of the substellar mass function from previous RV surveys and recent ones (i.e. the California Legacy Survey) show a local minimum at msin(i) $\sim$ 16--30 $M_{\rm J}$ \citep[e.g.][]{Sahlmann2011,Kiefer2019,Currie2022b}.      The minimum in the companion mass function may be proportional to the primary mass, indicating that companion mass ratio ($q$) could be a key discriminator \citep{Grether2006}.  Theory also suggests that the semimajor axes ($a_{\rm p}$) and mass ratios ($q$) of companions also help distinguish between bona fide planets and brown dwarfs \citep{Kratter2010,Currie2011}.   Binary companions to more massive stars with $q$ $\gtrsim$ 0.025 are exceptionally rare \citep{Kraus2008,Reggiani2016}.  Distributions of protoplanetary disk radii peak at $\sim$200 au and fall to low frequencies by $\sim$300 au \citep[e.g.][]{Andrews2007}: companions at wider separations are far less likely to have formed from a disk unless scattered to their current locations by unseen companions. Thus, we set the following limits for a planet vs. a brown dwarf: mass $<$ 25 $M_{\rm J}$, q $<$ 0.025, and $a_{\rm p}$ $\lesssim$ 300 au.
Our linked spreadsheet lists the current inventory of directly imaged exoplanets, protoplanets, higher mass-ratio/wider separation planet-mass companions, and controversial cases: \url{https://tinyurl.com/srb33b}.
As of 19 October 2022, we identify 22 directly imaged exoplanets that fit our criteria, three of which are imaged protoplanets.  We also list another 34 dozen companions that may instead be better considered as brown dwarfs instead of planets (including companions orbiting brown dwarfs), and 6 ``controversial" cases.  Some of these individual classifications will undoubtedly change pending new analysis as may the exact values used to separate planets from brown dwarfs.   Objects in each category will certainly be added over the next few years.
 
Below, we describe general properties of fully-formed directly imaged planets within 300 au and discuss three well-studied, emblematic cases (HR 8799, $\beta$ Pic, and 51 Eri). We also summarize our current knowledge of protoplanets and challenges with identifying imaged exoplanets.

\subsection{\textbf{Fully-formed Exoplanets}}
\begin{figure*}[h]
 \centering
  \includegraphics[angle=0,scale=0.9,trim=0mm 0mm 0mm 0mm,clip]{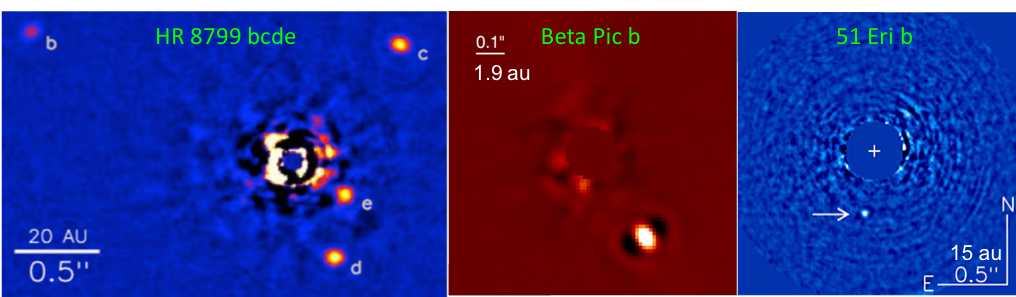}
 \vspace{-0.15in}
\caption{\small Well-studied directly-imaged exoplanetary systems: (left) HR 8799, (middle) $\beta$ Pic, and (right) 51 Eri.   Images draw from \citet{Marois2010a}, \citet{Lagrange19}, and \citet{Macintosh2015}, respectively.  Note that $\beta$ Pic c is not visible in the middle panel: the HR 8799 and 51 Eri images show the full complement of known planets in each system.}
\vspace{-0.1in}
\label{fig_DIarch}
\end{figure*}

\subsubsection{General Properties}
Most directly imaged planets known thus far have near-IR contrasts ranging between 10$^{-4}$ (e.g. $\kappa$ And b) and 10$^{-6}$ (51 Eri b).   All of them are self-luminous, made visible by thermal radiation released as the planets cool and contract. On the sky, the planets generally lie at angular separations of 0\farcs{}2 to 2\farcs{}0 ($\sim$ 5--40 $\lambda$/D for an 8m telescope).   

Stars hosting imaged exoplanets typically have ages of 10--100 Myr; many are found in young nearby stellar associations (often $<$100 pc, with ages $<$100 Myr) that share similar kinematics (proper motions, space motions) and formed in the same star-forming region, such as Sco-Cen or the $\beta$ Pic Moving group \citep[e.g.][]{Zuckerman2004a,Zuckerman2011,Song2003,Schlieder2012a,Schneider2012,Shkolnik2012}. (see Section~\ref{sec:surveysamples} for more details on such associations as sites for exoplanet searches). Most stars with imaged planets to date are B, A, or F stars, at least 50\% more massive than the Sun.   Most systems with imaged planets -- including systems with the first detections (HR 8799, $\beta$ Pic) --  also have Kuiper belt-like debris disks.

Planets detected to date share similar spectral types and temperatures with more massive field brown dwarfs, but often show evidence of low surface gravity in their spectra \citep{Barman:2011dq,Barman:2011fe}.  Brown dwarfs and exoplanets alike cool monotonically with age, beginning life as hot M-type objects, cooling to the L spectral type (with very red near-IR colors and silicate condensate clouds \citep{Kirkpatrick2005}, the cooler T spectral type  \citep[with blue near-IR colors and strong methane absorption at 1.6 $\mu$m and 2.2 $\mu$m,][]{Kirkpatrick2005} and eventually to the very cool Y spectral type \citep[]{Cushing2011, Kirkpatrick2012}.  Thus, there is an age / mass / temperature degeneracy for these objects, rendering mass estimates based on models very sensitive to the age assumed for the system.  

Most mass estimates for imaged planets depend on luminosity evolution models, challenges for which are described in Section~\ref{sec:interp-challenges}.   Some planets -- e.g. HR 8799 bcde, $\beta$ Pic b, 51 Eri b, and HD 206893 b -- have dynamical mass measurements or limits.   Masses inferred from luminosity evolution or derived from dynamics are typically $\sim$5--10 $M_{\rm J}$.  Some planets have masses near the deuterium-burning limit ($\kappa$ And b, HD 206893 b); 51 Eri b may have a far lower mass (as little as $\sim$2 $M_{\rm J}$).   Mass ratios for most imaged planets are $q$ $\sim$ 0.005--0.01: 51 Eri b may have the lowest mass ratio ($q$ $\sim$ 0.001).  Even leading extreme AO systems are typically not sufficiently sensitive to detect young Jupiter-mass planets, let alone Saturn-mass planets.   However, future capabilities will close these gaps (See Section~\ref{sec:future}).

\subsubsection{Emblematic Systems}
While planetary mass companions to brown dwarfs had previously been detected as early as 2004 \citep[most~notably~2M1207b,~][]{Chauvin2004}, the near simultaneous announcements of HR 8799bcd and $\beta$ Pic b are widely regarded as the first bonafide directly-imaged exoplanet detections \citep{Marois2008, Lagrange2009}.   Planets in both systems were detected from ground-based facility AO systems in the near-to-mid IR in thermal emission, not scattered light.
Just over 7 years later, \citet{Macintosh2015} announced the first exoplanet discovered with extreme AO: 51 Eri b.  



\textit{\textbf{HR 8799}} -- HR 8799 is a nearby ($\sim$40 pc) mid-A field star with an estimated age of $\sim$40 Myr \citep{Baines2012}.
The star hosts a massive, resolved Kuiper belt-like debris disk and a warm debris population interior to 10 au consistent with an asteroid belt \citep{Sadakane86,su09,Matthews2014, Booth2016}: possible signposts of massive, perturbing planets.  

In 2008, \citeauthor{Marois2008} announced the direct imaging discovery of HR 8799 bcd followed by a fourth planet discovery in 2010 \citep[HR 8799 e;][]{Marois2010a}, located at projected separations of 15--70 au (Figure \ref{fig_DIarch}, left panel).   Soon after HR 8799 bcde's announcements, other studies identified one or more planets in archival or separately-obtained data \citep{Lafreniere2009,Fukagawa2009,Metchev2009,Soummer2011,Currie2011,Currie2012b}.
A decade of Keck Observatory monitoring showed that HR 8799 bcde orbit close to the 1:2:4:8 resonance \citep{Konopacky16}.  Current analyses suggest that the planet orbits are nearly coplanar with the disk, with a small inclination of $\sim$27$^{\circ}$ \citep{Konopacky16}.
However, HR 8799 e may not orbit on a plane strictly coplanar with HR 8799 bcd \citep{Lacour19}. 


Originally, masses inferred from the planets' luminosities spanned a wide range of values (5--13 $M_{\rm J}$) \citep{Marois2008b,Marois2010b}.  However, dynamical stability modeling strongly favors masses below 10 $M_{\rm J}$ (7 $M_{\rm J}$) for HR 8799 cde (HR 8799 b) \citep{Marois2010b,Currie2011,Sudol2012}.  Using \textit{Hipparcos} and \textit{Gaia}, \cite{Brandt21} measured a dynamical mass for HR 8799 e of 9.6$^{+1.9}_{-1.8}$ $M_{\rm J}$, consistent with these limits (see Section~\ref{subsec:synergies}).  
Additional planets may explain the shape of the inner edge of HR 8799's cold belt \citep{Booth2016}, especially if they are near or below Jupiter's mass \citep{Gozdziewski18}. Any planets interior to HR 8799 e must be below 3--4 $M_{\rm J}$ at 7--10 au and 5--6 $M_{\rm J}$ at 4--7 au \citep{Zurlo2016,Wahhaj2021,Brandt21}.   The HR 8799 planetary system resembles a scaled-up version of our own outer solar system \citep{Marois2010b}.

HR 8799 bcde have been benchmark objects for understanding the atmospheres of young jovian planets.  Their photometry and low-resolution spectra differ from those of older, field substellar objects thought to have similar temperatures, identifying features diagnostic of clouds, chemistry, and gravity
\citep{Currie2011,Barman:2011fe,Galicher2011,Marley2012,bonnefoy14}.
Higher resolution spectra probe molecular abundances connected to formation mechanisms \citep{Konopacky2013,Barman2015}.   Section~\ref{sec:characterization} discusses HR 8799 bcde's spectra and atmospheres in more detail.


\begin{figure*}[h!]
 \centering
 \includegraphics[angle=0,width=.95\textwidth,trim=0mm 0mm 0mm 0mm,clip]{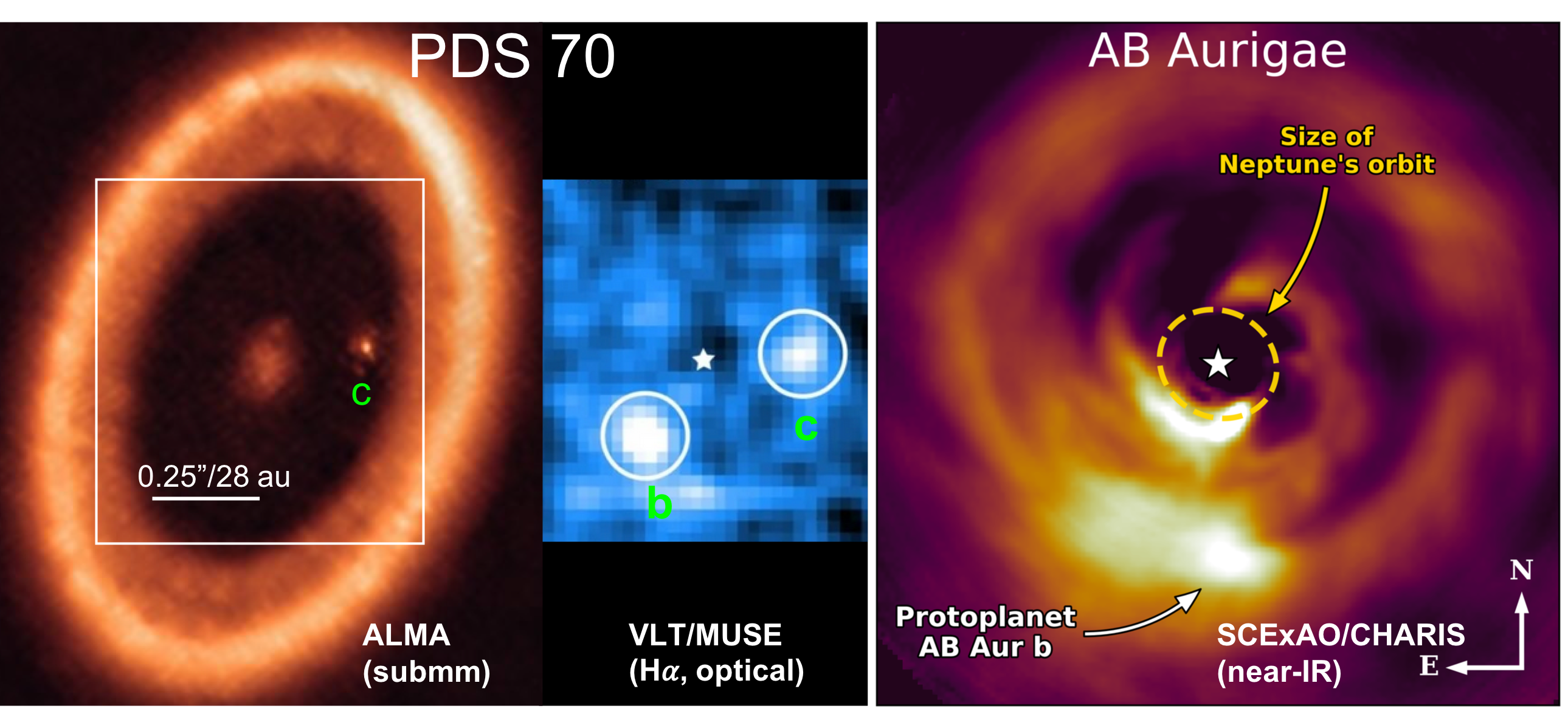}
 \vspace{-0.175in}
\caption{\small (left) ALMA image of PDS 70 showing a cleared protoplanetary disk cavity and a circumplanetary disk around PDS 70 c and H-$\alpha$ image showing the detection of PDS 70 b and c \citep{Benisty2021,Haffert19}. (right) SCExAO/CHARIS image (rescaled by stellocentric distance) of AB Aurigae showing a bright, concentrated emission source revealing the location of a protoplanet \citep{Currie2022}.}  
\vspace{-0.1in}
\label{figpds70}
\end{figure*}

\textit{\textbf{$\beta$ Pic}} -- The A6V star $\beta$ Pictoris is the eponymous member of a collection of kinematically-associated stars known as the $\beta$ Pic Moving Group (age$\sim$20 Myr).   The star hosts an edge-on debris disk, imaged since the mid-eighties from the ground and space \citep[e.g.][]{SmithTerrile1984}.  The disk consists of a planetesimal ring at 70--100 au responsible for the bulk of the dust through collisions and an outer extension comprised of small dust grains blown out by radiation pressure \citep[][]{lagrange00}.

\citet{Lagrange2009} first identified $\beta$ Pic b from data obtained in 2003 with VLT/NaCo; follow-up data obtained one year later confirmed the planet \citep{Lagrange2010}.   Thanks to $\beta$ Pic b's small semi-major axis ($\sim$ 9 au), over 75\% of its orbit has been monitored \citep{Lagrange20, Nowak2020b}.   More recently, a second planet ($\beta$ Pic c) was identified from RV data and then recovered with high-contrast inteferometric imaging with VLTI/GRAVITY \citep{Lagrange19,Nowak2020a}.   


$\beta$ Pic b may partially explain the observed warp in the edge-on debris disk and the many evaporating exocomets identified over the past 30 years \citep{lagrange00, Kiefer14}.   Planets besides $\beta$ Pic bc could explain why the inner 70 au is relatively devoid of debris dust; localized features in the disk also suggest additional planets could be present.  From RV and astrometric data, \citet{Brandt21bp} estimate a mass of 9.3 $^{+2.5}_{-2.6}$ $M_{\rm J}$ and 8.3 $\pm$ 1 $M_{\rm J}$ for $\beta$ Pic b and c, respectively \citep[see also][Section~\ref{subsec:synergies}]{snellen2018,Nielsen20}, although these values heavily depend on the assumed uncertainties on the RV data. RV and direct imaging data combined exclude additional planets more massive than 2.5 M$_{Jup}$ from 0.1 au to hundreds of au \citep{Lagrange20}.  $\beta$ Pic b is more luminous and hotter 
than the HR 8799 planets ($T_{\rm eff}$ $\sim$ 1700--1800 K), just slightly redder than field early L-type dwarfs, and is likely cloudy/dusty with a low gravity \citep{Bonnefoy2013,Currie2013, Chilcote:2017fv}; $\beta$ Pic c 
probably has a temperature intermediate between its sibling and the HR 8799 planets \citep{Nowak2020b}.

\textit{\textbf{51 Eri}} -- 51 Eridani is a 29 $pc$-distant early F star, a member of the $\sim$20 Myr-old $\beta$ Pic Moving Group, and a member of a wide hierarchical triple system that includes an M-dwarf binary GJ 3305 \citep{Feigelson06}.
51 Eri has a detected infrared excess, modeled with a cold dust belt located approximately between 5 and 80 au \citep{Patel14, Riviere14}. 

The GPI campaign (GPIES) team discovered 51 Eri b, a faint planet at $r_{\rm proj}$ $\sim$ 13 au \citep[Figure 6, right panel;][]{Macintosh2015}.  The planet is the first discovered using extreme AO and the first incontrovertible T dwarf planet, showing strong methane absorption in H band.  The planet is likely more eccentric than either $\beta$ Pic b or HR 8799 bcde \citep[$e$ $\sim$ 0.4--0.62]{Maire19,deRosa20}.  
If confirmed, an eccentric orbit could indicate the presence of an additional massive body or could be due to gravitational perturbations from GJ 3305AB.  Assuming a hot-start luminosity evolution, current data rule out other planets more massive than 4 $M_{\rm J}$ beyond 5 au and more massive than 2 $M_{\rm J}$ beyond 9 au \citep{samland17}. 

The mass of 51 Eri b is not well constrained: values derived from comparing 51 Eri b's luminosity and age to evolutionary models favor $\sim$2 $M_{\rm J}$, while atmospheric modeling may favor larger values, up to $\sim$ 9 $M_{\rm J}$ \citep{Macintosh2015,samland17}.  Different characterization studies also find slightly diverging atmosphere properties, illustrating the challenge associated with characterizing very faint exoplanets with direct imaging \citep{samland17,Rajan2017}.  Analysis of \textit{Gaia} and Hipparcos astrometry set an upper limit of 11 $M_{\rm J}$ for 51 Eri b \citep{Dupuy2022a}. The planet likely has a temperature of $\sim$700 K and either lacks clouds or is only partially covered by clouds \citep{Rajan2017, samland17}.
\subsection{\textbf{Protoplanets} }

Direct images of planets in active assembly (\textit{protoplanets}) around stars that still retain gas and dust-rich protoplanetary disks clarify how and where planets form.   The large distances to the nearest star-forming regions ($\sim$150 pc) mean that 
protoplanets orbiting their host stars at solar system scales are located at very small angular separations.  However, protoplanets can be  bright ($L$ $\sim$ 10$^{-2}$--10$^{-3}$ $L_{\rm \odot}$), especially if they are surrounded by their own circumplanetary disks \citep[e.g.][]{Zhu2015}.

The 5 Myr-old 0.87 $M_{\rm \odot}$ star PDS 70\index[obj]{PDS 70} hosts the first incontrovertible detections of jovian protoplanets: PDS 70 b and PDS 70 c \citep{Keppler2018,Keppler2019,Muller2018,Haffert19} (Figure \ref{figpds70}, left).  Both protoplanets are located within the PDS 70 disk cavity, at angular separations of $\rho$ $\sim$ 0\farcs{}18 and 0\farcs{}24 and estimated semimajor axes $\sim$20 and 34 au \citep{Wang2021}.   Dynamical arguments strongly favor masses less than 10 $M_{\rm J}$ for PDS 70 b, while PDS 70 c's mass is more poorly constrained.  Masses inferred from SED modeling range between 1 and a few jovian masses \citep[][]{Stolker2020}.  PDS 70 b is slightly eccentric
($e$ $\sim$ 0.17 $\pm$ 0.06), while PDS 70 c's orbit is consistent with being circular. The protoplanets' IR data are best fit by model atmospheres with substantial dust/extinction.   

PDS 70 bc show $H_{\rm \alpha}$ emission consistent with accretion at rates of 1--2$\times$10$^{-8}$ $M_{\rm \odot}$ $yr^{-1}$, slightly less than the stellar accretion rate \citep{Haffert19}, but lack evidence for Br-$\gamma$ accretion \citep{Wang2021}.  PDS 70 c shows direct evidence for a circumplanetary disk with an estimated mass of 0.007--0.03 $M_{\rm \oplus}$.  Thermal IR data may suggest  that PDS 70 b is surrounded by a circumplanetary disk \citep{Stolker2020,Wang2021}.

Recently, data from Subaru/SCExAO and the Hubble Space Telescope over 13 years reveal evidence for a wide-separation ($\sim$93 au) embedded protoplanet around the 1--3 Myr-old, 2.4 $M_{\odot}$ star AB Aurigae\index[obj]{AB Aurigae} \citep{Currie2022}.  AB Aur b is consistent with a protoplanet responsible for the millimeter dust cavity and CO gas spirals both seen by ALMA \citep{Tang2017}.  It appears spatially extended, plausibly due to light from the central source reprocessed by the star's protoplanetary disk.   The best-fit composite model explaining AB Aur b's optical to near-IR emission includes a 2.75 $R_{\rm J}$, 9 $M_{\rm J}$ source with a surface gravity of log(g) = 3.5, emitting at a much hotter temperature than PDS 70 bc ($\sim$2000--2500 $K$), and accreting at a rate of $\dot{M}$ $\sim$ 1.1$\times$10$^{-6}$ $M_{\rm J}$ $yr^{-1}$.   The source is detected in $H_{\rm \alpha}$ although it is unclear whether this emission is due to accretion.  Embedded in a massive disk with numerous spiral arms at over three times Neptune's disk instead of in a fully cleared cavity like PDS 70 bc, the properties of AB Aur b may point to a planet formation mechanism by disk instability (see Section 6).

Prior to the discovery of PDS 70 bc, other studies claimed detections of protoplanets located within the gaps of or embedded in disks around young stars.   The 2-solar mass, protoplanetary disk-hosting star HD 100546 has a protoplanet candidate at $\sim$ 50 au and another at $\sim$ 13 au, just interior to the gap in the protoplanetary disk \citep[HD 100546 bc;][]{Quanz2013,Currie2015}.  HD 100546 b has been detected in multiple data sets but evidence for orbital motion is not yet clear \citep{Rameau2017,Sissa2018}; HD 100546 c has been imaged in a single data set and inferred through spectroastrometry but not yet imaged in subsequent data \citep{Currie2015,Brittain2014,Sissa2018}.  Interpreting both candidates -- whether a planet or disk feature -- is challenging \citep{Currie2015,Currie2017,Rameau2017,Sissa2018}.  Both candidates require further study and confirmation.

Two studies presented detections of protoplanets around the young, Sun-like star LkCa 15 through a combination of sparse aperture masking interferometry (SAM) and $H_{\rm \alpha}$ differential imaging \citep[LkCa 15 bcd;][]{Kraus2012,Sallum2015}.   However, later direct imaging observations showed that the SAM detections correspond to disk features \citep{Currie2019}: LkCa 15 b technically remains a candidate for now due to its single epoch $H_{\rm \alpha}$ detection (though see \citealt{Mendigutia2018}).    Other claimed protoplanet detections have been revealed to likely be misidentified disk signals instead of planets \citep[e.g.][]{Rich2019}.   

\subsection{\textbf{Challenges with Interpreting Detections}}
\label{sec:interp-challenges}

\subsubsection{Confirming Companionship}
Direct imaging observations reveal many point sources that are unrelated background stars instead of bound companions, especially for systems in the Galactic plane \citep[e.g.][]{Galicher2016}.   Confirming candidate planets as bound -- i.e. sharing common proper motion with and orbiting their stars -- often requires multi-year observations, depending on the star's proper motion. In some pathological cases, background stars can have a non-zero proper motion and thus are more easily confused with bona fide planets \citep[e.g.][]{Nielsen2017}.   In the absence of full confirmation, an object's near-IR spectrum can provide strong evidence that it is a directly-imaged exoplanet \citep[e.g. for 51 Eri b;][]{Macintosh2015}.

\subsubsection{Estimating Accurate Masses}
Masses for imaged planets and planet candidates are typically not directly measured but are instead estimated using luminosity evolution models that map between an object's brightness and mass as a function of age.  Accurate mass estimates therefore are affected by uncertainties in these models and require precise system ages.  \textit{Hot start} models for planet luminosity evolution assume a high initial entropy, resulting in bright planets for the first 1--100 $Myr$ \citep{Baraffe2002,Burrows2001,Burrows2003}.  \textit{Cold start} models assume a low initial entropy, resulting in planets that are substantially fainter for the first 100 Myr \citep{Marley07}.   Originally, hot start models were used to describe planets formed by disk instability, while cold start models described planets formed by core accretion \citep{Marley07}\footnote{See Section~\ref{subsec:testing-formation} for definitions of the core accretion and disk gravitational instability formation pathways}.   However, recent models show that core accretion-formed planets can be compatible with hot start-like luminosity evolutions, and jovian planets may form with a range of initial entropies \citep{Mordasini2017,Berardo2017,Spiegel2012}.  Aside from 51 Eri b, all planets imaged thus far are only consistent with a hot-start luminosity evolution.   

Precise ages can be exceptionally challenging to derive for isolated stars \citep[e.g.][]{Soderblom2014}. 
Members of young moving groups or other nearby associations can be age-dated using a variety of methods: e.g. fitting group Hertzsprung-Russell (HR) diagram positions with stellar evolutionary models, Lithium abundances.  Thus, moving group or association membership can yield a star's age.  However, even stars with a motion similar to bona fide moving group members could instead be interlopers, especially if they have dissimilar space positions. For isolated Sun-like stars, stellar rotation and activity can give age estimates, albeit with significant scatter \citep{Barnes2007, Delorme2011, Angus2019}. For early-type stars,  HR diagram positions have provided approximate ages: optical interferometry can now provide more precise estimates by resolving the stars themselves \citep[e.g.][]{Jones2016}.

Revisions in the stellar age often affects the interpretation of an imaged companion.  
For example, based on the primary's claimed membership in the Columba association, $\kappa$ And b was originally thought to be 12--13 $M_{\rm J}$ \citep{carson13}.  In contrast, the primary's HR diagram position resembles an older system (220 Myr), which would imply the companion is $\sim$ 50 $M_{\rm J}$ \citep{Hinkley2013}.  However, through optical interferometry, \citet{Jones2016} showed that the star is likely (nearly) coeval with Columba even if it is not a member ($\sim$47 Myr), implying the object is a planet-mass companion.   
Later, near-IR spectra of $\kappa$ And b were found to be consistent with a planet interpretation \citep[][]{Currie2018}.  
 In contrast, the imaged companion to GJ 504 was announced as a 3--8.5 $M_{\rm J}$ planet with an age of 100--510 Myr \citep{Kuzuhara2013}.   Based on interferometric, RV, and high contrast imaging data, \citet{Bonnefoy2018} derived a mass range of 1--23 $M_{\rm J}$, while other analysis of the star suggested an age of about 2.5 Gyr and a mass well into the brown dwarf regime \citep{DOrazi2017}.

\subsubsection{Planet or disk feature?}
Distinguishing between highly-structured disk signals and bona fide protoplanets is a steep, chronic challenge.   Advanced algorithms needed to detect protoplanets can attenuate both disk and planet signals: forward-modeling is required to ensure that a claimed planet detection is not a partially subtracted piece of the disk \citep{Currie2017}.  For systems with hot dust (T $\sim$ 1000--2000 K) near the star that intercepts and reprocesses emission, the disk's scattered light spectrum may strongly resemble spectra of bona fide protoplanets \citep[][]{Mulders2013}.  Orbital motion over multi-year timescales can better establish that a signal is an orbiting planet and not a static disk feature \citep{Keppler2018}.  
$H_{\rm \alpha}$ emission can also pinpoint actively accreting protoplanets \citep{Haffert19}; however, accretion onto protoplanets embedded in disks may be unidentifiable, since optical extinction likely renders  $H_{\rm \alpha}$ undetectable.  

Fomalhaut b may represent the first of yet another class of objects with a challenging interpretation \citep{Kalas2008}.  The object was initially identified as a directly imaged planet, made visible by both reflected light from a circumplanetary disk and thermal emission, and responsible for sculpting the star's Kuiper belt-like debris disk.  However, later analysis showed that Fomalhaut b's spectrum is completely explained by scattered starlight and its orbit likely crosses the ring \citep{Janson2012,Currie2012,Galicher2013,Kalas2013}: the object is lower in mass and likely made visible purely by circumplanetary dust.   Recently, from analyzing archival and unpublished data, \citet{Gaspar2020} proposed 
that Fomalhaut b may be fading and dispersing, consistent with a massive planetesimal collision.  Other analyses find no clear evidence for these two trends \citep{Currie2012}, although they were conducted only over a subset of available data.   Future observations of Fomalhaut b and/or reanalyses of recent data may clarify its true nature.

\subsection{\textbf{Synergies with Other Techniques\label{subsec:synergies}} }

The limitations of direct imaging affect our ability to interpret data for individual objects and to draw conclusions from large-scale population studies.  For individual objects, mass estimates depend on both the age of the planet and the chosen luminosity evolutionary model (see Section~\ref{sec:interp-challenges}).   
At a population level, the angular resolution of the telescope  and the achievable contrast of the instrument limit the range of detectable planetary masses and semi-major axes with direct imaging. Combining direct imaging with other planet detection techniques partially mitigates these limits.

Optical interferometry provides a means to directly detect bright, modest planets at small angular separations.   As a prime example, GRAVITY interferometer coherently combines the light of all four VLT telescopes, yielding the equivalent angular resolution of a 130-m telescope \citep{GRAVITY2017}.  While the GRAVITY field-of-view is small ($\sim$50 mas), indirect detection techniques (e.g. RV) cna help predict the position of an exoplanet candidate \citep[e.g.][]{Nowak2020b}.  GRAVITY has yielded exceptionally high-SNR spectra and ultra-precise astrometry of known planets \citep{Lacour19}; the combination of GRAVITY and indirect techniques have now resulted in new planet discoveries \citep{Hinkley2022b}.

Relative astrometric measurements of a planet around its host star cannot alone be used to precisely measure the mass of the orbiting planet; the semi-major axis of the relative orbit encodes the system's total mass, not the individual components' masses. The system's mass ratio, and thus the mass of the planet, can be measured if the semi-major axis of the orbit of the star around the system barycenter is known. This orbit can be measured either using absolute astrometric measurements from catalogues such as {\it Hipparcos} and {\it Gaia}, or spectroscopic observations to measure the Doppler shift of the star's spectral lines over the course of the orbit.  In some cases, high-resolution spectroscopy can measure the Doppler shift of the planet's spectral lines \citep{Snellen2014,Wang2021}, also yielding the mass ratio.  Combining the star's orbit around the barycenter with the relative astrometry between star and planet yields the mass ratio, and a dynamical mass for the planet.  In multi-planet systems with high-precision astrometry, the the planets' mutual gravitation can be detected by fitting for deviations from Keplerian motion, yielding their mass estimates \citep{Lacour2021}.

The $\beta$ Pic system is the current benchmark example of combining planet detection techniques to measure model-independent masses.  Relative astrometry of $\beta$ Pic b and c combined with the star's astrometric and RV measurements provides dynamical masses for the planets \citep{snellen2018,Lagrange19,Dupuy2019}.  Measurements of non-Keplerian motion due to the mutual gravity of $\beta$ Pic b and c are also now possible given the precision of recent interferometric monitoring campaigns \citep{Nowak2020b}. 

Combining astrometry and/or RV and direct imaging data can yield dynamical masses for other directly-imaged planets, as well as for numerous brown dwarf companions 
\citep{grandjean19,Brandt20b,Brandt21}.  The upcoming {\it Gaia} data releases for the full and extended mission will be used extensively in future detection and characterization studies of directly-imaged planets, yielding further dynamical masses for directly imaged companions. The extended mission will provide precision astrometric measurements of the host star over $\approx$10 years.

Combining direct imaging with other detection techniques can improve planet occurrence rate measurements and can better determine the planet frequency distribution
over a range of masses, orbital periods, and host star properties.
Long-term Doppler surveys are largely complete to giant planets within $\sim$5 au, and partially complete to giant planets out to $\sim$10 au, the current inner limit of sensitivity of direct imaging surveys.  Thus, combining results from both techniques 
reveals giant planet demographics out to $\sim$100 au (Section~\ref{sec:occurrence}).  RV studies of young stars (e.g. \citealt{Lagrange19,grandjean2021}) allow occurrence rates to be compared for stars of similar ages, 
tracing the extent of giant planet migration over system lifetimes and its effect on planet frequency.  The final \textit{Gaia} planet catalog \citep{Perryman2014} will also help place demographic measurements of young, wide-separation giant planets into context, as astrometry will be better able to probe intermediate separation giant planets ($\sim$1-10 au, $\gtrsim$1 M$_{\textrm Jup}$) around younger, higher-mass stars compared to RV.  These target stars more closely match the hosts of imaged planets.  A Gaia-selected survey of accelerating stars has now led to the first joint direct imaging and astrometry discovery of an exoplanet \citep{Currie2022b}.   Combining Gaia observations of younger, higher-mass stars with direct imaging will illuminate trends in occurrence rate as a function of stellar mass and age. 

Direct imaging also shares an important overlap with planet detection by microlensing. 
Both techniques are sensitive to planets beyond the snow line, with microlensing probing lower planet-star mass ratios (e.g. \citealt{Suzuki2016}), while the host mass of imaged planets can be directly determined, and the planet mass inferred from models.  In addition, singly lensed short-period microlensing events can represent either a free-floating planet or a wide-separation bound planet \citep{Sumi2011}.  Constraints on the wide-separation giant planet population can be determined from imaging, thus more definitively constraining the free-floating planet occurrence rate.

\section{\textbf{Atmospheric Characterization of Directly Imaged Exoplanets} \label{sec:characterization}} 

Direct imaging 
enables
characterization of young (age $<$200 Myr) jovian exoplanets at wider separations with negligible irradiation compared to most transiting planets. 
 In this section, we first summarize from an empirical standpoint what we have learned about the atmospheres of the current cohort of young directly imaged giant planets from their photometry and spectroscopy. We then consider the theoretical side of the picture, in particular, the state-of-the-art in how we model these complex, cool atmospheres.

\subsection{\textbf{Empirical Constraints from Time-Averaged Observations}}


\subsubsection{Photometry} 
\label{subsec:photometry}
\begin{figure}[h]
\vspace{-0.1in}
 \epsscale{1.0}
  \plotone{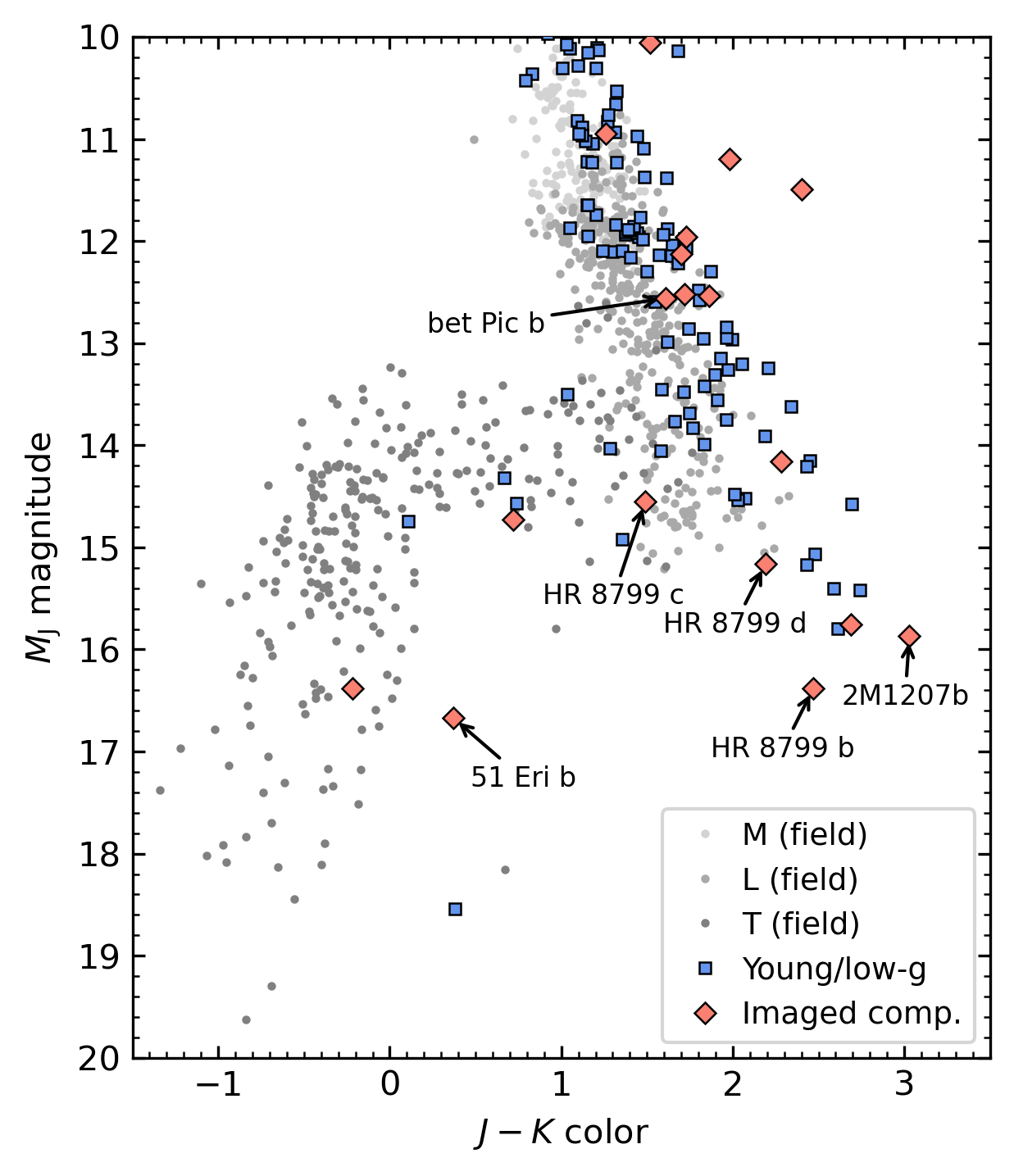}
 \vspace{-0.35in}
\caption{\small Near-infrared color-magnitude diagram of field M, L and T-dwarfs (gray circles), isolated objects exhibiting evidence of youth or possessing low surface gravity (blue squares), and directly imaged substellar/planetary-mass companions to stars and brown dwarfs (red diamonds).  We annotate selected imaged exoplanets/planet-mass companions. Young, low-gravity objects form a sequence displaced redward from that of older, higher-gravity field objects. Data obtained from the UltracoolSheet.}  
\vspace{-0.05in}
\label{fig:vlm-cmd}
\end{figure}

Photometry in the major near-to-mid IR passbands -- $J$, $H$, $K$, $L_{\rm p}$, $M_{\rm s}$ -- covers the bulk of emission for young superjovian planets and provided the first empirical diagnostic of young, directly-imaged planet atmospheric properties.  Figure \ref{fig:vlm-cmd} displays a typical near-IR color-magnitude diagram for selected directly-imaged planets and other planet-mass companions compared to older, more massive field brown dwarfs and younger brown dwarfs whose ages and masses partially overlap with those of most imaged planets. Imaged planets and planet-mass companions span the full luminosity range characteristic of mature late-M, L, and T field brown dwarfs (vertical axis).  However, young brown dwarfs, planets, and planet-mass companions typically have redder colors than field objects.   

The most pronounced differences between planet and field brown dwarf photometry occur 
near the transition from methane-poor L-type dwarfs to methane-absorbing T-type dwarfs.   The first imaged planet-mass companion -- 2M 1207 B -- and some of the first imaged exoplanets (e.g. HR 8799 b) are particularly discrepant, appearing to populate a previously empty
part of these diagrams consistent with a reddened extension of the L dwarf sequence to lower luminosities \citep{Chauvin2004,Marois2008b,Currie2011,Barman:2011fe}.  Over the full 1--5 $\mu m$ spectral range, photometry for these objects appears redder and more blackbody-like (Figure \ref{fig:atmoschardemo}, top-left).  More recent studies show that a number of other young imaged planets/planet-mass companions -- e.g. HD 95086 b, TYC 8998-760-1 c, HD 206893 b, 
2M 2236+4751 B -- also populate this region \citep[e.g.][]{Rameau2013,Bohn2021,Delorme2017b,Bowler2017}.
 
Differences between field and young objects with T spectral types are less clear.   Few young, T-type planet-mass companions have been identified in the last decade: e.g. 51 Eri b \citep{Macintosh2015}, GU Psc B \citep{Naud2014}. While GU Psc B follows the sequence of field brown dwarfs in near-IR color-magnitude diagrams, the exoplanet 51 Eri b is redder than field brown dwarfs. Four additional companions orbiting primary stars with intermediate ages \citep[$\le$1 Gyr; HN Peg B, ROSS458C, BD204-39B, Coconuts-2b,][]{Luhman2007, Goldman2010, Zhang2021b} show similar but less pronounced deviations with respect to field dwarfs in the same luminosity range. Some directly-imaged L/T transition planets and planet-mass companions have color-magnitude diagram positions discrepant from field objects in thermal IR passbands probing methane absorption (3.3$\mu m$, $M_{\rm s}$) but have positions similar to field objects in other passbands (e.g. $L_{\rm p}$, 4.05 $\mu m$) \citep{Skemer2012,Skemer2014,Currie2014a}.

\subsubsection{Spectroscopy\label{subsec:spectroscopy}}
\begin{figure}[h]
 \epsscale{1.0}
  \plotone{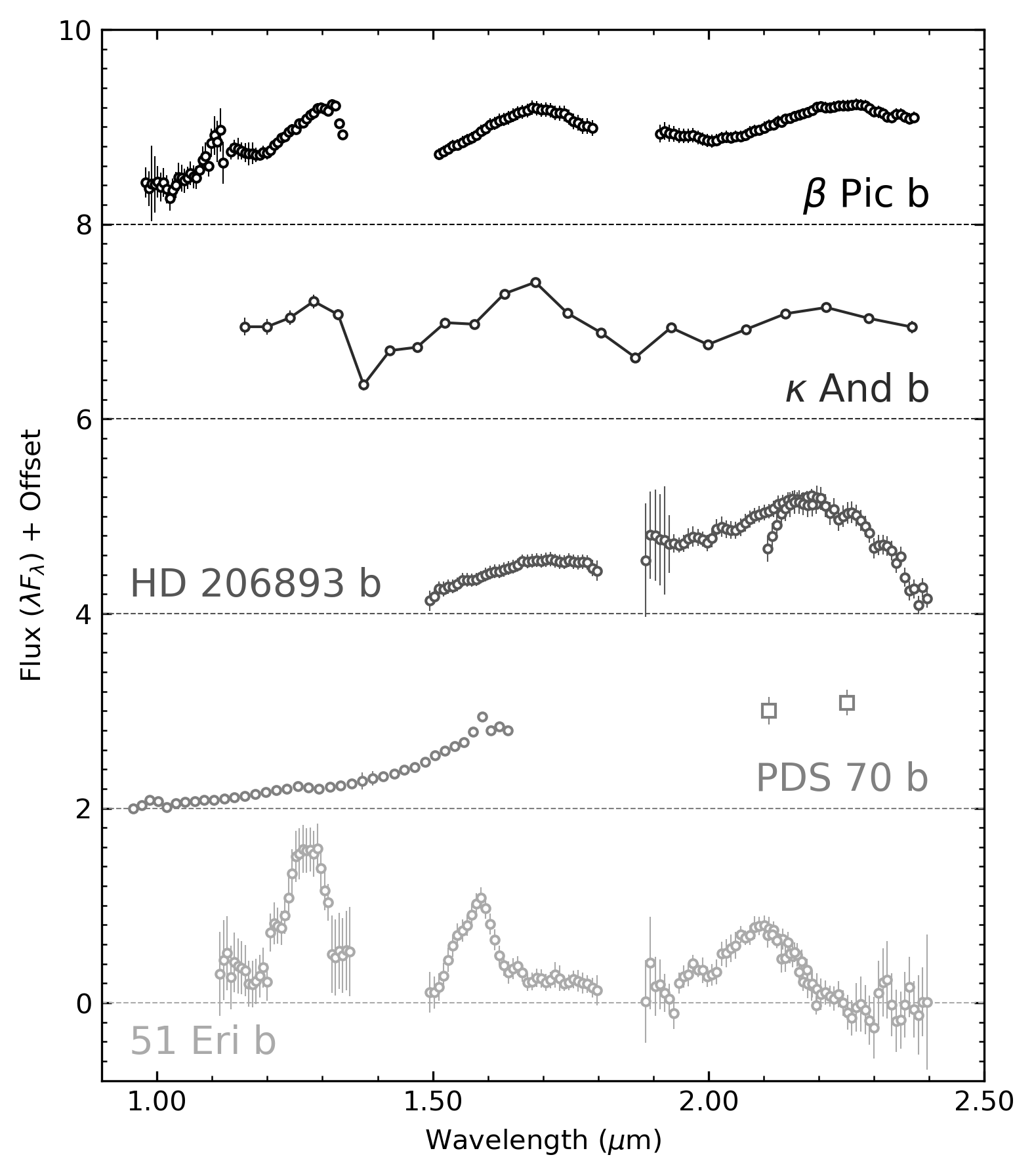}
 \vspace{-0.35in}
\caption{\small Near-infrared spectra of imaged exoplanets obtained using low-resolution integral-field spectroscopy: from top-to-bottom,  $\beta$ Pic b \citep{Chilcote:2017fv}, $\kappa$~And~b from Subaru/CHARIS taken in a lower-resolution single-shot mode \citep{Currie2018}, HD 206893 b \citep{Ward-Duong2021}, PDS 70b from VLT/SPHERE \citep{Mesa2019}, and 51 Eri b \citep{Rajan2017} from Gemini/GPI. The spectra are normalized by the flux between 2.05--2.15\micron~(1.6\micron~for 51 Eri b for clarity). Dashed lines indicate the offset applied to each spectrum.}
\vspace{-0.05in}
\label{fig:spec-gallery}
\end{figure}



Over the past decade, IFS instruments, especially those used in combination with extreme AO systems (P1640, SPHERE, GPI, SCExAO-CHARIS) have provided critical low-resolution (R$\sim$20 to 80) near-IR (1-2.5$\mu$m) spectra of most directly-imaged exoplanets, a representative sample of which is displayed in Figure \ref{fig:spec-gallery}.  
Near-IR spectra provide key diagnostics of brown dwarf and planet atmospheres. Spectral indices derived from low-resolution data like H$_{\rm 2}$0 yield coarse estimates of spectral types \citep{AllersLiu2013}.  The $H$-band continuum index and (possibly) the $H_{\rm 2}$-K index  is a diagnostic of surface gravity \citep{AllersLiu2013,Canty2013}.   

Following trends from photometry, the spectra of young directly-imaged planets and other planet-mass companions show significant differences with the spectra of field brown dwarfs. Among the most notable are the following:
\begin{itemize}
    \item \textit{Chemistry} -- The L/T transition traces the onset of methane absorption. 
    However, spectra of some L/T transition exoplanets and planet-mass companions -- e.g. the HR 8799 planets, 2M 1207 B -- show a lack of methane absorption compared to field dwarfs with similar temperatures and CMD positions \citep{Barman:2011fe,Barman:2011dq,Bonnefoy2016,Greenbaum2018}.   Thermal IR spectra confirm that the HR 8799 planets also have weaker absorption in the 3.3 $\mu m$ methane filter \citep{Doelman2022}.
    
    \item \textit{Gravity} -- The $H$-band spectra for L dwarf and L/T transition directly imaged exoplanets / planetary-mass objects such as HR 8799 b, $\kappa$ And b, ROXs 42Bb
    show highly peaked $H$-band spectra and/or red $K$-band spectra compared to field dwarfs \citep[][]{Lucas2001, Barman:2011fe,Currie2018,Currie2014b, Bowler2014,AllersLiu2013, Liu2013, Gauza2015}.  The $H$ and $K$-band shapes probe collisionally-induced absorption (CIA) of hydrogen; lower gravities result in weaker CIA and thus peaked $H$-band peaks and redder $K$-band slopes.  
    
    \item \textit{Dust} -- At least some of the HR 8799 planets have spectral properties and molecular absorptions that are well matched by those of young free-floating objects at the L-T transition (see also Section \ref{subsec:variability}).  However, no object yet reproduces the available 1-2.5$\mu$m spectra of HR8799b, which may be due to the lack of identified young free-floating T-type objects. A handful of early-T-type objects such as the AB Doradus member 2MASS J13243553+6358281  \citep{Gagne2018e} (see Section \ref{subsec:photometry}) can reproduce the spectral bands of HR 8799 b provided that an extra layer of extinction by sub-micron dust particles is applied to these empirical template spectra to match the spectral slope of the planet \citep{Bonnefoy2016}.  HD 206893 b presents an even more extreme example, with an even flatter, more blackbody-like spectrum yielding an extremely red spectrum from 1 to 2.5 $\mu$m, likely due to substantial atmospheric dust \citep{Milli2017,Delorme2017, Ward-Duong2021}.
    
\end{itemize}

   Spectra of the coolest and lowest mass imaged exoplanet known to date (51 Eri b) display a clear and so-far unique detection of a methane absorption at 1.6$\mu$m in the spectrum of an imaged exoplanet \citep{Macintosh2015, Rajan2017}, coincident with an enhancement in the K-band flux. Similar K-band flux enhancements have already been noted in young mid- to late-T dwarfs but the enhancement is particularly extreme in the case of 51 Eri b and is due to the reduced collision induced absorption of H$_2$ \citep{Borysow1997} in 51 Eri b's lower-pressure atmosphere. 

Protoplanets display far more featureless, blackbody-like spectra.   PDS 70 bc's spectra reveal an extremely red spectral continuum devoid of the strong water-band feature (1.3-1.4$\mu$m) expected given the observed luminosity of the planets \citep{Muller2018, Mesa2019}.  Higher-resolution K-band spectra show a lack of molecular absorption in PDS 70 b's spectrum \citep{Cugno2021}.  The circumplanetary disks and/or cocoon surrounding each planets \citep[see Fig 7 and][]{Christiaens2019, Benisty2021} produce significant foreground extinction \citep[e.g.,][]{Hashimoto2020} and may produce spectroscopic properties similar to what is seen in 
the near-infrared spectra of enshrouded class I protostars \citep[e.g. ][]{Connelley2010}.  The near-IR spectrum of AB Aur b is reproduced by a 2000-2500 $K$ blackbody but likewise lacks clear evidence for molecular absorption common in fully-formed substellar objects with similar temperatures \citep{Currie2022}.

Aside from a few isolated cases (e.g. HR 8799 bcde), the spectroscopic properties of imaged exoplanets are more poorly constrained at wavelengths longward of 2.5$\mu$m.
The 3-5 $\mu$m range is particularly interesting since it contains both methane and $^{12}$CO absorption features diagnostic of carbon chemistry (CO-CH$_{4}$) and cloud structures \citep[e.g.][]{Miles2018,Miles2020}.   
The $L_{\rm p}$-band spectra of hotter young M7 to L3 companions presented in \citet{Stone2016} show no significant difference with those of field dwarf counterparts.  The $L_{\rm p}$ spectrum of $\kappa$ And b is well matched with that of young, nearby free-floating brown dwarfs \citep{Stone2020}.
A few free-floating planet analogues have  high-quality 3-14.5 $\mu$m IRTF, AKARI, and Spitzer spectra \citep{Cushing2008, Sorahana2012}. Analysis of these data would benefit from an improved knowledge of the age and nature of these objects. 

While most spectra for directly imaged planets to date has been at resolution $R<$100, higher-resolution spectra are yielding promising results.
Near-IR IFS instruments at medium resolving powers (R=2000--6000) fed by standard adaptive-optics modules are available on multiple 8-m class telescopes (e.g. Gemini/NIFS, VLT/MUSE and SINFONI, Keck/OSIRIS).
Medium-resolution IFS data enables innovative strategies for evaluating and subtracting the stellar halo with the spatial diversity \citep[e.g.,][]{Thatte2007, Seifahrt2007, Wilcomb2020}. Most current medium resolution spectroscopic results are for
young late-M and early-L young companions straddling the deuterium-burning boundary at $\rho$ $>$0.5" \citep[e.g., ][]{McElwain2007, Schmidt2008, Patience2010, Bowler2011, Bonnefoy2014, Daemgen2017, Wilcomb2020}. However, ADI observations on these instruments \citep{Lavigne2009, Barman:2011fe, Meshkat2015} and improved data reduction strategies have recently allowed the extraction of 
high quality medium-resolution spectra of emblematic exoplanets HR8799 bc and HIP65426 b \citep{Konopacky2013, Barman2015, Petrus2021} deeply buried in the speckle noise.  The planets' spectra show fainter and narrower molecular bands (FeH, CO) and atomic lines (Na I, K I), sensitive to the surface gravity and primordial composition of the objects (C/O ratio, metallicity). They also reveal emission lines ($H_{\alpha}$/0.656$\mu$m, $Pa_{\beta}$/1.282$\mu$m, $Br_{\gamma}$/2.166$\mu$m) that trace active accretion \citep[e.g., ][]{Haffert19, Bonnefoy2014, ZhangY2021} and that can be used both for characterizing the accretion processes \citep[e.g., ][]{Aoyama2019, Thanathibodee2019} and detecting new planets \citep{Uyama2017, Haffert19}. 

\begin{figure*}[ht]
 \epsscale{1.6}
  \plotone{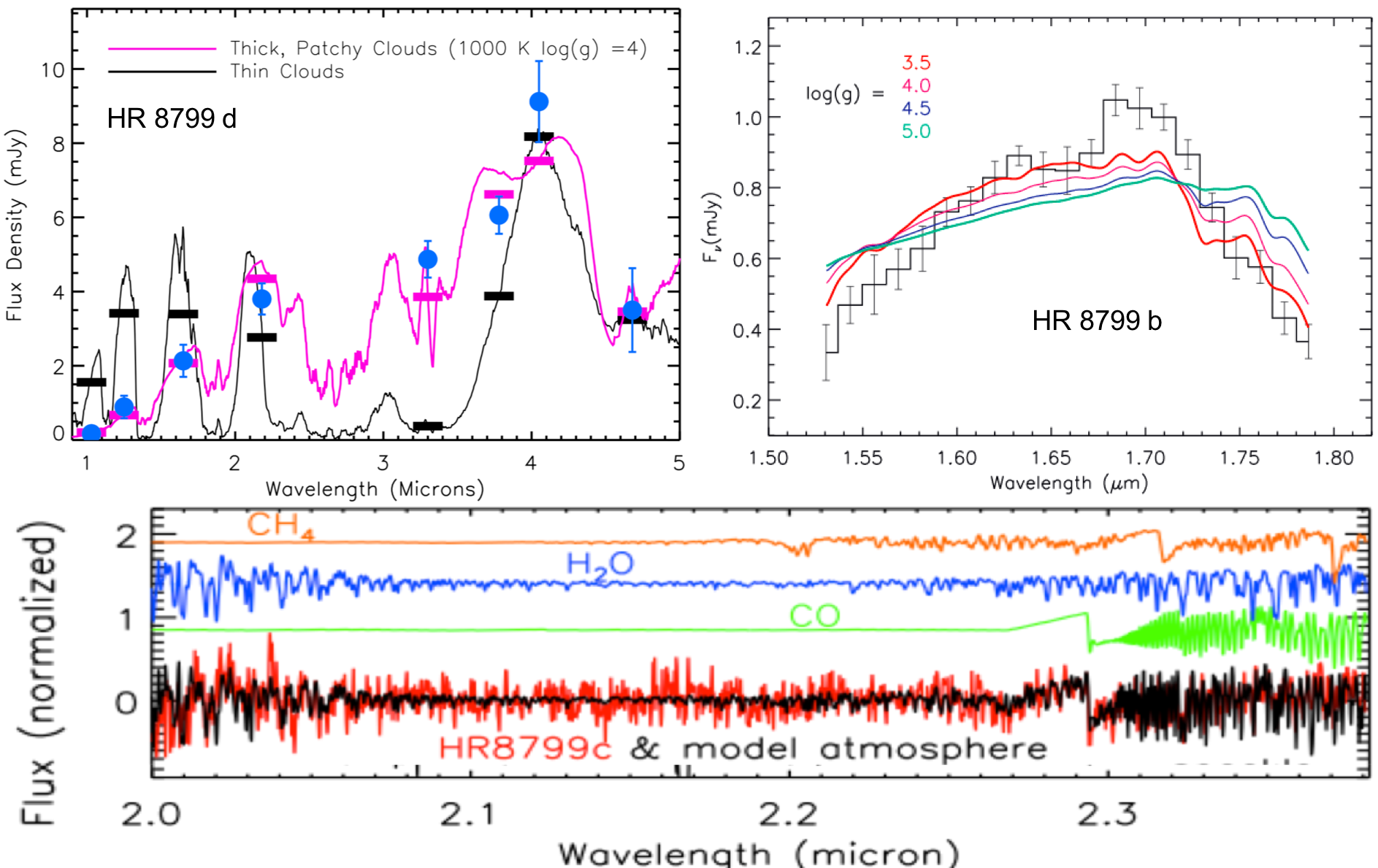}
 \vspace{-0.1in}
\caption{\small Demonstration of key properties inferred from direct imaging observations.  (top-left) Fit to HR 8799 d 1--5 $\mu m$ photometry showing that models with thick clouds better reproduce the planet data \citep[adapted from ][]{Currie2011}. (top-right) Keck/OSIRIS low-resolution H-band spectrum for HR 8799 b: a sharply-peaked H band spectrum is a signpost of low surface gravity \citep{Barman:2011fe}.  (bottom) Keck/OSIRIS medium-resolution HR 8799 c spectrum, revealing lines of $CO$ and $H_{\rm 2}O$ \citep{Konopacky2013}.   
}  
\vspace{-0.1in}
\label{fig:atmoschardemo}
\end{figure*}

Novel data processing techniques such as ``molecular mapping", in combination with medium-resolution spectra, 
can also partly deblend the many ro-vibrational lines contained in planet spectra, which form the broad absorption bands observed at lower resolution (GPI, SPHERE). These molecular patterns can be 
extricated by cross-correlating a template spectrum to the planet spectrum in order to reveal 
faint molecular signatures \citep{Konopacky2013, ZhangY2021}. ``Molecular mapping", applied to each individual spaxel of an IFS datacube can simultaneously (i) detect exoplanets whose continuum emission would have been blurred by the speckle noise and residual stray light introduced by the instrument optics, (ii) characterize unambiguously  their molecular content, and (iii) measure their Keplerian velocity \citep{Hoeijmakers2018, PetitditdelaRoche2018, Petrus2020, Cugno2021}. This method, as well as other supervised and non-supervised data processing techniques \citep{Ruffio2019, Rameau2021} applied on such IFS data will critically boost the detection and characterization capabilities on the next generation of  IFS operating on the 8-m class telescopes (GTC-FRIDA, VLT/ERIS, VLT/SPHERE+ and GPI 2.0) and the ELTs \citep[e.g.,][]{Houlle2021}.

High-resolution echelle spectroscopy (R$\geq$20,000) will go further to resolve individual molecular lines and atomic line profiles and provide a measurement of the objects' rotational broadening ($v~sin(i)$). High SNR spectra have been collected on a few young late-M and early-L companions \citep{McLean2007, Rice2010} and free-floating analogues, but these observations usually require significant telescope time. Cross-correlating lower SNR companion spectra or spectral residuals with templates can access the companion's Keplerian and rotational velocitiess, check for specific molecular signatures, and 
disentangle companions 
from the dominant speckle noise produced by the host star. This technique is being used on a steadily increasing number of companions \citep{Snellen2014, Schwarz2016, Wang2018,  Bryan2020, Xuan2020, Wang2021} and is a promising method for identifying
closer-in and fainter exoplanets (Section \ref{sec:future}).

\subsection{\textbf{Inferred Theoretical Properties from Time-Averaged Observations} }
The empirical comparisons described in the previous two sections, combined with atmospheric models spanning a wide parameter space in temperature, gravity, clouds, and chemistry, provide novel constraints on the atmospheric properties of directly-imaged exoplanets.

Figure \ref{fig:atmoschardemo} summarizes 
our current understanding of the key atmospheric properties of young, directly imaged giant planets. 
Young exoplanets likely are dustier and have thicker clouds than field brown dwarfs of the same effective temperatures (top-left panel) \citep{Currie2011,Barman:2011fe,Bonnefoy2016}.  For a fixed observed effective temperature, thicker clouds translate into hotter temperature profiles, meaning that the temperature is higher at a given atmospheric pressure.  The $\tau$ = 1 surface is more uniform with wavelength, resulting in a flatter, more blackbody-like spectrum (magenta line) compared to models with thinner clouds (black line).   This effect is most pronounced at the L/T transition -- e.g. for planets like HR 8799 bcde.   The redder colors of many L-type exoplanets are likely explained by clouds \citep[e.g.][]{Currie2013}; the redder color of 51 Eri b may also be due to at least partial cloudiness \citep{Rajan2017}.
Planets and planet-mass companions at the L/T transition also show evidence for non-equilibrium carbon chemistry, producing  
a lack of methane absorption in K band, flat spectra at 3--4 $\mu m$ and weak emission at 5 $\mu m$ \citep{Galicher2011,Konopacky2013,Skemer2014}.

Both clouds and chemistry are strongly influenced by 
surface gravity \citep{Marley2012}.   Lower gravities yield pressure-temperature profiles more characteristic of hotter dwarfs and move the depth at which carbon-based chemical reactions are quenched deep in the atmosphere, leading to an elevated abundance of CO \citep{Barman:2011fe,Marley2012}.  In the coming years, more directly-imaged planet detections spanning a wider range of mass, age, and temperature will better clarify how the clouds, chemistry, and gravity of young jovian planets evolve (see Section~\ref{sec:future}).

\subsection{Time-Resolved Atmospheric Properties: Rotation Measurements and Variability\label{subsec:variability}}

The photometric and spectroscopic observations described in the previous two sections are time-averages -- observations either capturing short, 1-2 hour snapshots or averaging over multiple observations covering different rotational phases.  However, directly imaged exoplanets, like their higher-mass brown dwarf cousins, are dynamic, rapidly rotating objects.  Measuring spectral line broadening ($v~sin(i)$) for field brown dwarfs and monitoring their photometric variability reveals rotation periods of  $\sim$3-20 hours \citep{ZapateroOsorio2006}.
Recently, \citet{Tannock2021} have even found 3 brown dwarfs with rotation periods as short as $\sim$1 hour!  Rotation measurements are scarcer for directly imaged exoplanets and their analogues, but current measurements suggest that directly imaged exoplanets are also fast rotators.  Spectral-line broadening measurements of $\beta$ Pic b indicate a rotation period of 7-9 hours, \citet{Snellen2014}.  
Rotational periods derived for 27 planetary mass objects (both as companions and free-floating) from the compilation of \citet{Bryan2020} range from 4 to 22 hours.

If an object is a rapid rotator and also has asymmetric top-of-atmosphere structure (e.g. patchy thin and thick clouds), it will display significant variability, appearing considerably brighter at some rotational phases than others.  Large variability surveys of brown dwarfs find variability is common \citep{Radigan2014a, Wilson2014, Radigan2014b, Metchev2015}.  Variability amplitudes extend up to $\sim$20$\%$ in the near-IR.  Young directly imaged giant planets are likely to be even more variable.  From a survey of 30 young, low-surface gravity brown dwarfs with estimated masses $<$25 M$_{\rm J}$ (e.g., free-floating analogues to directly imaged exoplanets), \citet{Vos2019} find that 30$\%$ of low surface-gravity L0-L7.5 dwarfs are variable, compared to 3$\%$ of field L dwarfs. Exoplanet analogues with L6-L7 spectral types appear to be particularly variable, with peak-to-trough amplitudes $>$5$\%$ in a number of notable cases \citep{Biller2015, Lew2016, Vos2018, Biller2018, Bowler2020b, Zhou2020, Zhou2022}.  These objects have spectra that are nearly identical to those of the archetypical HR 8799 planets \citep{Bonnefoy2016}: thus, their variability properties are likely similar to those of these bonafide exoplanet companions. Young, planetary mass T dwarfs are likely more variable than their old, field counterparts as well.  Two out of the three highest amplitude T dwarf variables known, specifically SIMP 0136 and 2M2139, have recently been confirmed as planetary-mass members of the Carina Near association \citep{Gagne2017,Zhang2021}.

Variability studies, especially spectroscopic variability studies, are valuable probes of the top-of-atmosphere structure (TOA) of brown dwarfs and directly imaged exoplanets. A unique map cannot be obtained from lightcurve monitoring.  However, indicative maps can be generated from the spectral mapping technique, which uses Markov-Chain Monte Carlo techniques to fit elliptical spots \citep{Kostov2013, Karalidi2015, Karalidi2016} and/or planetary-scale waves \citep[a~sinusoidal~ bright-dark~pattern~confined~to~a single~latitudinal~band,][]{Apai2017} to multi-wavelength lightcurves.  Quickly changing features on several field brown dwarfs 
may indicate that the beating between multiple planetary-scale waves better describes the lightcurves observed for field brown dwarfs than elliptical, starspot-like spots \citep{Apai2017,Apai2021}.  

Different wavelengths also probe different depths in brown dwarf and giant exoplanet atmospheres.  Mid-IR data probe higher-up, cooler cloudtops; the near-IR probes deeper into cloud decks \citep[e.g.][]{Manjavacas2021}.  Notably, TOA structure appears to vary in between this level, seen as phase shifts between the near-IR and mid-IR in the substellar object lightcurves, where successive maxima / minima in near-IR are offset in time relative to their mid-IR counterparts. Such phase shifts have been observed in several brown dwarfs \citep{Yang2016} and in at least one free-floating planetary mass object \citep{Biller2018}.

The studies described in the preceding paragraph have mostly targeted higher mass field brown dwarfs.   Next generation telescopes and instruments such as JWST will enable similar studies for free-floating planetary mass objects and bright exoplanet companions such as $\beta$ Pic b and HR 8799 bcde.  Giant exoplanet companions are likely to be equally variable compared to highly variable free-floating planetary mass objects, but are much more difficult targets for variability studies, given the technical challenges of obtaining high-fidelity, highly-stable spectrophotometry for close companions. Two studies have attempted variability monitoring for the HR 8799 planets using VLT-SPHERE \citep{Apai2016,Biller2021}, reaching sensitivity to variability with amplitudes down to 5$\%$ for periods $<$10 hours, but not yielding any detections to date. 

\subsection{\textbf{State-of-the-Art Modeling of Imaged Exoplanets}}
\label{subsec:theo}
\subsubsection{Recent developments of 1D models \label{subsec:1Dmodels}}

Atmospheric models for directly imaged exoplanets and brown dwarfs share a heritage with models for very low mass stars \citep{Chabrier:2000hq,Baraffe2002,Burrows2003,Allard2012,Baraffe2015} and solar system planets, building a self-consistent, one-dimensional atmospheric structure connecting a fully convective interior with a radiative atmosphere \citep[``radiative-convective" model, see review by][]{Marley2015}. 
However, the cooler temperatures and lack of core H fusion for brown dwarf and exoplanet atmospheres require a number of different ``ingredients" compared to stars, most specifically, a treatment of clouds and hazes in these atmospheres and the ability to handle disequilibrium chemistry and vertical mixing. 

The very red colors observed for young planet-mass objects cannot be described by simple 1D models without additional cloud or convection prescriptions \citep{Marley2012}.  
The most common parameterization used in modeling brown dwarf and giant exoplanet clouds utilizes a sedimentation efficiency parameter $f_{sed}$, which tunes cloud particles sizes necessary to balance vertical transport of condensible gases \citep[e.g.~the~Eddysed~grid~presented~in][]{Ackerman2001}. The Sonora model grid is an update to the 
original Eddysed grid models, reaching cooler effective temperatures and a wider range of C/O values than the earlier grids.  To date, only the cloudless model grid is available \citep{Marley2021}, but grids incorporating a similar $f_{sed}$ cloud parameterization should be available soon.
The ATMO 2020 models \citep{Phillips2020} also provide a key update to earlier grids, in particular, the COND and DUSTY models of \citet{Baraffe2002}, incorporating a new H-He equation of state, updated molecular opacity, and an improved treatment of the collisionally broadened potassium resonance doublet.
Finally, models like Exo-REM introduce a self-consistent cloud model, with cloudy and clear columns,  a similar $f_{sed}$ method of cloud parameterization \citep{Charnay2018}, and grids that consider non-solar compositions \citep{Nowak2020b}.

Canonical models utilize silicate dust clouds to 
describe
young exoplanet atmospheres, especially at the L/T transition.  However, other work explores alternatives.  For example,
\citet{Tremblin2016} propose that a thermo-chemical instability producing an extra source of diabatic convention can alter an atmosphere's temperature-pressure profile, providing another path to produce the observed red colors without clouds \citep[see also:][]{Tremblin2017,Tremblin2019,Tremblin2020}.




\subsubsection{Inversion techniques: achievements and challenges\label{subsec:inversion}}

The models discussed in Section~\ref{subsec:1Dmodels} are full self-consistent radiative-convective equilibrium models: forward models where the full 1D structure of the modeled atmosphere is solved iteratively (similarly as with stellar models). These models produce grids of synthetic planet spectra exploring the impact of a limited set of model free-parameters. The comparison of these models to observed spectra is known as ``forward modelling". 

Bayesian inference methods such as the Markov-Chain Monte-Carlo method or the Nested sampling algorithm can be coupled to precomputed grids of forward models to interpret spectra. The method can account for correlated and uncorrelated noise in the data and can estimate posterior probability distributions on the model parameters and evidence degeneracies between those parameters, thus identifying the sub-families of models that best represent the data \citep{Rice2010, samland17, Stolker2020, ZhangZ2021}. Recently proposed random forest algorithms are a more computationally rapid 
alternative to the Bayesian framework and can provide 
information regarding the degree in which datapoints at given wavelengths
constrain a specific free-parameter of the model  \citep[e.g.,][]{Oreshenko2020}.

\begin{figure}[h]
 \epsscale{1.0}
  \plotone{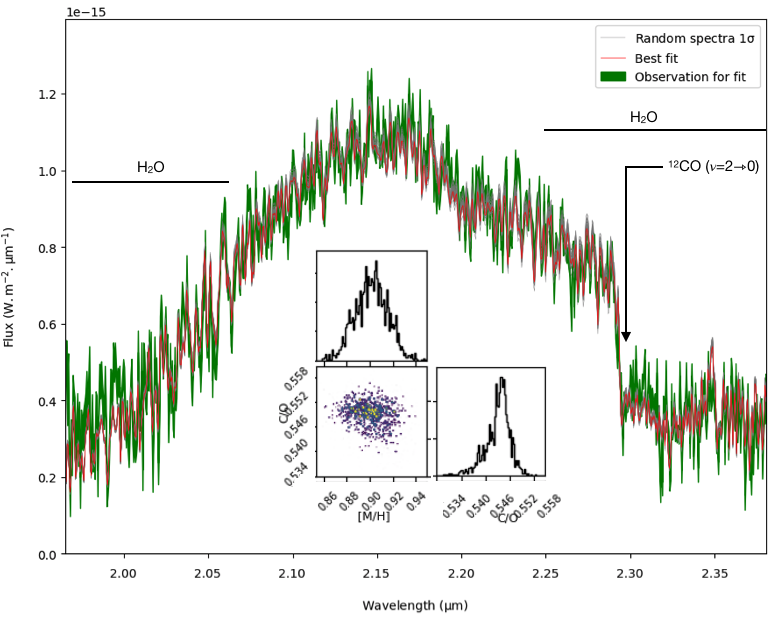}
 \vspace{-0.35in}
\caption{\small Keck/OSIRIS medium-resolution (R$\sim$4000) K-band spectrum of HR8799 b from \citet{Barman2015} fitted with the Bayesian forward modeling code \texttt{ForMoSA} \citep{Petrus2020,Petrus2021} and the most recent self-consistent 1D models Exo-REM \citep{Charnay2018, Blain2021}, exploring different abundance values (C/O, M/H). The fit provides a consistent representation of the spectral continuum and $^{12}CO$ absorption, confirms the solar C/O ratio of the planet inferred in \citet{Barman2015} from the same data, and matches well the C/O value measured for the primary star \citep{JiWang2020}.}  
\vspace{-0.05in}
\label{fig:spec8799}
\end{figure}

Modern self-consistent forward models track both the effect of cloud condensates on the sequestration of atomic species in the gas phase and the consequences of non-equilibrium carbon chemistry, if present.  If not properly treated, the presence of clouds and non-equilibrium chemistry
can lead to biased determination of bulk atmospheric compositions.
Nonetheless, for self-consistent models that carefully take into account these processes, ``forward modelling" can provide accurate fits to  medium-resolution spectra of companions consisting of several thousands of datapoints (see Figure \ref{fig:spec8799}). It however requires re-interpolating the grids of models and relies on a number of pre-determined sets of physical ingredients in the models. 

 The complementary ``retrieval" approach starts from the spectra and works backwards. In this case, a parameterized pressure-temperature profile is adopted and other fundamental parameters (mass, effective temperature, cloud properties, abundances, etc.) are then retrieved given the observed spectrum by using Bayesian or random forest techniques \citep[][]{Madhusudhan2009}.  Retrievals have been used extensively to interpret transiting exoplanet spectra \citep[see][for~a~recent~review]{Madhusudan2018}.  In the last few years, retreivals have also been applied to direct spectroscopy of brown dwarfs \citep{Line2017, Burningham2017,Burningham2021} and extrasolar giant planets \citep{Lavie2017, Molliere2020}.   However, retrieval approaches lose clear physical interpretations of the spectra. This approach is still in its infancy and will gradually benefit from improvements made to forward models.

\subsubsection{The need for 3D models}
The majority of modeling approaches for directly imaged exoplanet atmospheres are one-dimensional.  However, high-amplitude, quasi-periodic variability detected in both brown dwarfs and free-floating planetary mass objects indicate significant asymmetric top-of-atmosphere structure, often attributed to patchy thin and thick cloud cover \citep{Apai2013}.  (See Section~\ref{subsec:variability} for additional details on these observations.) Coupled with their relatively rapid rotation rates which likely drive significant winds, these are dynamic and quickly varying atmospheres -- and certainly not one-dimensional.  Full 3D general circulation modeling (GCM) including a treatment of radiative transport is still on the horizon for brown dwarfs and directly imaged giant exoplanets, but a number of groups have implemented modeling approaches that take into account the significantly asymmetric top-of-atmosphere structure and strong atmospheric circulation of these objects.  \citet{Freytag2010} use the CO5BOLD code to perform local, 2D radiation hydrodynamics simulations, incorporating dust clouds in the atmospheres of late M stars and brown dwarfs and finding that convectively excited gravity waves are an essential mixing process in these atmospheres.  For very cool T and Y dwarf atmospheres with patchy salt and sulfide clouds, \citet{Morley2014} introduce both clear and cloudy columns in their models along with a cloud-covering fraction for each model atmosphere.  \citet{Showman2013} present three-dimensional, global, numerical simulations of convection in the interior of brown dwarfs and directly imaged giant planets.  These authors consider how the temperature perturbations driven by this convection could drive top-of-atmosphere asymmetry, in particular, allowing patchy clouds to form near the photosphere of L-T transition brown dwarfs. \citet{Zhang2014} use a shallow water model to model the atmospheric circulation driven by the rapid rotation of brown dwarfs and directly imaged giant planets.  Depending on the strength of the heat flux and radiative dissipation, they find two circulation regimes: 1) with strong internal heat flux and weak radiative dissipation, banded east-west jets will spontaneously form and 2) with weak internal heat flux and/or strong radiative dissipation, vortices will instead form, potentially producing some of the ``spotted" features found on brown dwarfs.  To constrain the effect of latent heating due to condensation on both cloud growth and top-of-atmosphere asymmetric structure, \citet{Tan2017} applied a idealized general circulation model (GCM) including a condensation cycle for silicate vapor.  They found that simulations with conditions appropriate for T dwarfs developed both localized storms and east-west jets and that circulation driven by the latent heating due to the condensation of silicate clouds can generate large-scale cloud patchiness.  In a series of recent papers, Tan and Showman have investigated a second possible driver of short-time evolution of clouds and thermal structures in brown dwarf, specifically, radiative cloud feedback, first within a 1D model \citep{Tan2019}, then extending to a local 3D model \citep{Tan2021a}, and finally extending to a full global geometry model \citep{Tan2021b}.


\subsection{Connecting atmospheric chemical abundances to exoplanet formation and migration histories}

As seen in Sections~\ref{subsec:spectroscopy} and~\ref{subsec:inversion}, medium and high-resolution spectroscopy, coupled with state-of-the-art forward- and retrieval modeling approaches can measure atmospheric chemical abundances for exoplanets, in particular, key tracer ratios such as C/O and $^{12}$CO / $^{13}$CO.
The composition of a given planet-forming disk will vary as a function of radius, as various molecules (water, CO, CO$_2$) condense out into ices and form snow lines at progressively greater distances in the disk.  Thus, an exoplanet atmosphere's chemical composition will carry the imprint of where in the circumstellar disk it first formed and through which parts of the disk it migrated before reaching its final orbital configuration \citep[e.g.][]{Oberg2011, Oberg2019, Madhusudhan2019}. The nature of this imprint is challenging to describe end-to-end, from the protostellar cloud to the final observed exoplanet atmosphere.  This requires a detailed understanding of the balance of which species remain in gas phase vs. which condense into solids, as well as the assumption that the final atmospheric gas composition of the planet will reflect the disk gas composition.

\citet{Oberg2011} made the first prediction for the atmospheric chemical signatures for different formation histories.  For the canonical core accretion model, in which a several Earth-mass rocky core is built up by the accumulation of dust-to-boulder sized planetesimals and then accretes disk gas and residual planetesimals\footnote{Planet formation models are discussed extensively in Section~\ref{subsec:testing-formation}.}, jovian planets should have enhanced C/0$\sim$1 if they accrete most of their atmospheres from gas in the disk outside of the water snowline.   
Using high-resolution Keck-OSIRIS spectroscopy, \citet{Konopacky2013} measured a super-stellar C/O ratio for HR 8799 c, consistent with the \citeauthor{Oberg2011} predictions for the core accretion scenario.   
Since this first measurement, C/O ratios have now been published for all four HR 8799 planets, $\beta$ Pic b, HIP 65426 b, and $\kappa$ And b \citep{Lavie2017, Molliere2020, Wilcomb2020, JiWang2020, Nowak2020a, Petrus2021, Ruffio2021}.
Discrepancies in values between early measurements and more recent measurements \citep[e.g.~between][]{Lavie2017,JiWang2020} likely occur  because higher spectral resolution measurements yield more accurate C/O ratio measurements.

Interpreting C/O ratio measurements requires a careful comparison to disk/formation model predictions relative to different snowlines. To fully interpret C/O ratio measurements, several of the observational studies considered above also incorporate significant disk modeling in their analysis \citep[cf.][]{Molliere2020,Nowak2020a,Petrus2021}.  
See \citet{Madhusudhan2019} for a more detailed discussion.

A number of other elemental or isotopic ratios
may yield insight into the formation mechanisms / locations of young, giant exoplanets.
\citet{Oberg2019} suggest that the abundances of nitrogen-bearing species may also trace the formation location of a given planet in the disk \citep{Oberg2019}.  The $^{12}$CO / $^{13}$CO ratio may also distinguish between ``brown-dwarf" like and ``planet-like" formation scenarios \citep{ZhangY2021, Zhang2021b}. 




\section{\textbf{Architectures of Directly Imaged Planetary Systems}}
\subsection{\textbf{Orbits and Dynamics}}





\begin{figure}[h]
\vspace{-0.15in}
 \epsscale{1.0}
 \plotone{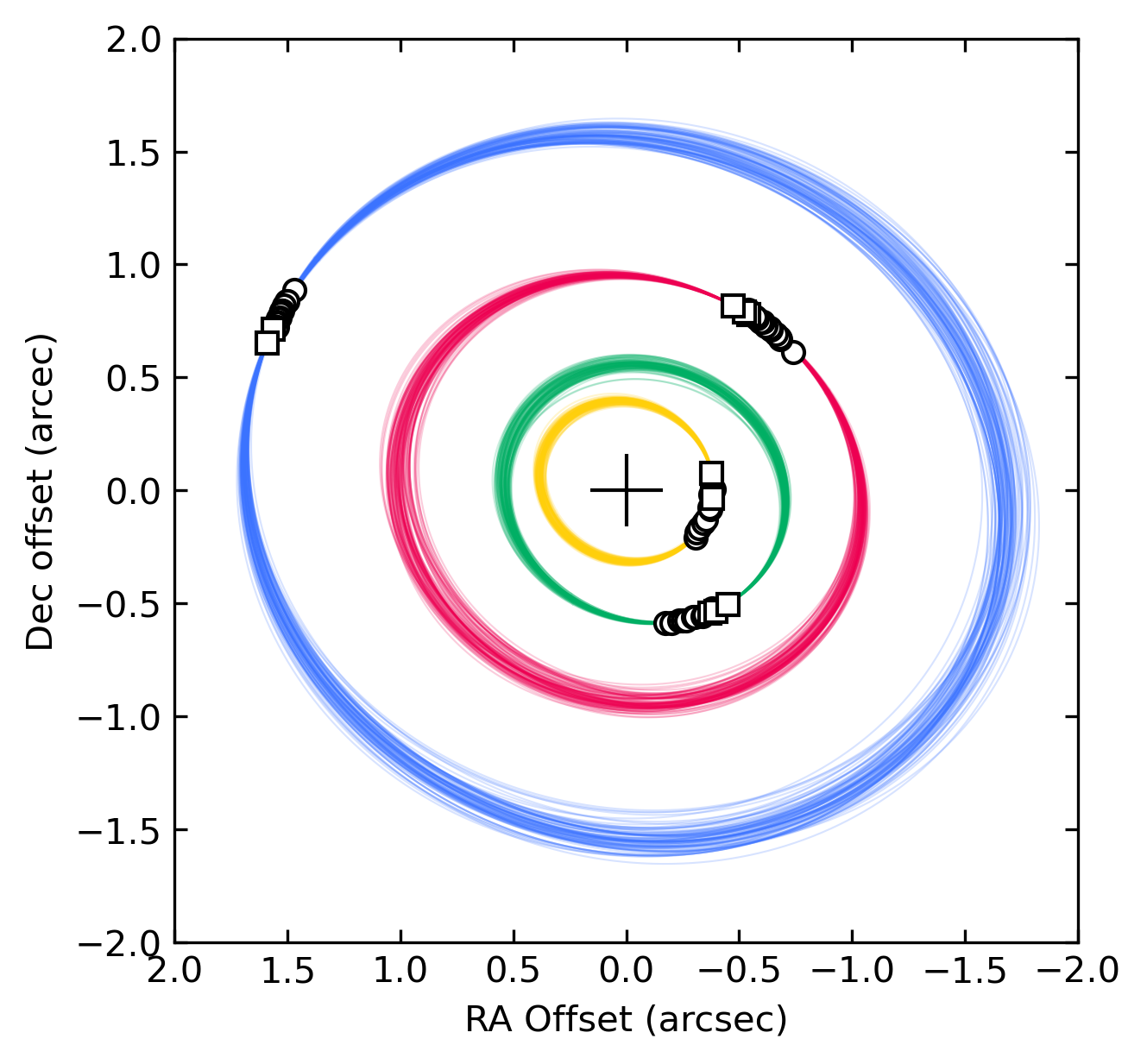}
 \vspace{-0.35in}
\caption{\small Visual orbits of the four planets in the HR 8799 system consistent with the astrometric measurements and assuming a dynamically stable co-planar configuration \citep{Wang:2018fda}. These orbits are a small subset of all visual orbits, stable or otherwise, that are consistent with the data.}  
\vspace{-0.075in}
\label{fig:orbit8799}
\end{figure}

Given their long orbital periods, directly imaged planets are not currently amenable to rapid 
orbital characterization. 
At least one years’ worth of astrometric measurements are typically required in order to place initial 
constraints on the parameters of the visual orbits of planets that have been detected thus far (e.g., \citealp{Pearce:2015je,Blunt:2017et}). In fact, no planet yet discovered via direct imaging has a “closed” orbit, i.e. with relative astrometric measurements covering the full phase of the orbit. Despite the challenges associated with the analyses of these incomplete datasets, a great deal of information can still be learned regarding the orbits
of imaged exoplanets, and the architectures of the planetary systems in which they reside.

The HR 8799 system is probably the best-studied system in terms of simulations of its dynamical configuration and long-term stability \citep{Fabrycky10, Currie2011, Sudol2012, Gozdziewski14, Gotberg16, Gozdziewski18, Wang:2018fda,Zurlo2022}. The four planets' astrometric measurements are consistent with a co-planar configuration in, or close to, a 1:2:4:8 period ratio (Figure~\ref{fig:orbit8799}; \citealp{Gozdziewski18,Wang:2018fda}). Integrating the system forward (or backward) demonstrates that such a configuration is dynamically stable, at least for many multiples of the current age of the system \citep{Gozdziewski18}.
While a resonance between the four planets is not necessary for long-term stability, resonances between pairs of the inner planets are required if their masses are $\gtrsim 6 M_{\rm Jup}$ \citep{Wang:2018fda}. In addition to testing the dynamical stability of the system, these simulations have also been used to investigate the parameter space in which additional planets could exist in a stable configuration \citep{Gozdziewski18}, either interior to the innermost known planet or in a wide orbit dynamically sculpting the outer debris disk  \citep{Booth2016,read2018}.

The dynamical stability of the other known multi-planet systems ($\beta$~Pic, PDS~70, and TYC 8998) have not yet been studied at the same level of detail. Their astrometric measurements are consistent with a dynamically stable configuration given the masses derived from either evolutionary models \citep{Wang2021,Bohn2020}, or from combined astrometric and radial velocity measurements \citep{Lagrange20,Brandt21}. Continued monitoring of these systems is warranted to investigate not only their dynamical stability given revisions to their orbits and estimated masses, but also the influence of the planets on the structure of any resolved circumstellar material in each system (see Section~\ref{subsec:planet-disk}).

Early astrometry of $\beta$ Pic b quickly revealed a near-edge on orbit for the planet configuration  \citep{Lagrange2010}, consistent with an edge-on inclination of the resolved debris disk \citep{SmithTerrile1984}. While geometrical arguments show that a long-period planet like $\beta$~Pic~b will likely not transit, a photometric event measured in 1981 showed a significant variation in the star's brightness \citep{Lecavelier97}, consistent with an eclipse of the star by the planet or its Hill sphere based on the ephemeris available at the time \citep{Chauvin12}. Continued astrometric monitoring ruled out a planetary transit but demonstrated that the Hill sphere would transit the star in mid-to-late 2017 \citep{Wang16}. An extensive photometric and spectroscopic campaign ensued to measure, or place stringent limits on, circumplanetary material as the Hill sphere transited. No significant event was detected: the 1981 event was probably not linked to the transit of an extended circumplanetary disk \citep{Lous18, vanSluijs19, Kenworthy21}. Despite this null detection, exo-cometary material not within the $\beta$~Pic~b Hill sphere has been detected transiting the star with TESS \citep[][Lecavleier et al 2022 in press]{Zieba19}, suggesting that the earlier photometric variability could also be due to material not associated with the planet.


\begin{figure*}[ht!]
  \includegraphics[width=0.9\textwidth]{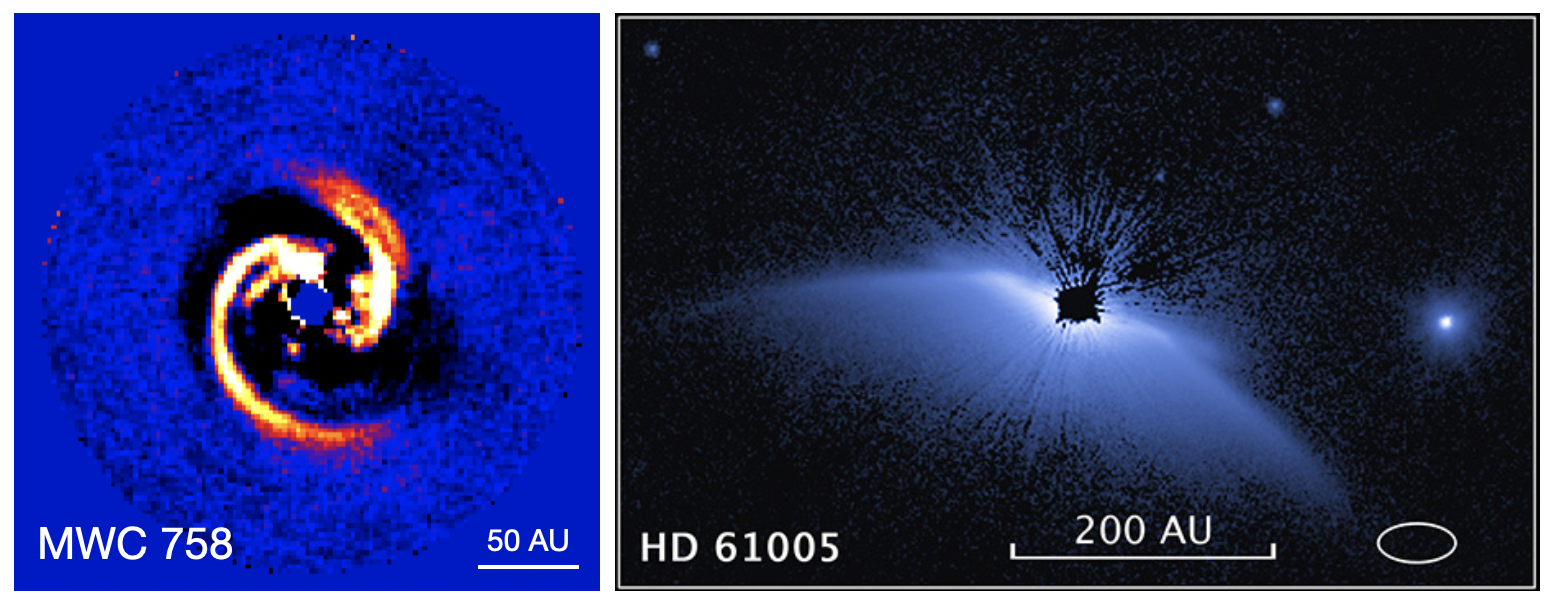}
 \vspace{-0.2in}
\caption{\small The protoplanetary disk around MWC 758 (left; SCExAO team, unpublished) and the debris disk around HD 61005 (right; NASA/ESA/G. Schneider) both have morphologies that can be explained by the gravitational influence of an orbiting planet. The spiral arms around MWC 758 can be generated by an external orbiting companion (e.g., \citealp{Dong15b}, c.f. \citealp{wagner18}), while the swept-back morphology of the HD 61005 debris disk can be explained by an inner eccentric planet \citep{Esposito16}.}  
\vspace{-0.1in}
\label{fig:disturbed-disks}
\end{figure*}

Monitoring efforts are also ongoing for the
the orbits of planets within single-planet systems 
(e.g., \citealp{Maire19,Kammerer21}). In addition to providing constraints on the basic orbit geometry, these monitoring efforts offer the potential for future dynamical mass measurements when combined with absolute astrometry or radial velocity measurements of the host star or planet (see Section~\ref{subsec:synergies}).   The orbit geometries of imaged planets within single-planet systems have been used to infer the presence of additional companions. \citet{Rameau:2016dx} compared the predicted range of periastron and apastron distances of HD 95086 b to the radial extent of the two circumstellar debris disks in this system between which the planet resides \citep{Moor:2013bg,Su:2015ju}. Their analysis suggested that HD 95086 b does not have a sufficiently large eccentricity to be responsible for truncating the outer edge of the inner debris disk. Instead, an additional massive planet is required on a shorter orbital period, one that lies below the current sensitivity limits for this star. In other systems, a significant non-zero eccentricity may provide evidence for interior massive companions due to their effect on the relative astrometry between the star and the outer imaged planet (e.g., \citealp{Maire19}). Continued astrometric monitoring will differentiate between a true non-zero eccentricity and a spurious detection caused by the perturbation of an inner companion.

Comparing the orbital geometry of planets to those of brown dwarfs and stars may can probe of the formation mechanisms.  For individual objects, \citet{Bryan2020b} has used the measured misalignment between the planet’s orbit, rotation and stellar spin axes 
as an empirical test of various formation mechanisms and dynamical processes.  Differences between planet and brown dwarf eccentricity distributions may also probe formation mechanisms \citep[see Section~\ref{subsec:testing-formation};][]{Bowler2020}. 

\subsection{\textbf{Planet-Disk Interactions}}
\label{subsec:planet-disk}

The many main sequence stars with IR excess found in the 1980s to 2000s by the IRAS, ISO, and Spitzer missions along with early resolved images of disks provided evidence for numerous, nearby optically thin, Kuiper belt-like debris disks that may be nascent planetary systems \citep[e.g.][]{Backman93,SmithTerrile1984}.  Pluto-sized objects in these exo-Kuiper belts stir collisions between planetesimals to produce copious debris; stirring may also be driven by massive jovian planets \citep{Kenyon2004,Kenyon2008,Kennedy2010}.   These same missions showed evidence for numerous \textit{transitional} protoplanetary disks whose spectral energy distributions (SEDs) suggested cleared cavities or gaps \citep[e.g.][, see Sect. 3.2]{Espaillat2007,Luhman2012}.

Resolved images of debris and transitional disks reveal tell-tale signposts of planets (Figure \ref{fig:disturbed-disks}).  Images of transitional disks resolve cavities inferred from SEDs and/or show spiral density waves plausibly due to massive jovian companions.  For example, the disk around MWC 758 shows a complex two-armed spiral structure extending from $\sim$ 25 au to over 100 au \citep{Benisty2015}; AB Aur's disk shows significant spiral structure on both tens of au and 100 au scales, some of which may be driven by AB Aur b \citep{Hashimoto2012,Tang2017,Currie2022}.   The PDS 70 disk has a cavity likely cleared by the system's two protoplanets \citep[e.g.][]{Keppler2018,Haffert19}.   Recent work suggests that actively-forming planets could be responsible for gaps, rings, and spirals seen in transitional disks \citep[e.g.][]{Dong15a, Dong15b}, although some structure (e.g. spirals) could instead be due to gravitational instabilities.  

Imaged debris disks may show sharp edges and pericenter offsets caused by the dynamical perturbations from a jovian planet located interior \citep{Kalas2005}.   Modeling the sharp edges of disks may help constrain the masses of planets \citep[e.g.][]{Kalas2008}.   Warps in debris disks are also likely planetary signposts: modeling the warps in some cases can constrain planet masses \citep{Lagrange2009}.

Finally, mm imaging of highly structured transitional disks may reveal indirect kinematic detection of planets or circumplanetary material.   \citet{Benisty2021} image a circumplanetary disk around PDS 70 c.   ALMA data for HD 163296 and HD 97048 may show an indirect kinematic detection of protoplanets that currently elude direct detection via local deviations from Keplerian rotation (kinks) 
\citep{Pinte18, teague18, Pinte19}.

 \section{Direct Imaging Surveys and Population Statistics\label{sec:surveys}}

Most of the known directly imaged exoplanets were discovered as part of large AO surveys.  New planet detections and characterization of existing planets via photometry and spectroscopy is only one part of the yield of such surveys; current surveys cover samples of 500-600 stars down to excellent contrasts and yield precise information as to where giant exoplanets reside in planetary systems.  While the new detection yield of these surveys has been relatively small, the deep contrasts probed enable robust studies of the frequency of wide ($>$20 au) young giant planets.    




\subsection{Survey Target Selection\label{sec:surveysamples}}
\begin{figure}[ht]
  \includegraphics[angle=0,scale=0.375,trim=0mm 10mm 0mm 5mm,clip]{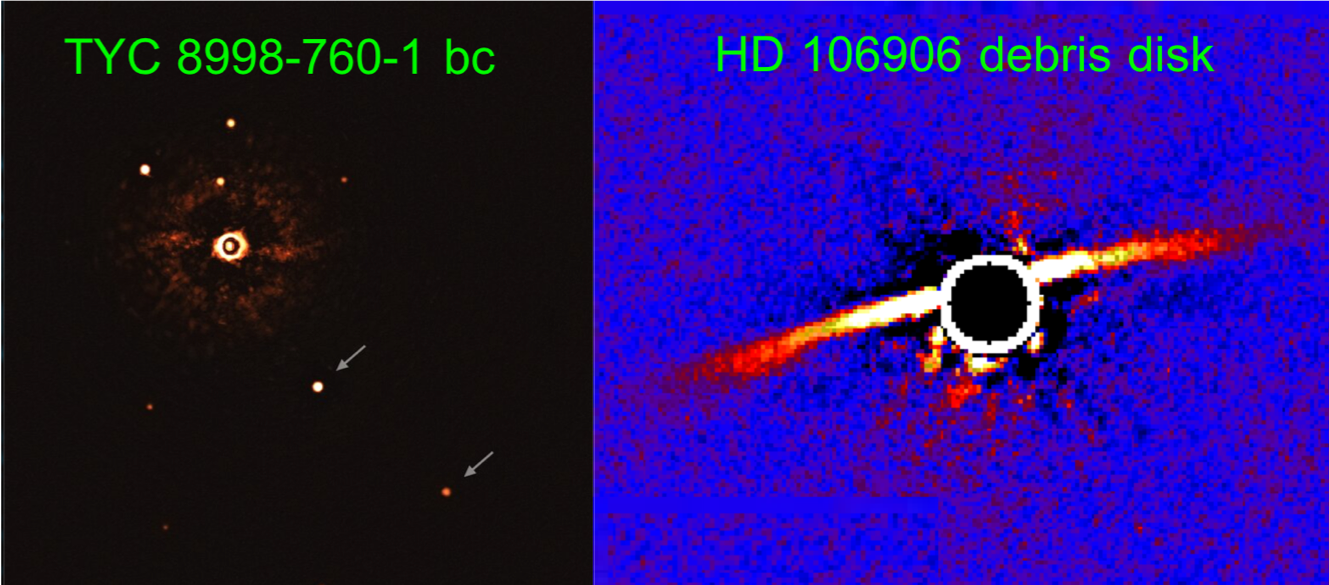}
 \vspace{-0.25in}
\caption{\small Selected high-contrast imaging results for systems in the Sco-Cen Association: (top) The directly-imaged wide separation companions TYC 8998-760-1 bc ($\sim$160 and 320 au; denoted by arrows), and (bottom) the HD 106906 debris disk \citep{Bohn2021,Lagrange2016}.   HD 106906 also has a much wider-separation planet-mass companion \citep{Bailey14}.}  
\vspace{-0.1in}
\label{fig:scocen}
\end{figure}

Direct imaging exoplanet surveys to date have focused primarily on stars drawn from the following samples:
\textbf{1) Young Moving Groups (MG):} as discussed in Section 3, these groups of stars share similar kinematics and originated in the same star-forming region.
    The ongoing Gaia mission is helping fill in the low mass stellar members of these groups 
    \citep[e.g.][]{Gagne2018a,Gagne2021}.
    Some notable moving groups that have been surveyed in depth by direct imaging surveys include: the $\beta$ Pic MG, the AB Dor MG, the Tucana-Horologium MG, and the Sco-Cen association.

\textbf{2) Young Star-forming regions (SFR):} these are regions with active ongoing star-formation and probe the earliest stages of planet formation ($<$10 $Myr$) \citep[see][]{Reipurth2008a,Reipurth2008b}.     Notable SFRs targeted for direct imaging include Taurus, Ophiuchus, Chamaeleon, Perseus, and Serpens.   
   SFRs are excellent regions to search for protoplanets.  They have generally been less surveyed than MGs due to their greater distances and confusion between exoplanet and disk signals. 
    There is also some overlap between SFR and MG designations, with boundary cases such as TW Hya and Upper Scorpius within the larger Sco-Cen association \citep{Song2012, pecaut12}.
    
\textbf{3) The Sco-Cen niche:}
        Direct imaging programs focusing on the 14$\pm$ 
3 Myr \citep{pecaut12} Scorpius-Centaurus association have proven particularly fruitful, yielding a much higher rate of exoplanet detections compared to other young moving group surveys.   At a distance of $\sim$100--150 $pc$, Sco-Cen is the closest OB association (massive star-forming region) to the Sun and contains over one thousand young stars, well over 100 of which show some indirect evidence for active planet formation \citep{Luhman2012}.  It is intermediate in properties between 10-100 Myr-old moving groups and $<$10 Myr-old embedded young star-forming regions.

Figure \ref{fig:scocen} shows some direct imaging results from this region. Sco-Cen includes multiple directly imaged planets and protoplanets orbiting less than $\sim$300 au from Sco-Cen host stars: HD 95086 b \citep{Rameau2013}, HIP 65426 b \citep{chauvin17}, PDS 70 bc \citep{Keppler2018,Haffert19}, and YSES 2b \citep{Bohn2021}.   The region also includes a large number of other wide-separation planet-mass companions or objects with masses near the deuterium burning limit and mass ratios well below $q$ $\sim$ 0.1 \citep[e.g.][TYC 8998-760-1 bc and HD 106906 b]{Bohn2020,Bailey14}.  Follow-up observations of many Sco-Cen members stars with ground-based extreme AO and ALMA spatially resolve numerous debris disks and transitional disks \citep[e.g.][]{Lagrange2016,Garufi2016,Draper2016}.  The disks often show structures (spirals, gaps, arc) that can indicate the presence of formed or forming planets (see Sect. 5.2).

\textbf{4) Debris disk-bearing stars:} debris disks may be signposts of imageable giant planets (see Section 5).
    The first two directly-imaged planetary systems -- HR 8799 and $\beta$ Pic -- both have luminous debris disks.  Debris disk-selected surveys like \citet{Rameau2013a, Meshkat17} target young, nearby stars whose debris disks have a high fractional luminosity (e.g. 10$^{-4}$--10$^{-3}$).  While \citep{Meshkat17} find evidence of a correlation between imaged planets and debris disks, debris disks may not correlate with planets detected by RV \citep{Yelverton20}.   
    
\textbf{5) A and B stars:} these massive stars have short lifetimes, so any companion to them must by definition be young ($<$200 Myr). Statistical results from several surveys show that substellar and exoplanet companions to A and B stars are indeed more common than similar companions to lower mass FGKM stars \citep[e.g.][~see~also~\ref{sec:occurrence}]{Nielsen2019}.
    
\textbf{6) Accelerating Stars:} unlike the categories of stars discussed above which are selected primarily due to signs of youth, this category of stars show indirect evidence for the dynamical pull from an unseen companion.  While such targeted surveys cannot provide unbiased demographic investigations of exoplanets, they can produce significantly higher yields than blind surveys \citep[e.g.][]{Crepp2016}.  Some surveys, specifically TRENDS, focused on stars with long-term RV trends indicating the presence of an unseen companion 
  \citep[e.g.][]{Crepp2012,Crepp2018}, 
    discovering a number of brown dwarfs, low-mass stars, and white dwarf companions.  A key challenge with this approach is that stars with sufficiently low chromospheric activity to yield good RV precision tend to be old and thus poorer direct imaging targets.  
    
    More recently, several groups have begun using astrometric data to select direct imaging targets. This method is immune from the age selection bias that plagues RV-selected target lists.  For an isolated star, the proper motion values measured by Hipparcos and Gaia should be identical.  However, the tug of an unseen faint companion will cause the position of the primary to diverge slightly from what would be expected for an isolated star (modulated by the orbital period of the companion).   
    Differences in proper motions measured between the Hipparcos and \textit{Gaia} missions thus reveal accelerations plausibly caused by unseen planetary companions \citep{Kervella2019, Brandt2018, Brandt2021hgca, Fontanive2019}. 
    While in its infancy, this approach is promising, already revealing brown dwarf companions and a handful of white dwarfs \citep{Currie2020b,Bonavita2020, Bonavita2022,Bowler2021b,Kuzuhara2022}.  It is capable of identifying hidden planetary companions: HR 8799, $\beta$ Pic, and HD 206893 all have proper motion anomalies caused by known planetary companions \citep{Brandt21,Nielsen2019,grandjean2021}.  Recently, the first discovery of an exoplanet has been gained from this approach \citep{Currie2022b}.  The modest astrometric precision of Hipparcos limits the current inventory of stars with detectable accelerations likely caused by planets.  However, future \textit{Gaia} data releases should significantly expand the list of known stars whose accelerations may be caused by imageable planets.
       

\subsection{Summary of Surveys}

Since the advent of ground-based AO-enabled imaging and space-based imaging with HST, there have been several generations of direct-imaging exoplanet surveys:
\begin{enumerate}
    \item {\bf Generation 1:} AO-enabled and HST efforts, 2000-2005 (i.e. the state-of-the-art at Protostars and Planets V, the last Protostar and Planets conference with a direct imaging chapter).  Typical sample size: $<$50 stars.  Typical contrast: $\Delta(mag) \sim 8$ at 0.5".  Examples: \citet{Masciadri2005} (VLT), \citet{Chauvin2003} (ESO 3.6 m, Tuc/Hor moving group), \citet{Lowrance2005} (HST).
    \item {\bf Generation 2:} 8-m class telescope + AO + speckle suppression techniques (but no coronagraphy).  Typical sample size: 50-100 stars.  Typical contrast: $\Delta(mag) \sim 10$ at 0.5".  Examples: \citet{Biller2007} (VLT and MMT), \citet{Lafreniere2007b} (Gemini), \citet{Chauvin2010} (VLT).
    \item {\bf Generation 3:} 8-m class telescope + AO + speckle suppression techniques + coronagraphy.  Typical sample size: 100-300 stars.  Typical contrast: $\Delta(mag) \sim 11-12$ at 0.5".  Examples: NaCo Large Programme: \citet{Desidera2015, Chauvin2015, Vigan2017}, NICI Science Campaign: \citet{Biller2013, Wahhaj2013,Nielsen2013}, International Deep Planet Search (IDPS): \citet{Vigan2012, Galicher2016}, SEEDS: \citet{Tamura2016,Brandt2014, Janson2013}, PALMS: \citet{Bowler2012a, Bowler2012b, Bowler2015}, LEECH: \citet{Skemer2014b, Maire2015, Stone2018}.
    \item {\bf Generation 4:} 8-m class telescope + {\bf extreme} AO + sophisticated speckle suppression techniques + coronagraphy.  Typical sample size: 500-600 stars.  Typical contrast: $\Delta(mag) \sim 13-14$ at 0.5", using custom-built extreme-AO coronagraphs with spectroscopic capability (due to their IFS detectors).  Examples: Gemini-GPIES \citep[][]{Nielsen2019} and SPHERE-SHINE \citep[][]{Vigan2021, Desidera2021, Langlois2021}.  Several smaller, targeted surveys are ongoing alongside the large Generation 4 surveys -- such as the Search for Planets Orbiting Two Stars SPOTS \citep{Thalmann2014, Bonavita2016, Asensio-Torres2018}, the B-star Exoplanet Abundance Study (BEAST) \citep{Janson2019, Janson2021}, the Young Suns Exoplanet Survey (YSES) \citep{Bohn2020a, Bohn2020, Bohn2021}, the SCExAO/CHARIS HGCA survey \citep{Currie2020b}, and the Imaging Survey for Planets around Young stars (ISPY) \citep{Launhardt2020}.
\end{enumerate}

In aggregate, these surveys typically detect $\sim$1 substellar or exoplanet companion per 100 stars \citep[e.g.~meta-analyses~by][]{Bowler2016, Bowler2018}, although targeted surveys (e.g. focused on accelerating stars) may produce higher yields.

Most of the direct imaging surveys described above have focused on planet detection in the near-IR, especially near $H$ band.   However, some surveys -- e.g. LEECH and ISPY \citep{Heinze2010,Skemer2014b,Launhardt2020} -- have searched in the thermal IR, in particular the $L_{\rm p}$ passband (3.78 $\mu m$).   For a given residual wavefront error, the Strehl ratio is significantly higher in the thermal IR than in the near IR.  As shown in Figure \ref{fig3}, contrasts are more advantageous in the thermal IR for both young and (especially) intermediate-aged jovian planets.
The planets $\beta$ Pic b and HR 8799 e were chiefly discovered from thermal IR data; discovery papers for other planets (e.g. PDS 70 b) also included thermal IR imaging \citep{Lagrange2009,Marois2010b,Keppler2018}. The thermal IR has lower background star contamination compared to the near-IR. Extreme AO systems can detect planets 100 times fainter than conventional AO counterparts in the near-IR \citep{Macintosh2014}.  However, thus far, extreme AO in the near IR has a smaller (factor of $\sim$ a few) advantage over a conventional AO system operating in the thermal IR 
\citep[e.g. see GPI/H band vs. Keck/thermal IR detections of 51 Eri b;][]{Macintosh2015}.


\subsection{\textbf{Occurrence Rates and Demographics \label{sec:occurrence}} }

The planet occurrence rate (number of planets per star) -- and how that occurrence rate varies as a function of stellar or planetary properties (e.g. host mass, planet mass, orbital semi-major axis, host star metallicity) -- strongly constrains models of planet formation and evolution, particularly as we build up larger sample sizes.  Early demographics studies focused on RV survey results, showing correlations in giant exoplanet occurrence with stellar metallicity \citep{fischer05} and host star mass \citep{johnson07}, and constraining occurrence rates as a function of planet mass and semi-major axis \citep{cumming08}.  In particular, \citet{cumming08} found a power-law dependence of planet period that corresponded to a rising occurrence rate of giant planets as a function of log semi-major axis out to $\sim$2.5 au.

Given the low yield of exoplanets from early direct imaging surveys, detailed occurrence rate measurements proved difficult.  Instead, Generation 1 and 2 survey analyses focused on constraining occurrence rates with assumed mass and semi-major axis distributions, or setting upper limits on the largest semi-major axis for which the \citet{cumming08} power law could be consistent with direct imaging yields \citep[e.g.][]{Masciadri2005, Lafreniere2007b, Nielsen2010}.  These early results showed that the occurrence rate of giant planets ($\gtrsim$1 M$_{J}$) in wide orbits ($\gtrsim$10 au) was relatively low ($\lesssim$10\%), and that the rising power law of \citet{cumming08} cannot extend beyond tens of au and still be consistent with the lower occurrence rates found by direct imaging surveys.

Starting with the first imaged planet discoveries in 2008 \citep{Marois2008,Lagrange2009},
it became possible for surveys to directly measure occurrence rates, rather than simply placing upper limits.  A planet of a given mass and separation should be easier to image around a low luminosity, low-mass star than a higher luminosity, high-mass star.  However, 
hosts (e.g. HR 8799, $\beta$ Pic) of directly-imaged wide-separation giant planets were typically more massive than the Sun ($\sim$1.5--2 M$_\odot$).  

As Generation 3 surveys reached higher contrasts than their predecessors, giant planets 
with masses down to $\sim$2 M$_{Jup}$ became detectable at separations above $\sim$50 au (beyond $\sim$1\arcsec{}) for nearby moving group stars.  Yields from these surveys remained low, however, typically 0-2 planets per survey.  Given the limited number of nearby young stars, the same planet was often detected by multiple surveys (e.g. $\beta$ Pictoris b, $\kappa$ And b).
Occurrence rate analysis confirmed the relatively low upper limits implied by Generation 1 and 2 surveys, with $\lesssim$10\% of nearby, young stars hosting a wide-separation giant planet.  Analysis of occurrence rate as a function of stellar mass from these Generation 3 surveys confirmed that wide-separation giant planets do not follow an extrapolation of the \citet{cumming08} power laws.  Wide-separation jovian planets were also found to be more common around higher-mass stars \citep[e.g.][]{Lannier2016}.   Wide-separation giant planet occurrence rates may also be higher for stars with Kuiper belt-like debris disks than diskless stars \citep{Meshkat17}.

\begin{figure}[!h]
 \vspace{-0.05in}
   \includegraphics[angle=0,width=\columnwidth]{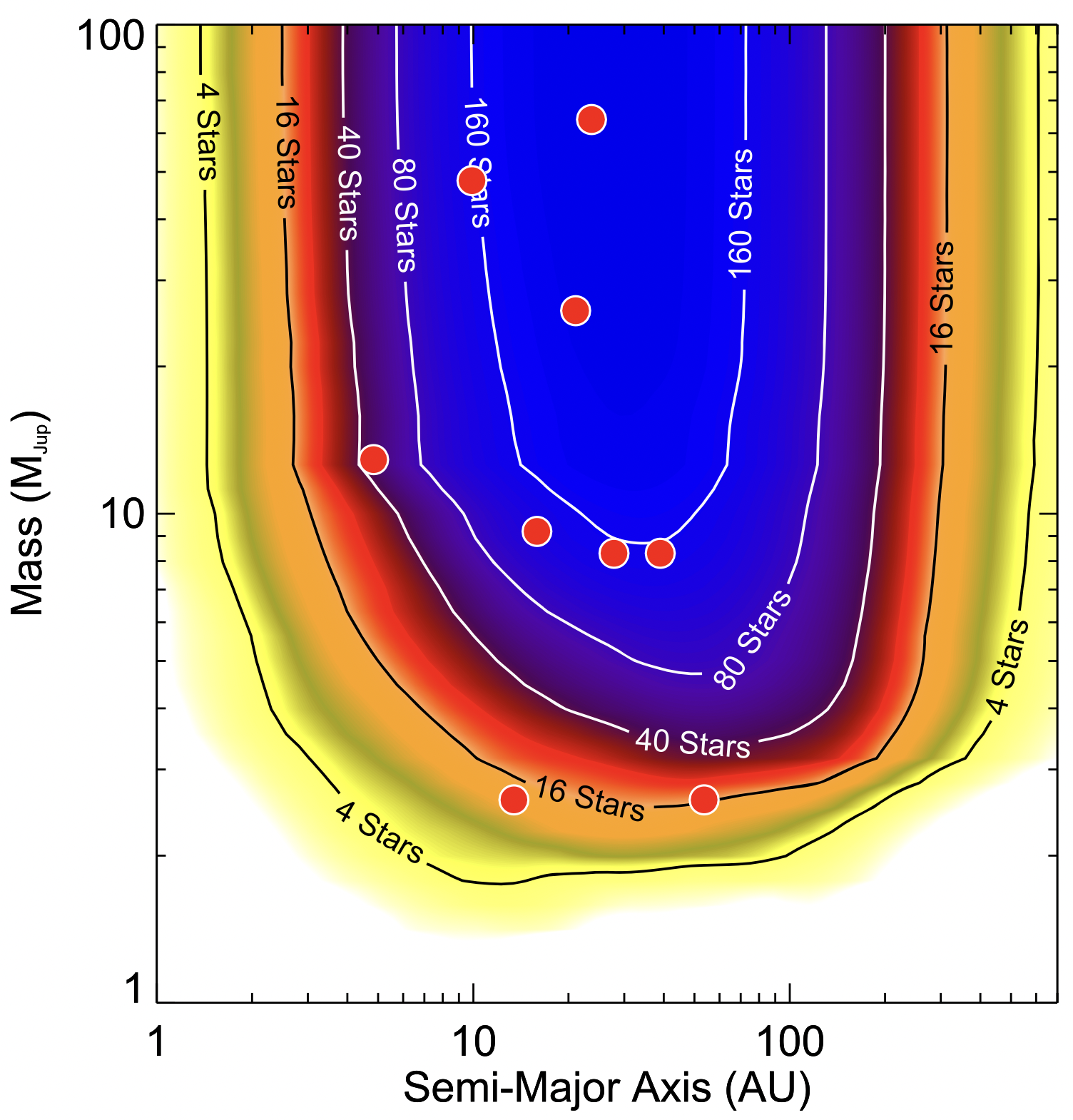}
  \includegraphics[angle=0,width=\columnwidth]{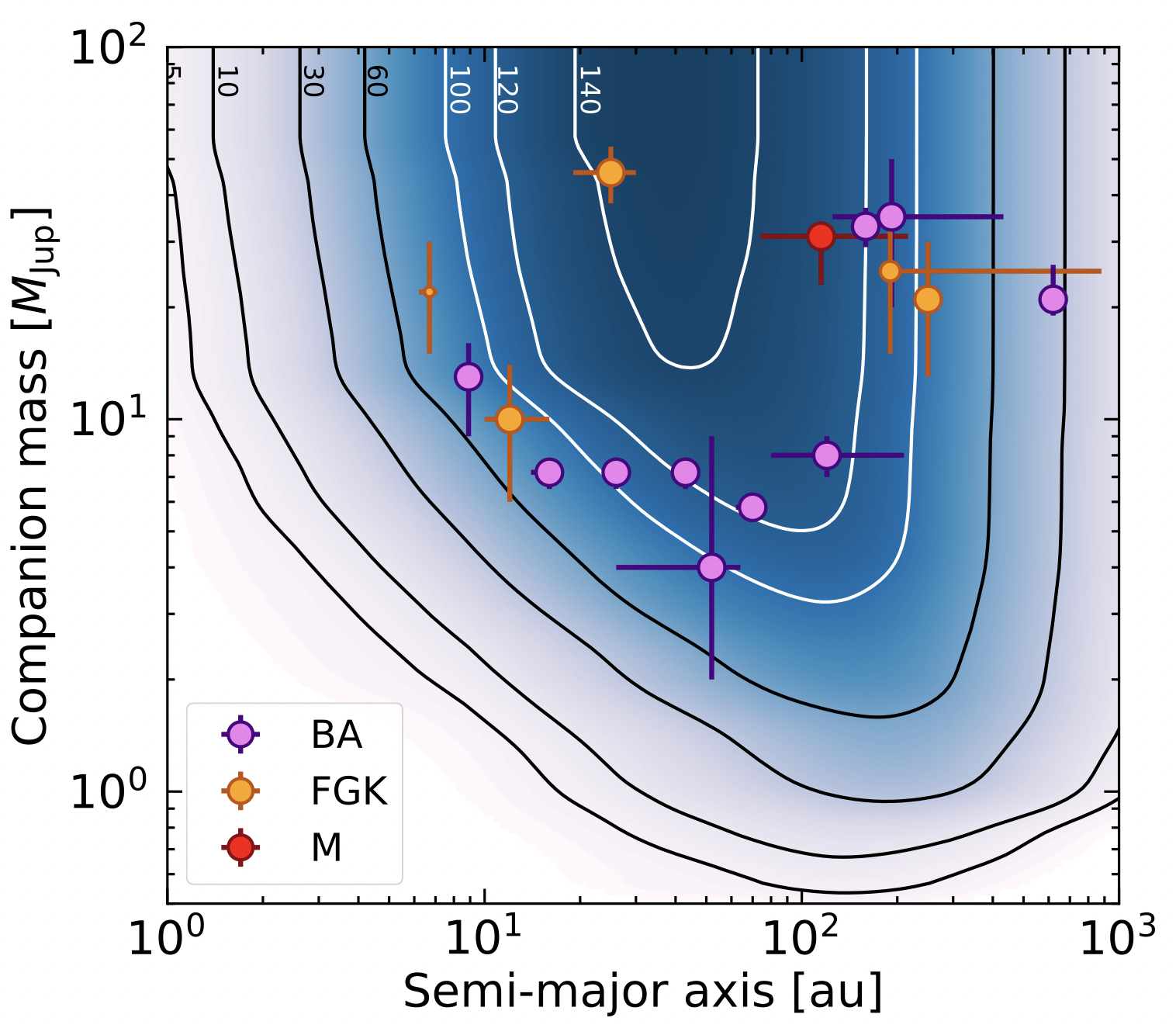}
 \vspace{-0.35in}
\caption{\small Tongue plots (also called depth of search plots, \citealt{lunine2008}) showing detections and sensitivities to substellar companions for the GPIES (top) and SHINE (bottom) surveys.  Unlike previous generations of surveys, these surveys could detect giant planets as close as $\sim$10 au from their parent star  (Figure 4 of \citealt{Nielsen2019}, Figure 1 of \citealt{Vigan2021}).}
\label{fig:tongue}
\end{figure}

Generation 4 surveys have significantly higher sensitivity to giant planets between $\sim$10-50 au compared to previous generations of surveys, due to improved corongraphs and AO.  These surveys had larger survey sizes, and detected planets at separations and contrasts not reachable with previous instrumentation (e.g. 51 Eridani b at $\sim$2 M$_{\textrm J}$ and $\sim$13 au).  Higher yields of substellar companions meant more detailed demographic studies became possible, with the first analysis of GPIES data detecting 3 brown dwarfs and 6 giant planets, and the first SHINE demographic analysis finding 8 giant planets and 7 brown dwarfs (Figure~\ref{fig:tongue}).  These surveys confirmed at $\sim$3$\sigma$ that wide-separation giant planets (2-13 M$_{\textrm J}$ at 10-100 au) had higher occurrence rates around higher-mass ($\gtrsim$1.5 M$_\odot$) stars compared to Solar-type stars \citep{Vigan2021,Nielsen2019}.

\begin{figure}[!h]
\includegraphics[angle=0,width=\columnwidth]{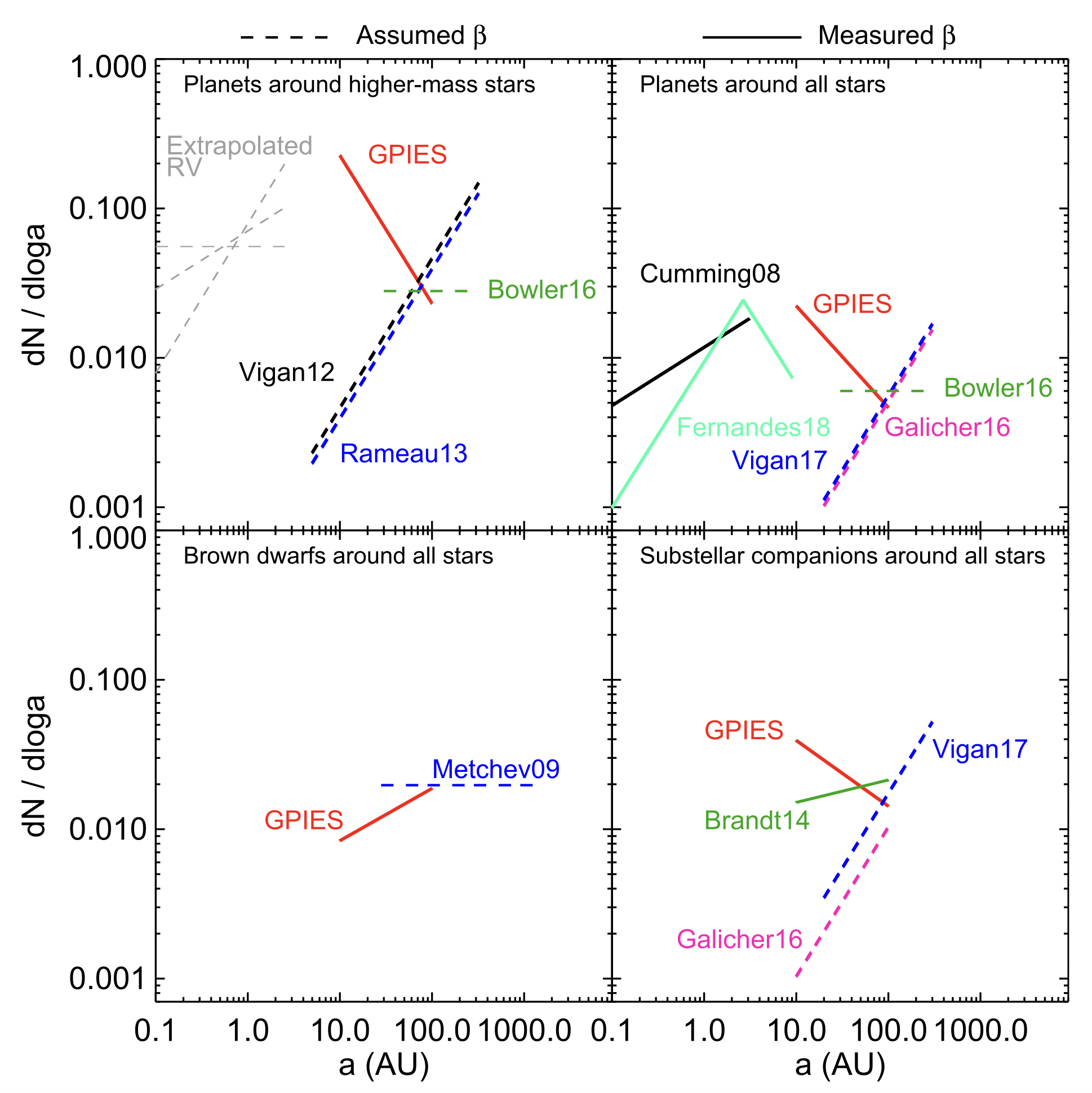}
 \vspace{-0.325in}
\caption{\small Occurrence rate of giant planets (5-13 M$_{\textrm{Jup}}$), brown dwarfs (13-80 M$_{\textrm{Jup}}$), and substellar companions (5-80 M$_{\textrm{Jup}}$) from multiple direct direct imaging and radial velocity surveys.  Giant planets around higher-mass stars (top left panel) and around all stellar types (top right panel) are consistent with a peak in the distribution near the snow line ($\sim$3-10 au), with falling occurrence rates toward larger and smaller separations.  Current demographics results provide robust evidence that giant planets are more common around higher-mass stars, and weaker evidence that giant planets and brown dwarfs follow different underlying distributions.  (Figure 8 of \citet{Nielsen2019})}
\label{fig:sma_pl}
\end{figure}

These higher planet yields also allowed constraints the occurrence rate of wide-separation giant planets as a function of semi-major axis, planet mass, and stellar host mass. Power law fits to the occurrence rates of wide-separation giant planets as a function of these parameters show that the giant planet occurrence rate at 10-100 au rises with increasing stellar mass, decreasing planet mass, and decreasing semi-major axis \citep{Nielsen2019}. Comparing the power-law fits for giant planets brown dwarfs suggests that planets and brown dwarfs are not drawn from the same population at wider separations, consistent with separate formation mechanisms \citep{Nielsen2019}.  When combined with results from RV surveys at smaller separations, the distribution of giant planets as a function of semi-major axis also appears (Figure~\ref{fig:sma_pl}) consistent with a peak at 3-10 au, near the snow line, with falling occurrence rates at smaller and larger separations.

RV population studies like \citet{Fernandes2019} and \citet{Fulton2021} are now sensitive to giant planets beyond the snow line.  
Both studies find a peak in the giant planet occurrence rate near the snow line.   The frequency of planets with masses of 5-14 M$_{\rm J}$ between 10-100 au extrapolated from the \citet{Fulton2021} RV study is in excellent agreement with the same measurement (in the same mass and separation bins) from the Gemini-GPIES survey \citep{Nielsen2019}.
%
Taken together, results from imaging and RV imply a giant planet occurrence rate (across all separations) of $\sim$20\% \citep{Fulton2021} for Sun-like stars, and potentially significantly higher for giant planets around higher-mass stars \citep{Nielsen2019}.




\subsection{\textbf{Testing Formation Mechanisms with Direct Imaging Exoplanet Demographics}}
\label{subsec:testing-formation}
The demographics of directly-imaged planets can provide powerful constraints on their likely formation mechanism. There are two main pathways by which giant planets may form:
\begin{enumerate}
    \item \textit{Core Accretion} -- a rocky core accretes first from planetesimals, followed by a gas envelope \citep[cf.][]{Mizuno1980, Pollack1996,Alibert2004,Ida2013}. The proto-planet's position within the disk will eventually determine its final mass; for instance, Jupiter, closer to the peak density of the Sun's protoplanetary disk, managed to accrete a considerably larger gas envelope than Neptune.  Pebble accretion can enhance the protoplanetary accretion rate, enabling the formation of massive planets far from their star \citep{Johansen2017}.
    \item \textit{Gravitational Instability within a Disk} -- If a disk is sufficiently massive to experience self-gravity, parts of the disk can undergo fragmentation and quickly form giant planets \citep[cf.][]{Kuiper51,Boss1998, Vorobyov2013,Forgan2013,Forgan2015}.  \citet{Kratter2010} note that in most cases, these fragment-formed companions will grow further into substellar or low mass stellar companions -- only the lowest mass members of this cohort remain planetary.  
\end{enumerate}

For very low mass primaries (e.g. brown dwarfs), binary-like formation can also produce planetary mass companions \citep{Offner2010}.  This is the probable formation mechanism for the brown dwarf - planetary mass binary 2M1207 B \citep{Chauvin2004}, for instance. 

\begin{figure}[h]
  \vspace{-0.10in}
  \includegraphics[angle=0,width=\columnwidth]{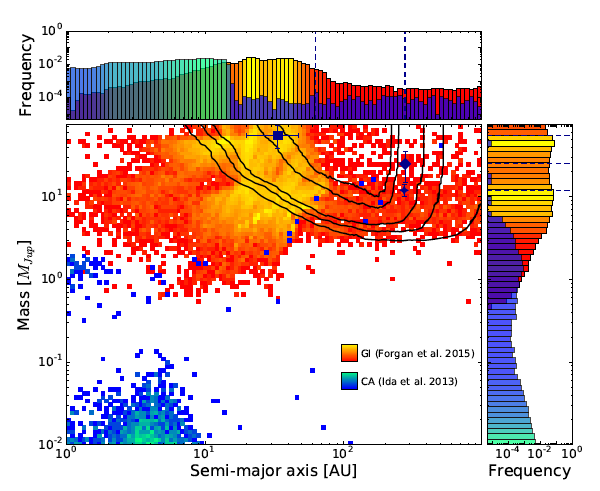}
 \vspace{-0.30in}
\caption{\small Plot reproduced from \citet{Vigan2017} comparing simulated population of companions formed via core accretion \citep[blue/green][]{Ida2013} and gravitational instability \citep[red/yellow][]{Forgan2015}.  Lighter colors represents higher occurrence rates.  Histograms of the distributions for each model population in semi-major axis (au) and mass (M$_J$) are shown on the top and side respectively.  Mean detection probabilities and the detected companions from the \citet{Vigan2017} sample are overplotted as black contours and points respectively.  Both of these simulated populations include multiple planetary embryos per system; both model populations undergo an evolution phase where they can scatter or be scattered by other objects in their system.  Disk instability predominantly forms massive companions (1-100 M$_{J}$) far from the star, whereas core accretion dominates within 10 au and predominantly produces lower mass companions.}
\label{fig:CA_GI_comparison}
\end{figure}

Giant planets observed at wide separations experience significant migration or scattering either during their formation or immediately afterwards \citep{Morbidelli2009,Johansen2017,Forgan2018, Emsenhuber2021}.  Exoplanet demographics from large surveys provide critical tests as to which of these formation and migration mechanisms are dominant, as different formation mechanisms place planets in very different parts of their stellar system. Gravitational instability
predicts giant planets generally at wider separations, with more higher-mass planets than lower-mass ones, and no strong dependence on stellar mass. Core accretion, on the other hand, predicts more giant planets closer to the star
and more lower-mass planets than higher-mass ones.  Figure~\ref{fig:CA_GI_comparison} shows a comparison of simulated planet populations formed via core accretion \citep{Ida2013} against a population formed via disk instability \citep{Forgan2015}.  Both of these studies simulate multiple planetary embryos per system; after an initial formation phase, both model populations undergo an evolution phase where they can scatter or be scattered by other objects in their system.  Obviously this is comparing just two simulation codes and a wide variety of similar populations exist in the literature \footnote{See Chapter 20 from this volume, "Planet Formation Theory in the Era of ALMA and Kepler: From Pebbles to Exoplanets".}.  However, while the exact population simulated by each team and model considered will vary, the hallmarks of each formation mechanism remain.

In the last decade, a number of large direct imaging surveys have used results from these types of population synthesis codes to inform their statistical analyses.
Informed by theoretical results, \citet{Janson2011, Janson2012, Rameau2013a} define regions in mass-semi-major axis space where gravitational instability formation would be possible. Given non-detections in these mass - semi-major axis regions in their surveys, they were then able to set upper limits on the frequency of giant planets formed via gravitational instability.  
\citet{Vigan2017} found that the disk instability population could describe the companions detected in their survey, but only if formation by disk instability was relatively rare. 
Comparisons of core accretion population synthesis models to direct imaging surveys before the advent of Gemini-GPI and SPHERE-SHINE (e.g. \citealt{Nielsen2010,Chauvin2015,Vigan2017}) did not yield meaningful constraints
as most of the predicted core-accretion planets were undetectable for these surveys.  

Statistical results from the Gemini-
GPIES and SPHERE-SHINE surveys \citep{Nielsen2019,Vigan2021} suggest that core accretion is the dominant mechanism forming directly imaged giant planets.
\citet{Nielsen2019} compared the power-law fit to 10-100 au giant planets detected in the GPIES survey to predictions
of both gravitational instability and core accretion.
While the brown dwarf companion population was well-described
by gravitational instability, \citet{Nielsen2019} found giant planets closer to their star in comparison to the brown dwarf companions, with more lower-mass planets than higher-mass planets, and with a higher frequency of planets around higher mass stars in the samples -- all hallmarks of core accretion. Fitting both disk instability
and core accretion populations\citep{Forgan2018,Mordasini2017} simultaneously to the detected
companions and contrast limits of a sample of FGK
stars observed as part of the SPHERE-SHINE F150 sample,
\citet{Vigan2021} qualitatively found that the contribution
of the core accretion simulated population is larger than the
contribution of the gravitational instability simulated population,
suggesting that core accretion is the dominant formation
mechanism for this sample as well. 

Directly imaged giant exoplanets are numerous enough to study the demographics of companions themselves, rather than only from detection statistics across a survey. For instance, \citet{Bowler2020} considered eccentricities of 27 long-period giant planets and brown dwarf companions between 5 and 100 au, derived from their own orbital fits to literature and additional Keck astrometry. They tentatively found ($\sim$2-$\sigma$) that their popuation dominated by giant planets (2-15 M$_J$) had somewhat lower orbital eccentricities than a brown dwarf-dominated population (15-75 M$_J$), which displayed a broad distribution of eccentricities with a peak in eccentricity between 0.6 and 0.9.   These population-wide differences in eccentricity may suggest different formaiton mechanisms for the two populations: formation within a disk for directly-imaged exoplanets vs. cloud fragmentation (binary-like formation) for brown dwarf companions.

\section{\textbf{The Future of Direct Imaging} \label{sec:future}}
Over the next 5 years, new technology applied to exoplanet imaging instruments from the ground will close the gap between the planets we can image and those we have detected indirectly, while JWST will enable the detection of planets undetectable from the ground. By 2030, new facilities should provide the first reflected-light detections and spectral characterizations of mature planets.   Finally, by the 2040s, we should be able to directly detect rocky, habitable zone planets around the nearest M stars and Sun-like stars using ELTs and NASA/ESA flagship missions.

In this section, we project the near, medium, and long-term development of direct imaging through technological innovation.
We describe as well new facilities applying these innovations and the science questions we will begin to answer via
these new capabilities.

\subsection{\textbf{New Tools}}

The hardware enabling current high contrast imaging capabilities described in \S\ref{sssec:atmWFC} -- detectors, deformable mirrors, high performance computing -- will continue to advance and provide deeper contrast levels. Improved coronagraph designs 
will also open access to planets at separations near the telescope diffraction limit. For ground-based extreme AO these incremental gains are being implemented in instruments on current large telescopes, either as part of a continuous development process \citep[SCExAO,~MagAO-X,~ KPIC][]{Guyon2020,Close2018,Males2020,Echeverri2019,Mawet2021}, or major upgrades to existing facility instruments \citep[GPI-2.0,~ SPHERE+][]{Chilcote2020,Boccaletti2020}. The same hardware technologies are at the core of future ground and space instrument designs, and will provide science capability improvements beyond the already significant gains expected from re-deploying current instruments on larger telescopes.


Advances in low-noise detectors and optical fiber technologies will enable exoplanet spectroscopy at high spectral resolution. 
Early on-sky demonstration \citep[KPIC,~REACH,][]{Mawet2021,Kotani2020} are paving the way for more capable future systems \citep[MODHIS,][]{Dumas2020}. Polarization differential imaging systems optimized for high contrast imaging \citep[ZIMPOL,~VAMPIRES,][]{Schmid2018,Norris2020}, when deployed on largest telescopes, will be able to reveal partially polarized starlight reflected by exoplanet atmospheres. 

Upcoming transformative advances in wavefront control, calibration, and starlight suppression will yield order-of-magnitude level improvement in contrast, or enable new measurement capabilities.   These include: 
\begin{itemize}
    \item {\bf Sensor fusion and predictive control} will optimize how real-time wavefront sensing telemetry is used to estimate wavefront errors. In predictive control, current and past measurements are combined to extrapolate future optimal AO commands, reducing time lag errors and improving sensitivity \citep{males2018}. By combining information from multiple sensors operating at separate wavelengths or in different optical configurations, the precision and accuracy of wavefront estimates should also be further improved \citep{Guyon2020}. 
    \item {\bf PSF calibration from real-time telemetry} holds the potential to numerically subtract starlight from science images at the photon noise level \citep{Gilles2011,Guyon2020}.
    \item {\bf Coherent modulation} can provide a real-time estimate of starlight, which can then be numerically removed from science images. The modulation can be performed spatially in a self-coherent camera \citep{Baudoz2006}, or temporally with coherent differential imaging.
    \item {\bf Photonic technologies} provide new ways to filter, transport and combine light \citep{gatkine2019}. Applications include fiber-fed spectroscopy, wavefront sensing, and nulling interferometry \citep{Jovanovic2020,Norris2020,Martinod2021}.
    \item {\bf Laser beacons} can improve wavefront control performance beyond the fundamental limits imposed by starlight photon noise. The approach has been successfully used for non-extreme AO ground-based observations, and is now being considered for both ground-based and space based \citep{Douglas2019} high contrast imaging applications.
    \item {\bf Starshades} can provide high contrast imaging without requiring exquisite optical stability. They are particularly well-suited for long exposure wide spectral coverage spectroscopy of exoplanets with space-based telescopes \citep{Gaudi2021}.
\end{itemize}

\begin{figure}[ht!]
  \includegraphics[angle=0,width=0.975\columnwidth,trim=0mm 0mm 0mm 35mm,clip]{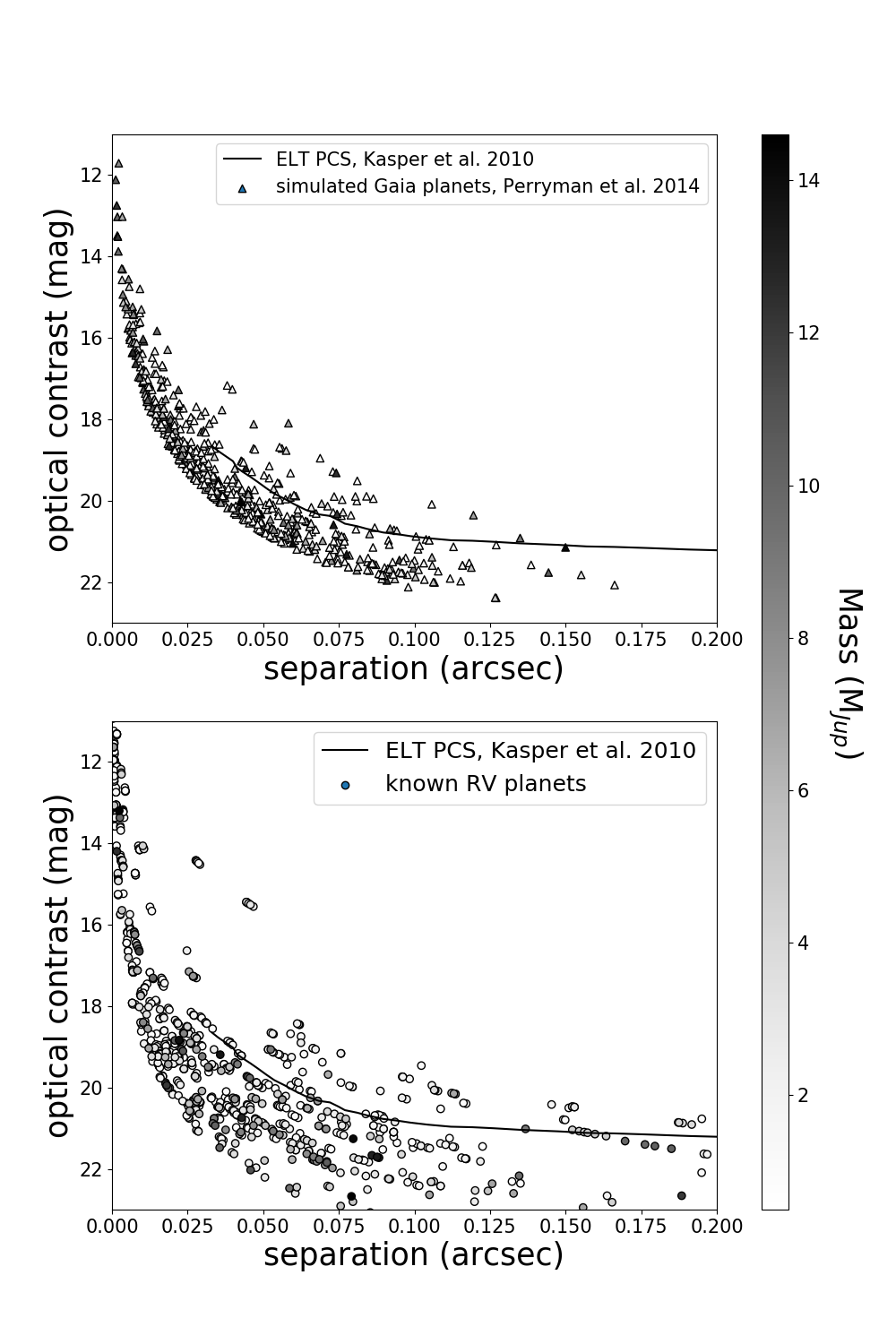}
 \vspace{-0.375in}
\caption{\small Detectability of mature giant planets in optical reflected light using an extreme AO planet-finding camera at an ELT.  \textit{Top:} simulated Gaia detected planets \citep[drawn~from~the~distribution~in][]{Perryman2014}, \textit{Bottom:} known RV planets.  The contrast curve shown is a prediction for the ELT PCS planet-finder \citep{Kasper2010}, but is representative of the performance expected for similar instruments on ELTs.  Known RV planet properties draw from the NASA Exoplanet Archive.   Contrasts are computed at the planets' brightest phases.
}
\label{fig:ELTS_RV_Gaia}
 \vspace{-0.15in}
\end{figure}

\subsection{\textbf{New Instruments and Facilities}}
\textit{James Webb Space Telescope (2022)} -- Launched in December 2021, the cryogenic, IR-optimized \textit{JWST} may significantly expand the diversity of imaged exoplanets and will break new ground with exoplanet characterization.  The 6.5m JWST will have poorer resolution and performance at the smallest angular separations than most leading exoplanet-imaging platforms.  However, JWST will have unrivaled sensitivity in the thermal/mid IR (3--20 $\mu m$) compared to ground-based instruments, which must contend with bright sky emission: JWST can discover cold sub-Jupiter mass planets at wide separations around the nearest stars \citep{Beichman2010, Carter2021, Hinkley2022}.   JWST Early Release Science data of the young exoplanet system HIP 65426b suggests that JWST is exceeding its nominal predicted performance by a factor of up to 10 \citep{Carter2022}, enabling the detection of planets down to masses of 0.3 $M_{\rm J}$ beyond 100 au.  Thermal IR IFS data will be free of telluric contamination.

\textit{Next-generation extreme AO on 6-10m telescopes (2022-2026)} -- Over the next few years, upgraded systems should improve planet-to-star contrasts by a factor of 10--100$\times$ at small separations, reaching contrasts floors nearing 10$^{-7}$, and improving performance for optically fainter stars.  SCExAO on Subaru will undergo major hardware and software upgrades, improving upon its current implementation of advances such as predictive control incorporating ones like real-time telemetry calibration, coherent modulation, and photonics technology \citep{Guyon2020,Currie2020a}.   MagAO-X \citep{Males2020} will leverage recent hardware and algorithm advances to push high contrast performance in the optical. Upgrades to GPI and SPHERE will likewise feature improved wavefront sensing capabilities \citep[e.g.][]{Chilcote2020,Boccaletti2020}.  

\textit{Roman-CGI} (2026) -- The Coronagraphic Instrument on the Roman Space Telescope will provide the first test of focal-plane wavefront control and advanced coronagraphy in a space-borne environment \citep{Spergel2013}.   CGI's benchmark performance is $\sim$10$^{-7}$: comparable to the best contrasts expected from ground-based extreme AO on 8-10m telescopes in the next 5 years.   However, laboratory tests simulating a space environment suggest that CGI could reach contrasts as deep as $\sim$10$^{-9}$: sufficient to image true Jupiter analogues in reflected light \citep{shi17,Spergel2013}.  The potential success of CGI's technological demonstration may justify a follow-on science program focused on detection and spectral characterization of reflected-light gas giant planets.

\textit{Planet imaging with \textit{Extremely Large Telescopes} (2030s)} -- The large apertures and 
excellent sky conditions at the sites of the upcoming \textit{European Extremely Large Telescope} (E-ELT; 39m; Cerro Armazones), \textit{Thirty Meter Telescope} (TMT; 30m; preferred site is Maunakea), and the \textit{Giant Magellan Telescope} (GMT; 24.5m; Las Campanas) make exoplanet imaging with ELTs a critical science case.  All three telescopes have approved or planned high-contrast imaging instruments covering the red optical to the mid-IR.   

Mid-IR instruments -- e.g. TMT/MICHI \citep{Packham2018}, E-ELT/METIS \citep{Brandl2021} -- use a single high-order AO correction plus a coronagraph to achieve expected contrasts of $\sim$10$^{-7}$ at 3--10 $\mu m$ capable of imaging warm Neptune or larger-sized planets around many nearby stars and exo-Earths around a dozen or so solar-type stars \citep{LopezMorales2019}.   Those operating in the red optical to near-IR -- e.g. TMT/PSI \citep{Jensen-Clem2021} and E-ELT/PCS \citep{Kasper2021} -- benefit from a second AO correction in series to achieve expected contrasts of $\le$10$^{-8}$, sufficient to image rocky, habitable zone planets around nearby M dwarfs.

\textit{Space-borne high-contrast imaging flagship missions (2040s+)} -- The Astro 2020 Decadal Survey ranked a large-aperture, direct imaging flagship mission -- similar to the proposed HabEx and LUVOIR concepts \citep{Gaudi2021,LUVOIR2019} -- as a top space-based priority.  The mission will employ advanced wavefront control and coronagraphy, building upon technical developments from Roman-CGI and the ground and yielding 10$^{-10}$ contrasts.  In some designs, it would fly with a starshade capable of delivering even deeper contrasts and higher SNR detections of exo-Earths identified from coronagraphy.   The European Space Agency is considering a mission -- LIFE -- to detect and spectrally characterize exo-Earths through high-contrast interferometry \citep{Quanz2021}.






\begin{figure*}[h]
  \includegraphics[angle=0,scale=0.6]{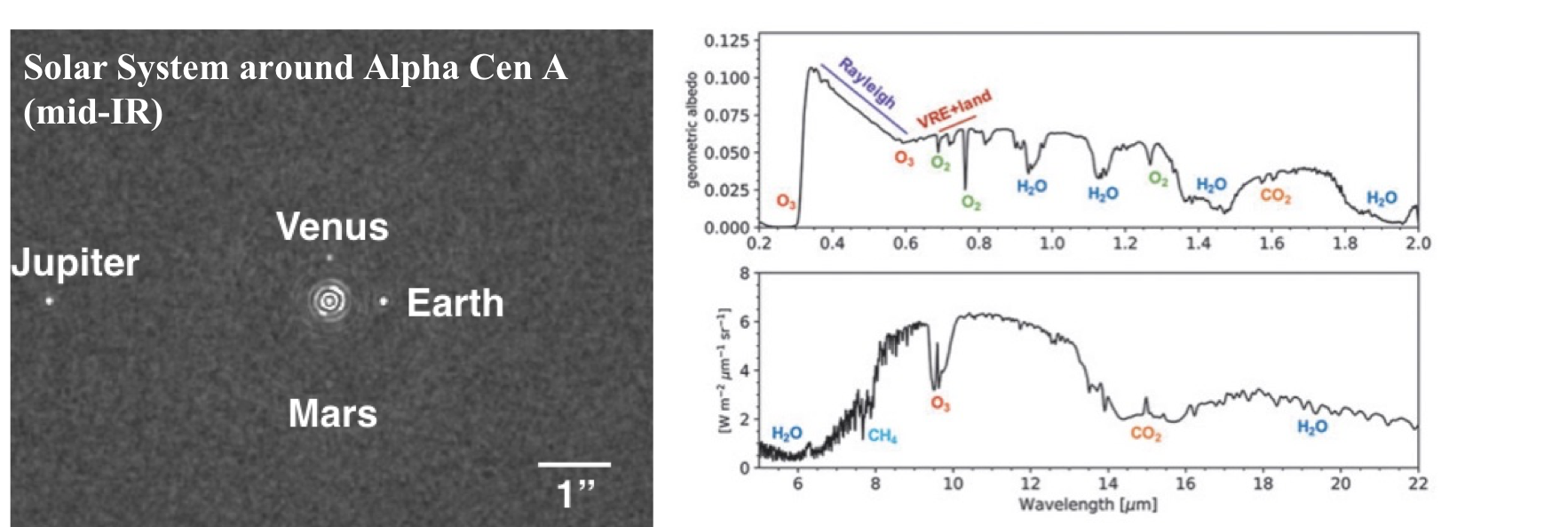}
 \vspace{-0.30in}
\caption{\small Detecting and characterizing exo-Earths with future facilities (from \citealt{LopezMorales2019}).  (left) Simulated image of solar system around $\alpha$ Cen A imaged at 10 $\mu m$ with an instrument like TMT/MICHI.   In several hours, an Earth and Jupiter are visible: in 100 hours, 4 planets are visible.  (right) Biosignatures in the optical and mid IR probed by a space coronagraphic mission like HabEx/LUVOIR and from the ground with ELTs, respectively \citep{Schwieterman2018}.}
\label{fig:exoearth}
 \vspace{-0.10in}
\end{figure*}

\subsection{Future Science Focuses}

\textit{The next 5 years: young giant exoplanet atmospheres at $\lambda$ $>$ 5 $\mu$m with JWST} -- JWST is already enabling the first exploration of directly imaged exoplanet atmospheres at $\lambda$ $>$ 5 $\mu$m.  Especially past 10 $\mu$m, directly imaged exoplanet as well as brown dwarf atmospheres are almost completely unexplored. Pre-JWST, the only existing $>$10 $\mu$m spectra for any substellar object are for a handful of L and T field brown dwarfs from the Spitzer cryogenic mission \citep{Cushing2006, Suarez2022}. 

The recently-presented JWST/NIRSPEC spectrum for L7, planet-mass companion VHS 1256 b demonstrates JWST's promise for atmospheric characterization \citep{Miles2022}.   The spectrum reveals numerous molecules -- H$_{\rm 2}$O, CH$_{\rm 4}$, CO, CO$_2$, and K -- at a wide range of wavelengths, including those poorly accessible from the ground.
The $\sim$10 $\mu$m spectral region is particularly diagnostic of atmospheric properties.  The presence of silicate emission at 10 $\mu m$ directly confirms silicate clouds in VHS 1256 b's atmosphere \citep{Luna2021, Miles2022}, the existence of which has previously been inferred by modeling, but not directly detected.  Future observations with JWST will confirm whether this feature (and hence direct evidence for clouds) is commonly found in other directly imaged exoplanet spectra.  Longer wavelengths may also better distinguish between different theoretical models \citep[e.g.][]{Skemer2012,Skemer2014}.  JWST may also detect thermal emission from sub-Jupiter mass planets $>$30 au from their host stars and sub-Saturn mass planets beyond $>$50 au \citep{Beichman2010, Carter2021}. 

\textit{The next 5 years: ground-based advances} -- Next-generation extreme AO systems will achieve contrasts capable of detecting young Jupiters and intermediate-aged superjovian-mass planets at Jupiter-like separations.   Should contrasts at or slightly below 10$^{-7}$ be reached, these new instruments might be capable of providing the first images of planets in reflected light.   

Direct imaging detections of significant numbers of super-Earth or Earth mass planets will have to wait for the advent of ELTs and new space telescopes.  However, imaging the known habitable zone planet Proxima Centauri b \citep{Anglada2016} might be possible with 8-m class telescopes and substantial dedicated observing time. For example, \citet{Lovis2017} propose targeting this planet by combining VLT/SPHERE with the ESPRESSO high resolution spectrograph. They estimate that 20-40 nights of observations would yield a 5-$\sigma$ detection of the planet and that 60 nights of observations would yield a 3.6-$\sigma$ detection of the biosignature gas O$_2$.

\textit{The next 10 years: Imaging and characterizing mature giant planets} --
With the advent of ELTs on the ground and dedicated space-based missions such as CGI, direct imaging of exoplanets will expand from detecting young planets via thermal emission in the infrared to detecting mature planets via reflected light in the optical. 
Indirect methods like astrometry and RV will provide a sample of solar-system age giant planets which can then be imaged via their reflected optical light using exoplanet imaging cameras on ELTS such as PCS on the E-ELT \citep{Kasper2021} and PSI on the TMT \citep{Guyon2018}. 
For example, Figure~\ref{fig:ELTS_RV_Gaia} compares a predicted ELT planetfinder reflected light contrast curve \citep{Kasper2010} with a simulated population of Gaia-detected giant planets from \citet{Perryman2014} and known RV planets.  Extreme AO planet-finding cameras on ELTs will directly image and spectrally characterize dozens of solar-system age giant planets.   


These new facilities and instruments will also enable deeper characterization of the current cohort of young, directly imaged exoplanets.  In particular, these instruments will incorporate high resolution spectroscopic capability alongside extreme adaptive optics and state-of-the-art coronagraphs \citep{Kasper2021}. This will enable high-precision spectroscopic and rotational studies, including the possibility of directly reconstructing top-of-atmosphere (TOA) cloud structure for a handful of young giant exoplanets \citep{Snellen2014, Crossfield2014b}. A particularly exciting future opportunity will be direct mapping of giant planet atmospheric structure via the Doppler imaging technique, which uses time-resolved, high-resolution spectroscopy (R$>$40000) to produce a high SNR composite line profile, then uses maximum entropy methods to reconstruct TOA structure from that line profile \citep{Vogt1987}.  To date, only one extremely bright brown dwarf, Luhman 16B, has been Doppler-mapped \citep{Crossfield2014a} and only 2-3 others are sufficiently bright for this technique with 8-m class telescopes \citep{Crossfield2014b}. Extremely large telescopes will enable Doppler imaging of dozens of brown dwarfs and a few directly imaged planets \citep{Crossfield2014b}. 

\textit{The next 20 years: Imaging and characterizing exo-Earth twins} --
ELTs will play an important role in imaging and characterizing habitable zone rocky exoplanets around a range of primary star masses.  Instruments like E-ELT/PCS and TMT/PSI will image and spectrally characterize a few dozen habitable zone exoplanets around M stars in red optical to near-IR reflected light at wavelengths probing key biomarkers \citep[e.g. O$_{\rm 2}$ at 1.27 $\mu m$;][]{LopezMorales2019}.  Very deep mid-IR data for the nearest $\sim$10--15 Sun-like stars can detect thermal emission bracketing the $\sim$10$\mu m$ ozone feature (Figure \ref{fig:exoearth}, left panel). 

Using a combination of advanced wavefront control, coronagraphy, and/or a starshade,
 space-based missions like HabEx, LUVOIR, and LIFE should be capable of imaging and spectrally characterizing exo-Earths around several tens of Sun-like stars.
The NASA missions will split their exo-Earth finding science program into a shallow coronagraphic search and a deep, focused follow-up of candidate habitable zone planets conducted with coronagraphy (LUVOIR) and/or a starshade (HabEx).   Survey designs are still in development and strongly depend on
the adopted value of $\eta_{Earth}$ -- the frequency of Earth-like planets around Sunlike stars.  Currently, values for $\eta_{Earth}$ are extrapolated from the Kepler transiting planet mission \citep{Dressing2015, Kopparapu2018}, although these $\eta_{Earth}$ values describe only the frequency of small, rocky planets, not necessarily habitable planets.

These missions will provide not just detections but spectra 
to identify biosignatures gases (and hence life) in the atmospheres of these planets (Figure \ref{fig:exoearth}, right panel).  No one biosignature gas by itself can clearly indicate the presence of life, as many biosignature gases, such as $O_2$, can have abiotic sources \citep{Meadows2018}.  Thus, confirming the presence of life on a planet requires detecting a combination of gases that do not appear together in equilibrium state -- disequilibria in this case is driven by the presence of life \citep{Meadows2018}.  To ensure coverage of a wide range of biosignature gases, these missions will deliver spectra over a broad wavelength coverage: e.g. from near-ultraviolet to near-IR for LUVOIR. 

The study of exoplanets has yielded many surprising results, such as the existence of hot Jupiters, the ubiquity of super-Earths, and the extremely red colors of young giant exoplanets, thus, caution is merited both in our predictions of what we may find with future instruments as well as in our interpretation of the data yielded by such instruments.  Considering this, it will likely require detections of key biosignature gases in a statistically significant sample of Earth-like exoplanets to be firmly convinced that we have truly detected life outside our own solar system.   

The detection of habitable zone planets around Sun-like stars through M stars through direct imaging will be necessary to provide our first steps towards truly providing a context for our own Earth.  
M star planets encounter a very different radiation environment than planets around FGK stars and are predicted to be tidally locked, which may pose significant obstacles to true habitability \citep{Shields2019}.  To understand the frequency of life, it is then key to detect and characterize both M star habitable zone planets and planets more like our own -- the only planet we know to date to truly be inhabited.  However, detection and characterization of true exo-Earth twins (defined here as planets with similar mass as the Earth in the habitable zone of Sunlike stars) pose a challenge for transit and RV techniques.  If viewed as an exoplanet, the Earth would only appear to transit once a year, thus requiring observations over many years or even decades to reach sufficient S/N in transit spectra to search for biosignatures.  While the sensitivity of current RV instruments such as ESPRESSO and HARPS is sufficient to detect the RV signal of an exo-Earth twin around a Sunlike star, that signal will generally be drowned out by noise from the stellar activity of the host star \citep{Haywood2020,Langellier2021AJ}.  
Thus, while many M star habitable-zone exoplanets will be discoved first via transit or RV, then characterized with direct imaging, 
direct imaging 
is more naturally suited for both the discovery and characterization of true Earth twins.\\

This review thus concludes where it began, with the prospects for
identifying, confirming, and characterizing a true Earth twin around a nearby Sun-like star via direct imaging.  The first direct images of exoplanets were published less than 15 years ago.  Since then, exoplanet direct imaging has advanced rapidly, constraining the orbits, masses, atmospheres, population demographics and formation mechanisms of young giant planets.  We forecast the rapid technical and scientific progress driven by the many and varied exciting new telescopes and instruments coming online in the next 20 years. 

With these new capabilities, direct imaging and spectroscopy are highly likely to yield the first detections of true Earth twins and following that, the first detection of life outside the solar system -- possibly the first substantive steps towards knowing whether or not we are alone in the universe.

\textbf{Acknowledgments.} 
This work has benefitted from The UltracoolSheet, maintained by Will Best, Trent Dupuy, Michael Liu, Rob Siverd, and Zhoujian Zhang, and developed from compilations by \citet{DupuyLiu2012}, \citet{DupuyKraus2013}, \citet{Best2021}, etc.  Discoveries and key insights into the nature of directly imaged extrasolar planets were made from many astronomical sites throughout the world, including Maunakea in Hawai`i, Cerro Paranal, Cerro Pachon, and Las Campanas in Chile, and Mt. Graham in Arizona.   We acknowledge the importance that these sites hold for many in nearby communities from a cultural and/or personal standpoint and support responsible stewardship of the land on which these and other observatories are located.


\bibliographystyle{pp7}

%
%
%

\end{document}